# GW Notes

October 2010 to March 2010

Notes & News for GW science

Editors:

P. Amaro-Seoane and B. F. Schutz

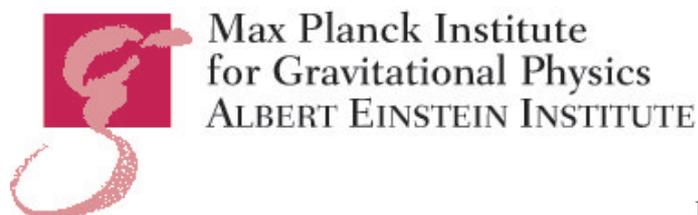

GW Notes was born from the need for a journal where the distinct communities involved in gravitation wave research might gather. While these three communities - Astrophysics, General Relativity and Data Analysis - have made significant collaborative progress over recent years, we believe that it is indispensable to future advancement that they draw closer, and that they speak a common idiom.

*Editorial*

Stellar cusps in galactic nuclei

The way gravity distributes stars around a massive black hole is a problem which has been addressed a number of times, by many authors, due to the obvious relevant applications it has not only to stellar dynamics, but also to low-frequency gravitational wave Astrophysics. Recently, Preto and Amaro-Seoane (2010) revisited the classical Bahcall-Wolf problem and extended the study to the situation in which a "strong" mass segregation occurs, i.e. when $N_h M_h^2 < n_l m_l^2$, where $M_h$ ($m_l$) is the mass of the heavy (light) components and $N_h$ ($n_l$) their total initial number.

Conducting a detailed study of different systems with different $M_h$, $N_h$, $m_l$ and $n_l$ by using a direct-summation program, they verified the Bahcall-Wolf solution, where $M_h$ are common and thus dominate the behavior of the system. In the case in which $M_h$ are rare, heavy stars sink to the center by dynamical friction and build there a steeper cusp, which is in agreement with both observations and also previous studies. The implications in terms of gravity waves are straightforward, since a steeper cusp leads to a higher number of candidates to be both extreme- and intermediate-mass ratio inspiral events.

For this GW Notes issue we have asked Miguel Preto (Heidelberg University) to expand this recent work and to advance some new results[1] on how stars distribute around massive black holes for our highlight article.

Pau Amaro-Seoane & Bernard F. Schutz, editors

---

[1] Preto & Amaro-Seoane 2010a, 2010b and Amaro-Seoane, Preto & Freitag 2010

*GW Notes highlight article*

*Stellar cusps*

DYNAMICAL EVOLUTION OF NUCLEAR STELLAR CLUSTERS
I. STELLAR DISTRIBUTIONS AROUND A MASSIVE BLACK HOLE

Miguel Preto
Astronomisches Rechen Institut (ZAH)
Mönchhofstr. 12-14
Heidelberg, Germany
e-mail: miguelp@ari.uni-heidelberg.de

**Abstract**

Mass segregated stellar cusps are the natural configuration for nuclear stellar clusters in steady state with a central massive black hole of mass $M_\bullet \lesssim \text{few} \times 10^6 M_\odot$. Detailed $N$-body and Fokker-Planck studies agree quite well with each other in the description of the *bulk* properties of the stellar distribution of such systems. For typical initial mass functions, the cluster is in the regime of *strong* mass segregation and therefore stellar-mass black holes dominate the spatial density in the innermost regions close to the massive black hole. As a natural consequence, the rates of extreme mass ratio inspirals detectable by LISA will be dominated by the inspiral of stellar-mass black holes.

How stars distribute around a massive black hole

## Contents

How stars distribute around a massive black hole

# 1 Introduction

Massive black holes (MBHs) are thought to be ubiquituous in the centers of galaxies. Although there has been no direct detection of a MBH yet, there is ample observational evidence for the presence of huge mass densities inside the central inner few parsecs of galactic nuclei. During the last decade, a great deal of effort has been made observationally to characterize the physical properties of galactic nuclei and of their putative MBHs (Ferrarese and Ford, 2005). For a very long time, the evidence for the existence of MBHs was essentially based on observations of active galactic nuclei (AGN), in particular of quasi-stellar objects (QSOs). In fact, accretion of matter onto an extremely compact object, or MBH, of mass $\gtrsim 10^6 M_\odot$ was invoked as the most probable mechanism that could power the high luminosities of cosmologically distributed QSOs (Lynden-Bell, 1969 and Rees, 1984). The number density of such active nuclei is much lower (by factors of $\sim 10^{2-3}$) than that of normal, inactive galaxies. Nevertheless, the existence of such active nuclei with MBH at the center led naturally to the prediction that MBHs should reside at least in the center of some galaxies at low redshift. With the advent of the Hubble Space Telescope (HST) not only it became clear that MBHs were pervasive at the center of most (if not all) galaxies, but also a number of phenomenological relations were unveiled: the masses of MBHs appear to be correlated with some physical properties of the host galaxy nucleus, namely its luminosity (Magorrian et al., 1998), its velocity dispersion (Ferrarese and Merritt, 2000 and Gebhardt et al., 2000). Such results constitute strong hints that there is not only an intimate link between the formation (and growth) of MBHs and their host nuclei, but also of a direct relation between the properties of QSOs at high redshift and inactive MBHs at low redshift. (Soltan, 1982) analysed the demography of the active population and showed that the observed energy density emmited by an active nucleus, in a given band, at a given redshit, can be used to estimate the remnant MBH mass density present in inactive nucleus at the present epoch. This argument only requires an assumption about the efficiency of conversion of accreted rest-mass energy into radiation. Such estimates are in remarkable agreement with the observationally derived MBH mass density in nearby galactic bulges (Aller and Richstone, 2002 and Yu and Tremaine, 2002).

Moreover, it has been shown observationally that nuclear stellar clusters (NSCs)—which are revealed by a clear upturn in central surface brightness—exist in the centers of a wide range of Hubble types (Böker, 2008). The frequency with which NSCs occur in galaxies varies among Hubble type from $\gtrsim 50\%$ in early-type spirals up to $\gtrsim 70\%$ in spheroidal galaxies. These estimates are currently lower limits. Furthermore, a number of NSCs that coexist with a MBH have been recently detected suggesting that NSCs around MBHs, like the one in the center of the Milky Way, may be quite common (Graham and Spitler, 2009). NSCs are more luminous than the average globular cluster in the Galaxy—but are as compact. NSCs are the densest and most massive stellar systems known: they are characterized by half-mass radii of $1-5$ pc and stellar masses of $10^{6-7} M_\odot$. Their mass density stands out clearly—at much higher values—from that of galaxy bulges. They show signs of very complex star formation histories, with evidence for several generations of stars and thus of recurrent episodes of star formation. The younger stars are often younger than $10^8$ Myr. NSCs also obey to similar scaling laws with the host galaxy bulge as MBHs, so in all likelihood they have had a common evolutionary history.

The morphological and material content of NSCs may be considerably richer than entailed by simplified models of cuspy, spherical stellar clusters. In the Galactic-center, there is ample observational evidence for the presence of at least one stellar disk (Paumard et al., 2006 and Lu et al., 2009), which is inclined with respect to the Galactic plane. The total stellar mass contained in this disk is weakly constrained, but probably $\lesssim 10^4 M_\odot$, so its dynamics is expected to be nearly-Keplerian. Its inclination suggests it may have formed via a starburst resulting from the infall and subsequent fragmentation of a gaseous cloud (or a collision of two clouds) (Nayakshin et al., 2007 and Bonnell and Rice, 2008). There is further observational evidence towards the occurrence of previous star formation in the Galactic center and of disks in other galaxies (*e.g.* in M31) (Goodman, 2003, Bender et al., 2005 and Levin, 2007).

The prospects for detection of gravitational waves (GWs) from extreme mass ratio inspirals (henceforth EMRIs) with future GW detectors such as the *Laser Interferometer Space Antenna* (LISA) urge us to build a detailed observational knowledge and clear theoretical understanding of sub-parsec structure of galactic nuclei. In fact, EMRI rates will depend strongly on the stellar density of compact remnants as well as on the detailed physics within $O(0.01\text{pc})$ of MBHs with $M_\bullet \lesssim \text{few} \times 10^6 M_\odot$, which is the distance from which these inspiralling sources are expected to originate (Hopman and Alexander, 2005). Mass segregation—which is the natural dynamical outcome for any realistic stellar mass function—will also strongly impact the expected EMRI rates, since it favors the accumulation of the heavier towards the center (Hopman and Alexander, 2006b, Alexander and Hopman, 2009 and Preto and Amaro-Seoane, 2010a).

The NSC in the Galactic-center is—and will remain in the foreseable future—the only nucleus for which one may hope to have detailed measurements of individual stellar orbits. The innermost stars in the Galactic center are a population of fast moving stars in highly eccentric, nearly Keplerian orbits around a MBH of $\approx 4 \times 10^6 M_\odot$ (Gillessen et al., 2009). The pericenter distance of some of these stars is inferred to be $\lesssim 10^4$ of the gravitational radius, $GM_\bullet/c^2$, of the MBH implying speeds of a few percent of light. The high pericenter velocities inspire a search for general relativistic perturbations to purely Keplerian orbits—namely, the (prograde) precession of pericenter and Lense-Thirring node precession. An extra layer of complexity is added by the likely existence of an unseen extended mass distribution composed of dark compact remnants such as white dwarfs (WDs), neutron stars (NSs) and stellar black holes (SBHs). In order to disentangle both effects from observations will require to fit simultaneously for post-Newtonian effects and Galactic perturbations (Preto and Saha, 2009).

## 2 Stellar distribution of NSCs around a MBH

### 2.1 Basic quantities and models

The Schwarzschid radius of a non-spinning black hole of mass $M_\bullet$ is

$$R_{\text{Schw}} = \frac{2GM_\bullet}{c^2} \approx 9.5 \times 10^{-8} \, \text{pc} \left(\frac{M_\bullet}{10^6 M_\odot}\right). \quad (1)$$

The radius of influence $r_h$ is the radius within which the MBH potential dominates the dynamics

$$r_h = \frac{GM_\bullet}{\sigma^2(r_h)} \approx 0.43 \text{pc} \left(\frac{M_\bullet}{10^6 M_\odot}\right) \left(\frac{100 \text{Km/s}}{\sigma_h}\right)^2,$$

where $\sigma$ is the 1D velocity dispersion. In the case of an isothermal sphere, $\rho(r) = \sigma^2/2\pi G r^2$, for which $\sigma(r) = $ const, this is equivalent to the condition (which, for simplicity, we adopt in the analysis of the models even though they are not necessarily isothermal)

$$M(< r_h) = 2M_\bullet.$$

A star that comes within a distance

$$R_t = \left(\nu \frac{M_\bullet}{m_*}\right)^{1/3} r_* \quad (2)$$

of the central MBH will be tidally disrupted, where $\nu \sim O(1)$. Combining (1) and (2), we see

$$\frac{R_t}{R_{Schw}} \propto M_\bullet^{-2/3},$$

and, as a result, for main-sequence (MS) stars the tidal disruption radius is overtaken by the Schwarschild radius only for black hole masses $M_\bullet \gtrsim 1.14 \times 10^8 \nu^{1/2} M_\odot$. For a MBH in the LISA range, *i.e.* $M_\bullet \lesssim$ few $\times 10^6 M_\odot$, MS stars are tidally disrupted before they can be captured with the emission of GWs; in contrast, for compact objects $R_{Schw} \gg R_t$, so they either promptly fall through the event horizon or undergo a slow inspiral with very many close pericentric passages and the corresponding emission of GWs—without being tidally disrupted. A *loss cone* can be associated to both $R_{Schw}$ and $R_t$; and, in each case, it consists of all orbits whose pericenter falls inside the corresponding radius.

The two-body relaxation time $T_{rlx}$ is a (somewhat loosely) defined quantity that provides a measure of the time required for local deviations from a Maxwellian distribution to be significantly decreased—assuming the underlying relaxation mechanism is the sum of a large number of independent, uncorrelated, weakly deflecting two-body encounters (Spitzer, 1987). For a spatially homogeneous distribution of stars of equal mass $m_*$, the relaxation time is defined by

$$T_{rlx} = 0.34 \frac{\sigma^3}{G^2 \rho m_* \ln \Lambda} \sim \frac{2 \times 10^{10}}{\ln \Lambda} \text{yrs} \left(\frac{\sigma}{100\text{km/s}}\right)^3 \left(\frac{1 M_\odot}{m_*}\right) \left(\frac{r}{1\text{pc}}\right)^\gamma \propto r^{\gamma - 3/2}, \qquad (3)$$

where $\sigma$ is the one-dimensional velocity dispersion, $\rho$ is the stellar mass density (the proportionality assumes a $\rho(r) \sim r^{-\gamma}$ profile and $\sigma^2(r) = GM_\bullet/r$), and $\ln \Lambda = \ln(b_{max}/b_{min})$ is the Coulomb logarithm. $b_{max}$ and $b_{min}$ are the maximum and minimum (Keplerian) impact parameters, respectively. For instance, $b_{max}$ is usually set equal to the size of the system, while $b_{min} = 2Gm/\sigma^2$ corresponding to a $90^o$ deflection due to a two-body encounter. In the gravitational potential of a MBH, *i.e.* within $r_h$, we set $b_{max} = r_h$ and two-body relaxation leads to the establishment of a steady-state distribution of orbital energies on a timescale $\approx T_{rlx}$ (Preto et al., 2004).

In order to understand how mass segregation proceeds in a galactic nucleus, it is useful to consider the (realistic) situation where a small fraction $f_\bullet \approx 10^{-3}$ of the stars are stellar black holes of mass $m_\bullet \gg m_*$ (take $m_\bullet = 10 M_\odot$ as a fiducial value). Since there are very few SBHs, they sink toward the center under the dynamical friction from the dominant light population of MS stars and do not self-interact. In such situation, assuming as a rough approximation that the SBHs orbits are circular, the typical timescale for their sinking to the center is given from Chandrasekar's formula by

$$T_{DF} = \frac{v_c^3}{4\pi G^2 \rho_* m_\bullet \ln \Lambda} \left[ \text{erf}(X) - \frac{2X}{\sqrt{\pi}} e^{-X^2} \right]^{-1} \sim \frac{T_{rlx}}{R},$$

where $v_c$ is the (local) circular velocity, $\rho_*$ is the mass density of the MS stars, $R = m_\bullet/m_*$ and $X = v/\sqrt{2}\sigma$. Note that in (3), $m_* = \langle m_* \rangle \sim m_\bullet/R$, so in case there are stars with a range of masses, the most massive should sink to the center on a dynamical friction timescale which is a small fraction ($\approx 1/R$) of the (local) relaxation time. The presence of a mass spectrum in a stellar population of a dense stellar cluster thus significantly speeds up its internal evolution, and such systems will show signs of mass segregation on a timescale significantly shorter than the relaxation time.

Encounters between stars with an impact parameter small than a few $b_{\min}$ lead to large deflection angles that are not accounted for by the Fokker-Planck description of diffusion in phase space. Such "large angle scattering" or "close encounters" are expected to be comparatively rare but, in contrast with the diffusion due to weak encounters, can eject stars from the cusp in a single encounter—rather than through a long random walk. The average time between strong encounters for a typical star can be estimated from a $n\sigma v$ argument

$$\langle T_{LA} \rangle \sim \frac{1}{n\pi(\alpha b_{min})^2 \sigma} \sim \frac{\ln \Lambda}{4\alpha^2} T_{rlx},$$

where $\alpha = O(1)$. The effect of strong encounters on the bulk evolution of the stellar cluster is expected to be negligible so long as $\ln \Lambda \gg 1$, which is the case in galactic nuclei where $\ln \Lambda \sim \ln(M_\bullet/m_*) \gtrsim 15$.

How stars distribute around a massive black hole

If a stellar mass object forms a very close binary system with a MBH, it will dissipate its orbital energy (and angular momentum) through the emission of GWs and will, as a result, slowly inspiral towards the MBH. Peters equations describe this secular dynamical process by equating the loss of orbital energy and angular momentum with that which is radiated away to infinity with the emmited GWs—assuming the latter is sufficiently well approximated by Einstein's quadrupole formula computed along Keplerian orbits. In case of a binary with extreme mass ratio $M_\bullet/m_* \gg 1$, Peters equations reduced to (Peters, 1964)

$$\left\langle \frac{da}{dt} \right\rangle = -\frac{64}{5} \frac{G^3 \mu_{red} M_\bullet^2}{c^5 a^3 (1-e^2)^{7/2}} \left(1 + \frac{73}{24}e^2 + \frac{37}{96}e^4\right)$$

$$\left\langle \frac{de}{dt} \right\rangle = -\frac{304}{15} \frac{G^3 \mu_{red} M_\bullet^2}{c^5 a^4 (1-e^2)^{5/2}} e \left(1 + \frac{121}{304}e^2\right),$$

and the brackets denote that the quantities are orbital-averaged. As a result, the characteristic time for the stellar mass object to change its orbital eccentricity due to GW emission alone is given by

$$T_{GW} = \left|\frac{e}{de/dt}\right| \approx \frac{15}{304} \frac{c^5 a^4 (1-e^2)^{5/2}}{G^3 \mu_{red} M_\bullet} \approx 6.5 \times 10^{11} \text{yr} \left(\frac{M_\odot}{\mu_{red}}\right) \left(\frac{10^6 M_\odot}{M_\bullet}\right)^2 \left(\frac{a}{0.01\text{pc}}\right)^4 \left(1-e^2\right)^{5/2}.$$

LISA will be sensitive to the inspiral of compact remnants onto MBHs in the mass range $10^4 M_\odot \lesssim M_\bullet \lesssim 10^7 M_\odot$. Let us assume MBHs in this mass range follow the $M_\bullet - \sigma$ relation

$$M_\bullet \propto \sigma^4,$$

then the influence radius $r_h$ scales as[2]

$$r_h \sim \frac{GM_\bullet}{\sigma^2} \propto M_\bullet^{1/2},$$

and the mean stellar density at $r_h$ scales as

$$\langle n_h \rangle \sim \frac{2M_\bullet/m}{r_h^3}.$$

Since the Jeans equation implies the velocity dispersion inside $r_h$ is $\sigma^2(r) \sim GM_\bullet/r$, the dependence of two-body relaxation timescale on MBH mass becomes

$$T_{rlx} \sim \frac{\sigma^3}{\rho} \propto M_\bullet^{5/4}.$$

We conclude that heavier black holes reside in nuclei of lower density. As a rule of thumb, nuclei harboring MBH with $M_\bullet \gtrsim 10^7 M_\odot$ have $T_{rlx} > T_H$, where $T_H \sim 13 - 14$ Gyr is the Hubble time, and are unrelaxed (their stellar distributions retain memory from the formation process and/or the latest strong perturbation); while those with $M_\bullet \lesssim 10^6 M_\odot$ have $T_{rlx} < T_H$ and should have had time to relaxed into a steady-state. The Milky Way nucleus, with $M_\bullet \sim 4\times 10^6 M_\odot$, stands on the borderline.

---

[2] Here we implicitly assume the cluster is isothermal, $\rho \sim r^{-2}$ and $\sigma(r) =$ const, for $r \gtrsim$ few $\times 0.1 r_h$.

Consider a NSC with a MBH of mass $M_\bullet$ at the center. The stellar distribution function (DF) $f(\mathbf{x},\mathbf{v})$ gives the probability of finding a star within a volume element $d^3\mathbf{x} d^3\mathbf{v}$ of $(\mathbf{x},\mathbf{v})$ space by: $f(\mathbf{x},\mathbf{v})d^3\mathbf{x} d^3\mathbf{v}$. According to Jeans theorem, the DF of a spherical cluster depends on $(\mathbf{x},\mathbf{v})$ only through the star's binding (specific) energy, $E = \Phi(r) - 1/2 v^2$, and its (specific) orbital angular momentum, $J = |\mathbf{x}\times\mathbf{v}|$, where $\Phi(r) = \Phi_*(r) + GM_\bullet/r$ is the total gravitational potential due to the cluster plus the hole. The loss cone, for tidal disruption of a MS star or capture of a compact object, can be defined in $(E, J)$ space by

$$J \leq J_{lc}(E) = r_d \sqrt{2(\Phi(r_d) - E)} \approx \sqrt{2GM_\bullet r_d},$$

where $r_d = R_t$ or $R_{Schw}$ for each case respectively.

The number of stars $N(E, J)$, with energy $E$ and angular momentum $J$, is related to $f(E, J)$ through

$$N(E, J) = \int d^3\mathbf{x}\, d^3\mathbf{v}\, \delta(E - \Phi(r) + 1/2\, v^2)\, \delta(J - |\mathbf{x}\times\mathbf{v}|)\, f(E, J) = 8\pi^2 J\, T_r(E) f(E, J),$$

where $T_r(E) = \oint dr/|v_r|$ is the radial period from apocenter to pericenter and back, $v_r$ is the radial velocity. Furthermore, if we assume an isotropic DF then $f = f(E)$ (Binney and Tremaine, 2008). In this latter case, the integral above simplifies

$$N(E) = \int d^3\mathbf{x}\, d^3\mathbf{v}\, \delta(E - \Phi(r) + 1/2\, v^2)\, f(E) = 4\pi^2 p(E) f(E),$$

where $p(E) = 4\int_0^{r_{\max}(E)} dr\, r^2 \sqrt{\Phi(r) - E} = -\partial q/\partial E$ is the fraction of phase space accessible to each (bound) star of specific energy $E > 0$, and the upper integration limit is set by $E = \Phi(r_{\max})$. The radial density profile $\rho(r)$ is uniquely determined from the knowledge of the DF $f(E)$ and the gravitational potential $\Phi(r)$ through

$$\rho(r) = 4\pi \int_0^{v_{\rm esc}(r)} dv\, v^2 f\left(\Phi(r) - 1/2 v^2\right) = 4\pi \int_0^{\Phi(r)} dE\, f(E) \sqrt{2(\Phi(r) - E)}. \tag{4}$$

Within the MBH gravitational potential well, $p(E) \sim \frac{\sqrt{2}\pi}{16}(GM_\bullet)^3 E^{-5/2}$ and $q(E) \sim \frac{\sqrt{2}\pi}{24}(GM_\bullet)^3 E^{-5/2} E^{-3/2}$.

To 0$^{\rm th}$ order, the motion of stars evolves under the influence of the global, smoothed gravitational potential $\Phi(r)$ created all the stars and the MBH. If the cluster is spherical, the energy $E$ and orbital angular momentum $J$ of each star are constants of motion. However, the graininess of the potential created by a large (but finite) number $N$ of stars produces fast, small-scale fluctuations that lead to the slow diffusion of $E$ and $J$, *i.e.* relaxation results from the interaction between individual particles superimposed on the mean, smooth potential created by the system as a whole. When relaxation is driven by uncorrelated, weak, two-body encounters, the Fokker-Planck equation is appropriate for its description. The time-dependent, orbit-averaged, isotropic, Fokker-Planck equation in energy space is defined, for each component (Spitzer, 1987 and Chernoff and Weinberg, 1990), by

$$p(E)\frac{\partial f_i}{\partial t} = -\frac{\partial F_{E,i}}{\partial E}, \quad F_{E,i} = -D_{EE,i}\frac{\partial f_i}{\partial E} - D_E f_i,$$

$$D_{EE,i} = 4\pi^2 G^2 m_*^2 \mu_i^2 \ln \Lambda$$
$$\times \left(\frac{\mu_j}{\mu_i}\right)^2 \sum_j^{N_c} \left[ q(E) \int_{-\infty}^{E} dE' f_j(E') + \int_{E}^{+\infty} dE' q(E') f_j(E') \right],$$

$$D_{E,i} = -4\pi^2 G^2 m_*^2 \mu_i^2 \ln \Lambda \sum_j^{N_c} \left(\frac{\mu_j}{\mu_i}\right) \int_{E}^{+\infty} dE' p(E') f_j(E').$$

In this equation, $i, j$ run from 1 to $N_c$ (the number of mass components), $\mu_i = m_i/m_*$ where $m_* = 1/N$ is a reference mass. The gravitational potential $\Phi$ is determined from Poisson' equation

$$\nabla^2 \Phi(r) = 4\pi G \rho(r).$$

As the stellar density $\rho(r)$ evolves over time, the stellar gravitational potential $\Phi(r)$ needs to be updated. In order to do this self-consistently, we adopt the operator-splitting method outlined by (Cohn, 1980) and (Chernoff and Weinberg, 1990). This method consists in successively updating the DF $f(E)$ through the diffusion equation and the gravitation potential $\Phi(r)$ through Poisson equation. In the diffusion step, we keep $\Phi(r)$ fixed while updating $f(E)$ and the diffusion coefficients. In the Poisson step, $\Phi(r)$ is updated while keeping the DF fixed as a function of the adiabatic invariant q. The Poisson step consists of several iterations until convergence; since the evolution of $\Phi(r)$ is slow, this step is only done at every five diffusion steps or so. By solving the Poisson equation as described, one is able to follow accurately the slow re-expansion of the stellar cluster.

**2.2 Bahcall & Wolf Cusp**

The distribution of stars around a massive black hole (henceforth MBH) is a classical problem in stellar dynamics (Peebles, 1972, Bahcall and Wolf, 1976 and Lightman and Shapiro, 1977). (Peebles, 1972) was the first to recognize that statistical thermal equilibrium, with $f(E) \propto e^{-E/\sigma^2}$, must be violated sufficiently close to the MBH either due to the tidal disruption at $R_t$ and the capture of stars by the MBH at $R_{Schw}$; or, alternatively, by physical collisions between stars inside some characteristic radius $r_{coll}$. Peebles also predicted that, as a result, a steady state with a (small) net inward flux of stars and energy would obtain. Nevertheless, stars well within the gravitational field of the MBH, but far from the tidal disruption radius, should have nearly-isotropic velocities. Therefore, if stars are all of the same mass, the quasi-steady solution takes a power-law form—an isotropic DF $f(E) \sim E^p$ in energy space, and $\rho(r) \sim r^{-\gamma}$, in physical space. These exponents are related with one another, $\gamma = 3/2 + p$, via Equation (**4**). Then, in order to derive the values of the exponents, Peebles used a simple (but erroneous) scaling argument. If $\rho(r) \sim r^{-\gamma}$, the number of stars within $r$ is $N(<r) \sim r^{3-\gamma}$. If stars were to be transported through the cusp with a characteristic timescale equal to the relaxation time, then the rate $\dot{N}$ at which stars are transported through a given radius $r$ would be given by $\dot{N} \sim N(<r)/T_{rlx}(r) \sim r^{9/2-2\gamma}$; in steady state this rate should be independent of radius, so $\gamma = 9/4$ (and $p = \gamma - 3/2 = 3/4$). In fact, in steady state stars are transported at a slower rate as those flowing inward balance those flowing outward almost precisely.

(Bahcall and Wolf, 1976)—henceforth BW76—subsequently studied the same problem, through a detailed kinetic treatment. By means of their more detailed analysis, they concluded that, in the case all stars are of the same mass, a quasi-steady distribution indeed takes the form of the power laws described above, but with different values for the exponents: $\gamma = 7/4$ and $p = \gamma - 3/2 = 1/4$. A subtle scaling argument can be used with hindsight to derive the correct result. First, let us identify the relaxation time $T_{rlx}$ with the characteristic timescale for energy transport in the cusp—instead of with the time scale for transport of stars throughout the cusp, as Peebles did. In this case, the rate at which energy is transported through the cusp is $\langle \dot{E} \rangle \sim (N(<r)/T_{rlx}(r))\langle E(r) \rangle \sim r^{9/2-\gamma}/r$; and by requiring the steady state energy flux to be independent of radius yields $\gamma = 7/4$ (and $p = 1/4$). This is the so-called *zero-flow solution* for which the net flux of stars in energy space is precisely zero, and results in the limit case where $R_t \to 0$ (or $E_t \to +\infty$). The *nearly zero-flow solution* obtains in the more realistic case where a finite inner boundary condition is imposed. These results concerning the transport of stars, or energy, through the cusp are not contradictory if one recognizes the timescale $T_S$ associated with the transport of stars within the cusp is different (and longer) than the timescale $T_E$ associated with energy transport, $T_S(r) \propto (r/R_t) T_E(r)$ (Shapiro, 1985). For the exact *zero-flow solution* $T_S \to 0$ as the number of stars drifting in and out, at each radius r, cancel out precisely.

Furthermore, BW76 have shown that Peebles solution ($\gamma = 9/4$) violates the inner boundary condition by implying a huge outward flux of stars from the center. This can be understood through the inspection of the FP flux terms. For a single mass population, the flow rate $F(E)$ of stars in $E$-space is

$$F(E) \propto f(E) \int_E^{+\infty} dE' p(E') f(E') - \frac{\partial f}{\partial E} \left[ q(E) \int_{-\infty}^{E} dE' f(E') + \int_E^{+\infty} dE' q(E') f(E') \right].$$

Substituting $f(E) \sim E^p$, $\partial f/\partial E \sim pE^{p-1}$, $p(E) \sim E^{-5/2}$ and $q(E) \sim E^{-3/2}$, and setting the flux to zero, $F(E) = 0$, one obtains

$$-\frac{E}{16} \frac{1}{p - 3/2} \left[ E^{p-3/2} \right]_E^{+\infty} + \frac{p}{24} \left( \frac{E^{-3/2}}{p + 1} \left[ E^{p+1} \right]_0^E + \frac{1}{p - 1/2} \left[ E^{p-1/2} \right]_E^{+\infty} \right) = 0,$$

where the $-\infty$ integration limit was set to zero since $E > 0$ for bound stars. In case $p = 3/4$, follows that $F(E) \sim -E^{1/4} \to -\infty$ as $E \to +\infty$. This implies a very large outward stellar flux and, indeed, turns the MBH into a source rather than a sink—violating the assumed inner boundary condition. On the other hand, convergence as $E \to +\infty$ requires $p < 1/2$. The only remaining solution which is independent of $E$ is $p = 1/4$ which corresponds to $F(E) = 0$ identically zero. Note that even though the central density of a BW cusp is is extremely high, it is populated by a relatively small number of stars given by $N(E) \propto E^{p-5/2}$. As a result, disruption rate typically peak at intermediate energies which, for galactic nuclei, is around $E_h$ corresponding to stars whose mean radius is around $r_h$. If the stellar cluster around a MBH deviates from spherical symmetry, and it is axisymmetric flattened due to rotation, its DF $f(E, J_z)$ will depend on the component of angular momentum $J_z$ projected along the axis of symmetry. Also in such case, a BW solution $f(E) \propto E^{1/4}$, upon averaging $f(E, J_z)$ over $J_z$, develops on a relaxation time scale (Fiestas and Spurzem, 2010).

### 2.3 Mass Segregated Cusps

The properties of stellar systems that display a range of stellar masses are only very poorly reproduced by single mass models. The so-called universal initial mass functions (IMFs) (Kroupa, 2001), with a mass that covers a range from brown dwarfs ($\sim 0.1 M_\odot$) to an upper limit ($\sim 120 M_\odot$), give rise to an evolved stellar population (either through an old coeval starburst or continuous star formation) that, to first order approximation, can be represented by two (well-separated) mass scales: one in the range $O(1 M_\odot)$ corresponding to low mass main-sequence stars, white dwarfs (WDs) and neutron stars (NSs); another with $O(10 M_\odot)$ representing stellar black holes (SBHs).

It is well known from stellar dynamical theory that when several masses are present there is mass segregation—a process by which the heavy stars accumulate near the center while the lighter ones float outward (Spitzer, 1987 and Khalisi et al., 2007). Accordingly, stars with different mass get distributed with different density profiles. By assuming a stellar population with two mass components, (Bahcall and Wolf, 1977)—hencefort BW77—generalized their early cusp solution and argued heuristically for a scaling relation $p_L = m_L/m_H \times p_H$ that depends on the star's mass ratio only. However, they obtained no general result on the inner slope of the heavy objects; nor did they discuss the dependence of the result on the component's number fractions. The assumption was that whenever the heavy stars dominate the stellar gravitational potential, they should find themselves distributed according to a 7/4 cusp. It is worth to stop here and see how this scaling relation comes about. Take, for simplicity, a 2-component stellar system and assume: (1) the stellar flux of both components is identically zero $F_{E,L} = F_{E,H} = 0$; (2) the solutions are isotropic DFs of the same form as the single mass case $f_L(E) \sim E^{p_L}$ and $f_H(E) \sim E^{p_H}$, but different exponents. Under these conditions, one can easily show

$$F_{E,L} - \frac{m_L}{m_H} F_{E,H} = m_L^2 \frac{D_{EE,L}}{E} \left( p_L - \frac{m_L}{m_H} p_H \right) + m_H^2 \frac{D_{EE,H}}{E} \left( p_L - \frac{m_L}{m_H} p_H \right) = 0.$$

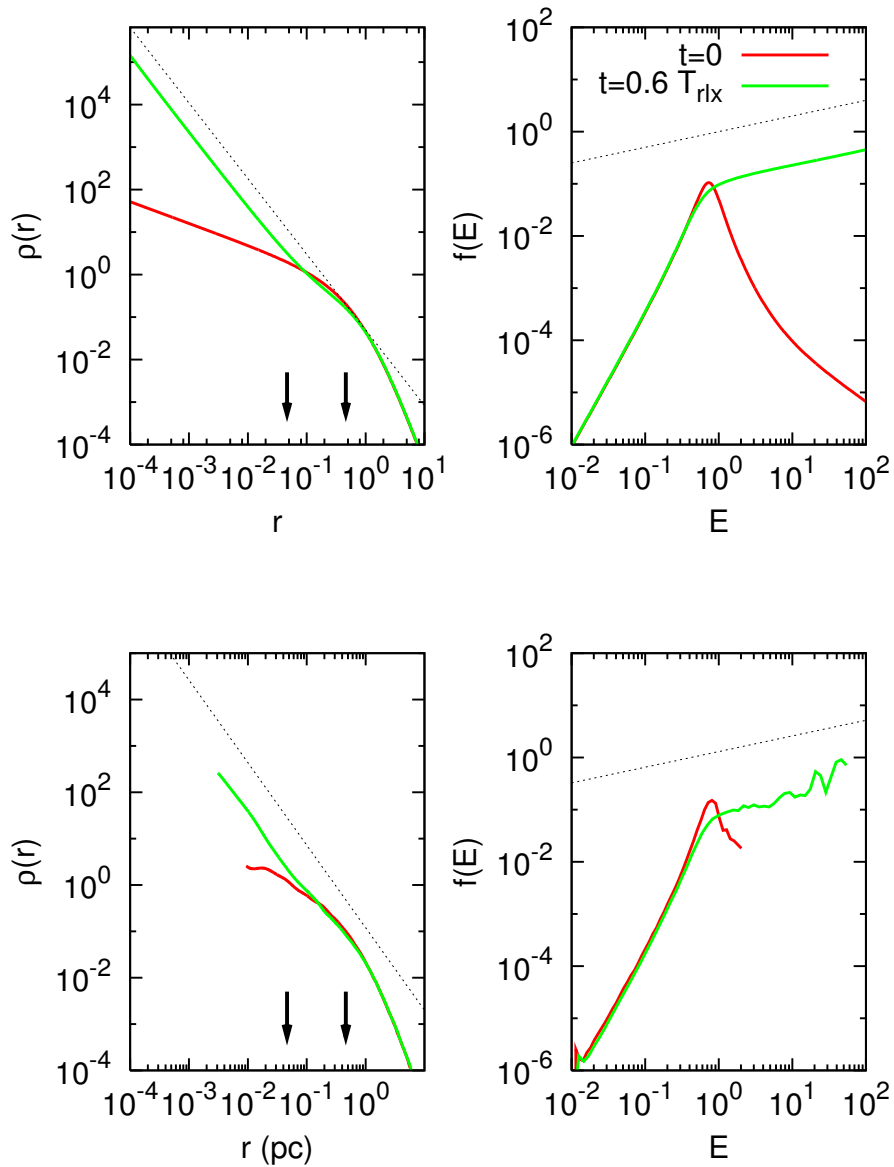

**Fig. 1** Robust N-body growth of a single-mass Bahcall & Wolf cusp. FP and NB evolution (top and lower panels, respectively) of the spatial density profile $\rho(r)$ (left panels) and the phase space density $f(E)$ (right panels). The asymptotic density cusps $\rho(r) \sim r^{-7/4}$ and $f(E) \sim E^{1/4}$ are reached after roughly $0.6T_{rlx}$, and remains stable afterwards. The arrows point to $0.1r_h$ and $r_h$.

How stars distribute around a massive black hole

Since the diffusion coefficients $D_{EE,L}, D_{EE,H} > 0$ for all $E > 0$, it follows immediately $p_L = m_L/m_H \times p_H$. This argument is immediately generalized to $N$ components of mass $m_i$ ($i = 1, ..., N$).

(Alexander and Hopman, 2009) have stressed that BW77 scaling relation cannot be valid in the limit where the number fraction of heavy stars is realistically small. In that situation, a new solution that they coined *strong mass segregation* obtains with density scaling as $\rho_H(r) \sim r^{-\alpha}$, where $\alpha \gtrsim 2$. They have shown that the solution has two branches and can be parametrized by

$$\Delta = \frac{D_{HH}^{(1)} + D_{HH}^{(2)}}{D_{LH}^{(1)} + D_{LH}^{(2)}} \approx \frac{N_H m_H^2}{N_L m_L^2} \frac{4}{3 + m_H/m_L}.$$

$\Delta$ provides a measure of the importance of the heavy star's self-coupling relative to the light-heavy coupling (in terms of the 1$^{st}$ and 2$^{nd}$ order diffusion coefficients); and it depends essentially only on the mass and number ratios, which is one parameter more than proposed by BW77. The *weak* branch, for $\Delta > 1$ corresponds to the scaling relations found by BW77; while the *strong* branch, for $\Delta < 1$, generalizes the BW77 solution. Stellar populations with continuous star formation and an IMF given by $dN/dM \propto M^{-\alpha}$, will be characterized by $\Delta < 1$ if $\alpha \gtrsim 1.8$ and $\Delta < 1$ otherwise; and, in particular, Salpeter and Kroupa's IMF generate evolved stellar populations with $\Delta < 1$ (Alexander and Hopman, 2009).

There is a straightforward physical interpretation for the strong branch of mass segregation. In the limit where heavy stars are very scarce, they barely interact with each other and instead sink to the center due to dynamical friction against the sea of light stars. Therefore, a quasi-steady state develops in which the heavy star's current is not nearly zero and thus the BW77 solution does not hold exactly anymore. Indeed, in the limit where the number fraction $f_H$ of heavy stars is vanishingly small, the stellar potential is dominated by the light component. In this case, the light stars should evolve as if in isolation and develop a $\gamma_L \sim 7/4$ density cusp. The scarce heavy stars will sink to the center due to dynamical friction against the background of light stars, and will not exert any significant back-reaction on them. Following AH09, let us assume the lights behave as a single mass population, $\rho_L(r) \sim r^{-\gamma_L}$, inside $r_h$, $f_H \ll 1$ and $R = m_H/m_L \gg 1$. The dynamical friction force acting on the heavies is $F_{df} \sim \rho_L/v_H^2 \sim r^{1-\gamma_L}$, and assuming a circular inspiral orbit, the resulting torque is $\dot{J} \sim F_{df} r \sim r^{2-\gamma_L} \sim r^{-1/2}\dot{r}$, so $\dot{r} \sim r^{5/2-\gamma_L}$. In steady state, the stellar current must be independent of radius $r$, so $\dot{N}_H(r) \sim \dot{r} r^2 n_H(r) =$ const. Therefore, the number density of heavy stars is $n_H(r) \sim r^{-\gamma_H} \sim r^{\gamma_L - 9/2}$. But, since $\gamma_L \to 7/4$, $\gamma_H = 9/2 - \gamma_L \to 11/4$! ($p_H = \gamma_H - 3/2 \to 5/2 \neq m_H/m_L \times p_L$).

Alternatively, one can look again at the FP flux $F_H(E)$, and by keeping only the terms that describe the dynamical friction effect of the lights on the heavies, this turns out to be

$$F_H(E) \propto m_L m_H f_H(E) \int_E^{+\infty} dE' p(E') f_L(E') = \text{const.}$$

In steady state, $F_E(H)$ should be independent of $E$; as a result, it scales as $F_H(E) \sim E^{p_L - 3/2 + p_H} = \text{const.}$ Since $p_L \to 1/4$, it follows that $p_H \to 5/4$ as before.

### 2.4 N-body validation

(Preto et al., 2004) and (Baumgardt et al., 2004a) were the first to report *N*-body realizations of the Bahcall & Wolf solution, thereby validating the assumptions inherent to the Fokker-Planck (FP) approximation—namely, that scattering is dominated by uncorrelated, 2-body encounters and, in particular, dense stellar cusps populated with stars of the *same mass* are robust against ejection of stars from the cusp. Ejections—due to strong encounters—are *a priori* excluded from the FP evolution, even though they could occur in a real nucleus. This was not *a priori* a trivial question since for a BW $\gamma = 7/4$ cusp, stellar densities are very high and the fraction of stars with speeds close to the escape velocity $v_{esc}$ from the cusp is also quite large at every radius $r \ll r_h$. Taking the DF $f(E) \propto (E/\sigma^2)^p$ of a BW cusp, it is easy to compute the fraction of stars $N(> v_{esc})/N_{tot}$ with velocity above any given fraction $\alpha \in [0, 1]$ of $v_{esc}$

$$\frac{N(> \alpha v_{esc})}{N_{tot}} = 1 - \frac{\int_0^{\alpha v_{esc}} dv\, v^2 \left(\frac{GM_\bullet}{r} - \frac{1}{2}v^2\right)^p / \sigma^{2p}(r)}{\int_0^{v_{esc}} dv\, v^2 \left(\frac{GM_\bullet}{r} - \frac{1}{2}v^2\right)^p / \sigma^{2p}(r)} = 1 - \frac{\alpha^3\, _2F_1(3/2, -p, 5/2; \alpha^2)}{_2F_1(3/2, -p, 5/2; 1)},$$

where $_2F_1$ is a hypergeometric function. Taking $p = 1/4$ there are $\approx 84\%$ of stars with velocity higher than $v_{esc}/2$ and $\approx 19\%$ higher than $0.9 v_{esc}$; if $p = 0$ ($\gamma = 3/2$), these values are $\approx 87\%$ and $\approx 22.6\%$ respectively. From (Yu and Tremaine, 2003) one would expect that a non-negligible fraction of stars can be ejected from the cusp in a relaxation time. It is reassuring that despite this fact, the *N*-body simulations show unambiguously that the stellar cusps are robust against stellar ejections. Nevertheless one should also note that, well within the sub-parsec region of galactic nuclei, velocity dispersions can reach values as high as $\sim 1000$ km/s, so the minimum impact parameter $b_{min}$ can be as small as $\sim 10^{-8}$ pc $\approx R_\odot$, meaning physical collisions between MS stars (or between a MS star and a compact object) might result from a close encounter—rather than the high velocity ejection of a MS star from the cusp (Yu and Tremaine, 2003). This can be seen in Figure **1**, which displays the FP and NB evolution of spatial density $\rho(r)$ and phase space density $f(E)$ for comparison. The starting model was a $\gamma = 1/2$ Dehnen profile, with a MBH of 5% of the total cluster mass. The density of stars around a MBH increases as the cusp grows and reaches a quasi-steady state after $\approx 0.6 T_{rlx}$ in close agreement with timescales derived from FP (Preto et al., 2004), remaining stable afterwards.

There has been a surprisingly small number of *N*-body studies of multi-mass systems around a MBH (Baumgardt et al., 2004b and Freitag et al., 2006) and, until very recently, none of them reported the occurence of strong mass segregation. (Hénon,

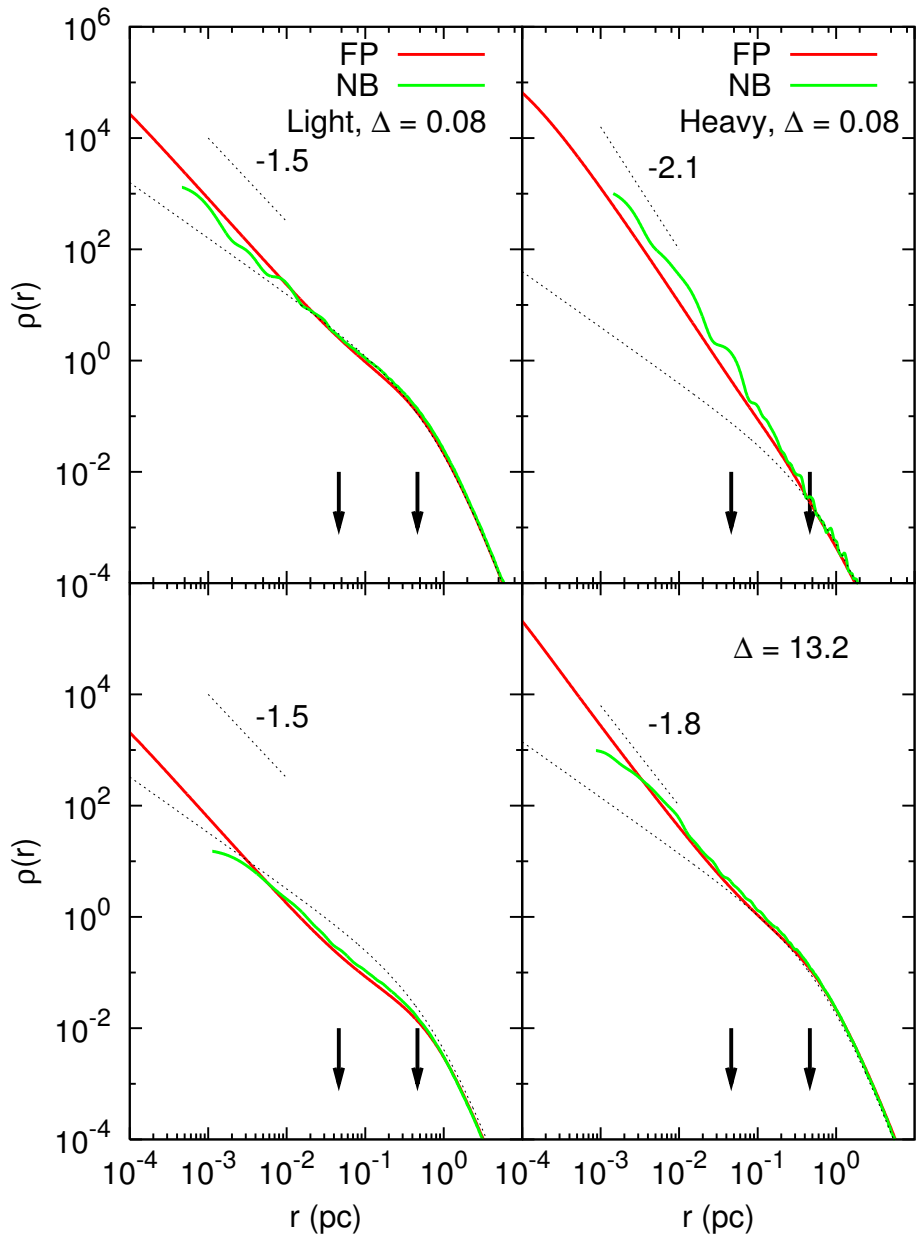

**Fig. 2** Mass density profiles, $\rho_L(r)$ (left panels) and $\rho_H(r)$ (right panels) at the end of the integrations, after $\approx 0.2 T_{\rm rlx}(r_{\rm h})$. Smooth curves are from FP calculations, noisy curves are from NB simulations. The mass ratio between heavy and light stars is $R = 10$; the number fraction of heavy stars $f_H = 2.5 \times 10^{-3}$ (top panels) and $f_H = 0.429$ (lower panels), corresponding to the strong and weak segregation regimes respectively. The initial condition is a Dehnen profile with central slope $\gamma = 1$ and a central MBH with 5% of the total mass of the cluster. The asymptotic slope $\gamma_H$ decreases from $\gtrsim 2$ to $\approx 7/4$ when moving from the strong to the weak branch of the solution. The asymptotic slope $\gamma_H \approx 3/2$ throughout, or just slightly below this value. The arrows point to the radius $r_h$ and $0.1 r_h$.

1969) has shown long ago that the presence of a mass spectrum leads to an increased rate of stellar ejections from the core of a globular cluster, but he did not include the presence of a MBH at the center. Hénon's work raised the question as to whether *multi-mass* stellar cusps, obtained from the solution of the FP equation, are robust against ejection of stars from the cusp. For all these reasons, it was fundamental to systematically verify the validity of the Bahcall-Wolf solution—as well as its Alexander-Hopman generalization—with *N*-body integrations.

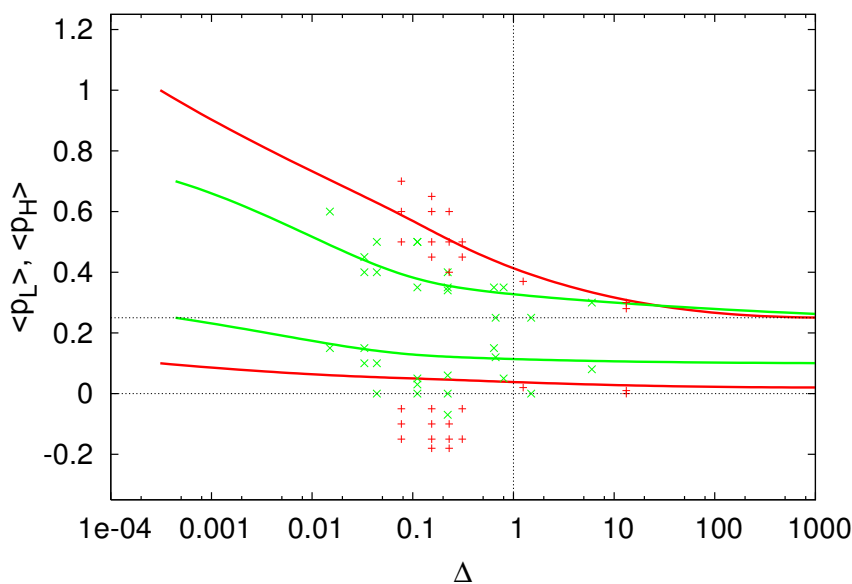

**Fig. 3** Two branches of mass segregation: weak ($\Delta > 1$) and strong ($\Delta < 1$) branches. The average logarithmic slopes of the phase space density for the heavy and light stars, $\langle p_H \rangle$ and $\langle p_L \rangle$, as a function of the Alexander & Hopman parameter $\Delta$. The green/red curves (FP) and points (NB) correspond to a mass ratio $R = 3/R = 10$. The agreement is quite good, even though NB calculations produce slightly stronger mass segregation.

(Preto and Amaro-Seoane, 2010a) have shown, for the first time, clear *N*-body realizations of the strong mass segregation solution for the stellar distribution around a MBH. Figure 2 displays the spatial density profiles $\rho_L(r)$ and $\rho_H(r)$ at late times, $t \sim 0.2 T_{rlx}(r_h)$, from FP and NB calculations. The agreement between both methods is quite good although there is the tendency, in the strong branch, for NB's asymptotic slope $\gamma_L$ to be slightly smaller than in FP—for which $\gamma_{L,min} = 1.5$. The slopes of the inner density profiles of the heavy component decrease as the solution evolves from the strong to the weak branch when $\Delta$ is increased, as expected. In the limit of $\Delta \gg 1$, $\gamma_H$ tends to evolve to a quasi-steady state close to the 7/4 solution, while for $\Delta \ll 1$, $\gamma_H \gtrsim 2$. The asymptotic inner density slopes, in both solution branches, of the light component extend out to $\sim 0.1 r_h$; in contrast, the heavy component shows a different behavior depending on the solution branch: on the weak branch, $\gamma_H$'s asymptotic slope also extends only up to $\sim 0.1 r_h$, while on the strong branch it extends virtually all the way to $r_h$. In the strong branch, the density of the heavy component exceeds that of the light for $r \lesssim 0.01 r_h$ (and will therefore dominate the interaction events with the MBH); in the weak branch, $\rho_H > \rho_L$ throughout.

Figure 3 summarizes the results of the extensive study of mass segregation in double-mass systems (Preto and Amaro-Seoane, 2010b). It displays the dependence of the average logarithmic slopes $p_L$ and $p_H$ on the Alexander & Hopman $\Delta$ parameter—for the cases with mass ratio $R = 3$ and 10. The smooth lines are measured from the FP calculations, while points are from NB simulations. It is clear from the figure that the solution is qualitatively different depending on whether $\Delta < 1$ or $\Delta > 1$. In the *weak* branch, $\Delta > 1$, BW77 scaling relations are valid: $p_H \sim 1/4$, while $p_L \sim 1/3 p_H \sim 0.083$ and $p_L \sim 1/10 p_H \sim 0.025$. In the *strong* branch, $\Delta < 1$, we see that $p_H > 1/4$ throughout and reaches a value $\approx 1$ ($\gamma_H \approx 2.5$) when $\Delta \approx 10^{-3}$—still below the limit 5/4 corresponding to $\Delta \to 0$. Note that it is very unlikely to find a galaxy with such a steep slope, since realistic stellar mass functions are not expected to have $\Delta \lesssim 0.01$. The slopes for the light stars, in the strong branch, also increase slightly with respect to the weak branch, but not enough to compensate that of the heavy star's distribution—BW77 scaling relations are broken in the strong branch, although not by a very large margin. Furthermore, for $R = 3$, $p_L \sim 1/4$ already at $\Delta = 10^{-3}$; but, for $R = 10$, $p_L \sim 0.1$, well below, 1/4 for the same $\Delta = 10^{-3}$ value even though the light stars constitute virtually a single-mass system at this $\Delta$. However, one should be careful in not overinterpreting the asymptotic limits $\Delta \to 0$ derived heuristically either through a scaling argument or by inspection of only one of the FP flux terms. In fact, both of these two derivations essentially assume the dynamics of heavy stars is due to dynamical friction on the sea of lights alone—and neglect both its self-coupling and its back-reaction on the light component. This is not necessarily an accurate approximation at all times and thorough the cusp; at late times, when a very steep cusp develops, the densities may become comparable, $\rho_H \sim \rho_L$ inside some radius $R_{eq}$. This should have an effect on the asymptotic slopes and constitutes a natural explanation the logarithmic slope $\gamma_L$ does not reach the 7/4 value especially when $R \gg 1$. This interpretation would imply that for lower mass ratio, say $R = 3$, it would be easier for the light component to reach $\gamma_L \sim 7/4$ characteristic of single mass cusps, than for the $R = 10$ case—this is indeed the result obtained from FP results shown in Figure 3. Another cautionary note regarding the interpretation of Figure 3 is that there are some sources of uncertainty on the measurement of the slopes: (i) there is some dependence of the slopes on the radial/energy range over which they are measured; (ii) the slopes show some slight variation with time; (iii) there is a weak dependence on the energy $E_{max}$ at which the inner boundary condition $f(E_{max}) = 0$ is imposed and also on the loss cone size, if present. In sum, although there are some differences in quantitative detail, these NB results broadly confirm the FP calculations and validate its inherent assumptions—at least in what concerns the description of the *bulk* properties of stellar distributions.

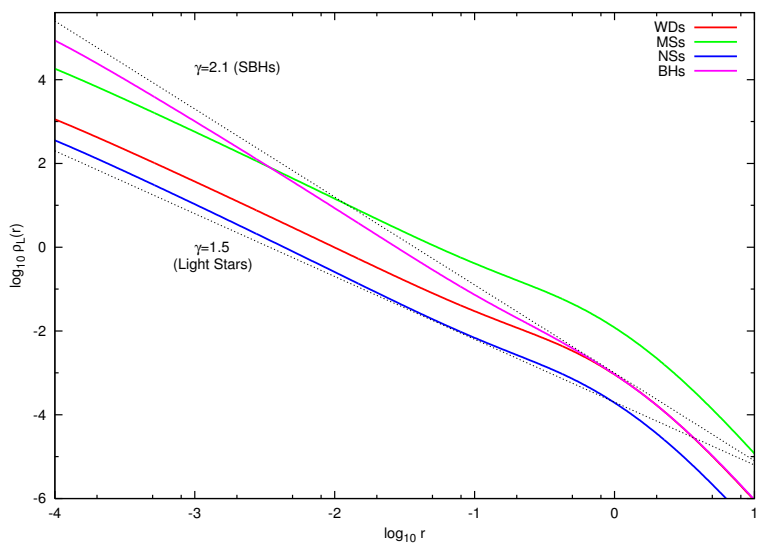

**Fig. 4**  Fokker-Planck results with 4 components. The asymptotic slopes of all components after the cusp has fully grown around the MBH. The SBHs have the steepest profile and, as a result, they become dominant in mass at the inner regions, even though MS stars dominate in number throughout the cusp.

Figure 4 shows the steady state density profiles of a four-component stellar system, established after $\approx 0.15 T_{\rm rlx}$, computed with FP. In this specific calculation, the four components are purported to represent WDs ($f_{WD} = 0.11$ and $R_{WD} = m_{WD}/m_{MS} = 0.6$), MSs ($f_{MS} = 0.873$ and $R_{MS} = 1$), NSs ($f_{NS} = 0.01$ and $R_{NS} = m_{NS}/m_{MS} = 1.4$) and SBHs ($f_{BH} = 0.007$ and $R_{BH} = m_{NS}/m_{BH} = 10$). The number fractions of compact remnants is sensitive to the IMF of high-mass stars. There are indications the IMF in galactic nuclei is top heavy (Maness et al., 2007), so in our extensive study of mass segregation, we adopt a range of values—$f_{BH} \in [10^{-3}, 10^{-2}]$, $f_{NS} \sim (3-10) \times 10^{-2}$ and $f_{WD} \sim 0.11$ (Preto and Amaro-Seoane, 2010a, 2010b); the mass distribution of, especially, SBHs is also weakly constrained so we adopt several mass ratios $R = 10, 15, 20$ (O'Leary et al., 2009). In this particular calculation, we have adopted a slightly large value $f_H = 0.007$ in order to facilitate the comparison with the NB integrations. The asymptotic slope of SBHs is $\gamma_{BH} \sim 2.1$, while $\gamma_{WD} \sim \gamma_{NS} \sim \gamma_{MS} \sim 3/2$. These distributions are clearly in the strong segregation regime. MSs dominate in number everywhere but, for $r \lesssim R_{eq} \sim (3-4) \times 10^{-3} \approx 0.005 r_h$, SBHs dominate in density. $R_{eq}$ evolves and increases in time as the cluster undergoes a slow global re-expansion after the cusp forms, so the region over which SBHs dominate the density will grow from the inside-out. A detailed analysis of this evolution and impact on EMRI rates is described in (Preto and Amaro-Seoane, 2010b).

The top panel of Figure 5 shows the density profiles of the same model, but computed with NB simulations. It is evident from the plot that the solution is qualitatively similar to the one obtained from the FP evolution—although, as in the double-mass case, the NB segregation is slightly stronger as the slopes of the components other than the SBHs are slightly below 3/2. The bottom panel depicts the temporal evolution of the logarithmic slopes of the spatial densities of all four components; that of the SBHs clearly stands out above all the others, a clear sign that SBHs have segregated to the center. There are fluctuations, due to numerical noise which is absent from the FP evolution, but the slopes seem to stabilize around $\gamma_{BH} \sim 2$, $\gamma_{MS} \sim \gamma_{NS} \sim 1.25$, although the WD slope has some indication to evolve secularly to lower value reaching $\gamma_{WD} \sim 1$ by the end of the integration—which, if scaled to the Milky Way nucleus, would correspond to roughly a Hubble time.

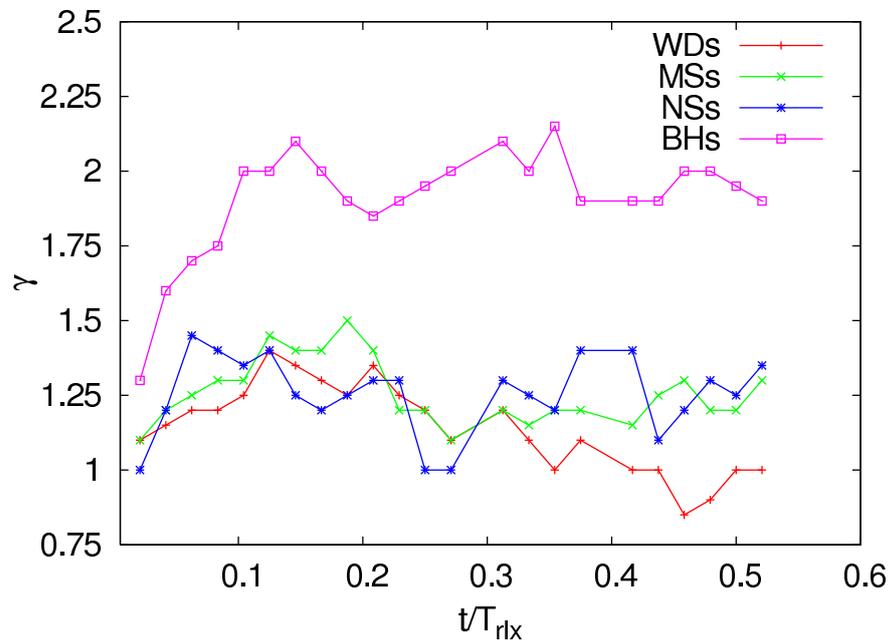

**Fig. 5** N-body results with 4 components. Evolution of the logarithmic slopes over a relaxation time scale. The slope of the SBHs stands out clearly from that of all other components—a signature of strong mass segregation in the nucleus.

## 3  Summary and discussion

Mass segregation is a robust outcome from the growth and subsequent evolution of stellar cusps around a MBH. Detailed comparisons between NB and FP integrations with two, or more, masses that are representative of the mass range covered by stellar populations are in very good agreement. NB simulations substantiate the robustness of cusps against stellar ejections and boost our confidence in the FP description of the *bulk* properties of cusps around MBHs. Which regime—weak or

strong—of mass segregation is present in a particular nucleus depends on the IMF of its stellar population. Assuming continuous star formation and a typical Kroupa IMF, one expects that strong mass segregation results; however, it could be that IMFs in galactic nuclei tend to be *top-heavy*, in which case it is more likely that the Alexander & Hopman $\Delta > 1$ and weak segregation would ensue.

As shown in (Preto and Amaro-Seoane, 2010a) and (Preto and Amaro-Seoane, 2010b), timescales for cusp growth in a Milky Way type nucleus—if driven by two-body relaxation—are expected to be shorter than a Hubble time, unless a very large core/hole is postulated to be present in the initial stellar distribution (*i.e.* a core radius $\gtrsim 2$ pc). Analysis of the number counts of spectroscopically identified, old stars in the sub-parsec region of our own Milky Way (Buchholz et al., 2009 and Do et al., 2009)—believed to be complete down to magnitude $K = 15.5$—reveals a deficit of old stars with respect to the high number a strongly segregated cusp would entail. Although the slope of the density profile is still weakly constrained, the best fits from number counts data seem to exclude slopes $\gamma > 1$ (Schödel et al., 2009), and there could be a core with a stellar density decreasing towards the center, $\gamma < 0$, although such a fit is only marginally better than one with $\gamma \sim 1/2$. It is difficult to devise plausible mechanisms for the formation of such large cores in the stellar distribution. For instance, the inspiral of an IMBH of mass $M_\bullet \sim 10^{3-4} M_\odot$ that forms an unequal-mass binary with the MBH and ejects stars through three body encounters would tend to progressively wipe out the stellar cusp. However, the core radius carved by such an event is $r_c \sim 0.02$ pc (Baumgardt et al., 2006) and thus a steady inflow of such IMBHs would be required in order to carve a large core 50 or 100 larger. Such large inflow of IMBHs have been proposed by (Portegies Zwart et al., 2006), one at roughly every $10^7$ years, but that does not seem very likely anymore in light of the fact that such IMBHs were hypothesized to be formed by runaway mergers of stars in the center of globular clusters. At solar metallicities, such mechanism seems very innefficient as mass loss, due to very strong winds, severely limits the growth of the stellar object being formed and the end result of a runaway merger is a $\sim 100 M_\odot$ Wolf-Rayet star. At lower metallicities, mass loss is lower and the remnant can be more massive $\sim 260 M_\odot$, but in any case it will not form an IMBH (Glebbeek et al., 2009). In sum, it looks very unlikely that sufficient IMBHs can be formed in order to generate such steady inflow to the Galactic center. Another possibility would be that SgrA* is a binary MBH, but this would most likely imply that there has been a more or less recent major merger involving the Milky Way—aside from the fact that there are strong constraints from the SgrA* proper motion (Reid and Brunthaler, 2004). This would contradict the apparent pure-disk nature of the Galaxy, as theoretical interpretations of stellar kinematic data of the Galactic Bulge seem to favor that the Bulge is part of the disk and not a separate component resulting from a merger (Shen et al., 2010). This actually connects with one of the major challenges for the theories of galaxy formation (Mayer et al., 2008). But see also (Callegari et al., 2010).

EMRIs of compact remnants will be detectable by LISA precisely for MBHs of SgrA* mass and below (Hopman and Alexander, 2005 and Amaro-Seoane et al., 2007). Estimates for event and detection rates by LISA costumarily assume that the stellar cusps are in steady state (Hopman and Alexander, 2006a, 2006b) and the Galactic center is taken as the prototipical case. But recent observations revealing a dearth of giants inside 1 pc from *SgrA*$^*$ and raise the possibility that cored nuclei are common. Although stellar cusps may re-grow in less than a Hubble time, the existence of cored nuclei still remains a possibility—especially for nuclei with MBHs in the upper part of the mass range—, since cusp growth time scales can still be quite long (*e.g.* 6 Gyr in Milky Way type nuclei). However, since EMRI detection rates by LISA are expected to peak around $M_\bullet \sim 4\times 10^5 - 10^6 M_\odot$ (Gair, 2009), and re-growth times are $\lesssim 1$ Gyr for $M_\bullet \lesssim 1.2\times 10^6 M_\odot$, we still expect that a substantial fraction of EMRI events will originate from segregated stellar cusps. The Milky Way nucleus is *not* necessarily the prototype of the nucleus from which LISA detections will be more frequent. (Amaro-Seoane et al., 2010) develop a general theoretical framework for exploring the dependence of EMRI rates on the properties of the system, such as the central MBH mass.

## Acknowledgements

I have benefited from discussions with Pau Amaro-Seoane, Tal Alexander, José Fiestas and Rainer Spurzem.

This work was partly supported by DLR (Deutsches Zentrum für Luft- und Raumfahrt). The simulations have been carried out on the dedicated high-performance GRAPE-6A clusters at the Astronomisches Rechen-Institut in Heidelberg[3], which was funded by project I/80 041-043 of the Volkswagen Foundation and by the Ministry of Science, Research and the Arts of Baden-Württemberg (Az: 823.219-439/30 and /36). Some of the simulations were done at the Tuffstein cluster of the AEI.

---

[3] GRACE: see http://www.ari.uni-heidelberg.de/grace

# References in the highlight article

> *Selected abstracts*
>
> *October 2009 to March 2010*

## Searches for Cosmic-String Gravitational-Wave Bursts in Mock LISA Data

**Authors:** Cohen, Michael I.; Cutler, Curt; Vallisneri, Michele

**Eprint:** http://arxiv.org/abs/1002.4153

**Keywords:** bursts; data analysis; gr-qc; Metropolis-Hastings; MLDC; parameter estimation; search algorithms

**Abstract:** A network of observable, macroscopic cosmic (super-)strings may have formed in the early universe. If so, the cusps that generically develop on cosmic-string loops emit bursts of gravitational radiation that could be detectable by both ground- and space-based gravitational-wave interferometers. Here we report on two versions of a LISA-oriented string-burst search pipeline that we have developed and tested within the context of the Mock LISA Data Challenges. The two versions rely on the publicly available MultiNest and PyMC software packages, respectively. To reduce the effective dimensionality of the search space, our implementations use the F-statistic to analytically maximize over the signal's amplitude and polarization, A and psi, and use the FFT to search quickly over burst arrival times $t_C$. The standard F-statistic is essentially a frequentist statistic that maximizes the likelihood; we also demonstrate an approximate, Bayesian version of the F-statistic that incorporates realistic priors on A and psi. We calculate how accurately LISA can expect to measure the physical parameters of string-burst sources. To understand LISA's angular resolution for string-burst sources, we draw maps of the waveform fitting factor [maximized over (A psi, $t_C$)] as a function of sky position; these maps dramatically illustrate why (for LISA) inferring the correct sky location of the emitting string loop will often be practically impossible. We also identify and elucidate several symmetries that are imbedded in this search problem, and we derive the distribution of cut-off frequencies $f_{max}$ for observable bursts.

## Reduced Hamiltonian for next-to-leading order Spin-Squared Dynamics of General Compact Binaries

**Authors:** Hergt, Steven; Steinhoff, Jan; Schaefer, Gerhard

**Eprint:** http://arxiv.org/abs/1002.2093

**Keywords:** astro-ph.HE; gr-qc; post-Newtonian theory; spin

**Abstract:** Within the post Newtonian framework the fully reduced Hamiltonian (i.e., with eliminated spin supplementary condition) for the next-to-leading order spin-squared dynamics of general compact binaries is presented. The Hamiltonian is applicable to the spin dynamics of all kinds of binaries with self-gravitating components like black holes and/or neutron stars taking into account spin-induced quadrupolar deformation effects in second post-Newtonian order perturbation theory of Einstein's field equations. The corresponding equations of motion for spin, position and momentum variables are given in terms of canonical Poisson brackets. Comparison with a nonreduced potential calculated within the Effective Field Theory approach is made.

## Modulation of LISA free-fall orbits due to the Earth-Moon system

**Authors:** Cerdonio, M.; De Marchi, F.; De Pietri, R.; Jetzer, P.; Marzari, F.; Mazzolo, G.; Ortolan, A.; Sereno, M.

**Eprint:** http://arxiv.org/abs/1003.5528

**Keywords:** detectors; geodesic motion; gr-qc; interferometers

**Abstract:** We calculate the effect of the Earth-Moon (EM) system on the free-fall motion of LISA test masses. We show that the periodic gravitational pulling of the EM system induces a resonance with fundamental frequency 1 yr$^{-1}$ and a series of periodic perturbations with frequencies equal to integer harmonics of the synodic month (9.92 10$^{-7}$ Hz). We then evaluate the effects of these perturbations (up to the 6th harmonics) on the relative motions between each test masses couple, finding that they range between 3mm and 10pm for the 2nd and 6th harmonic, respectively. If we take the LISA sensitivity curve, as extrapolated down to 10$^{-6}$ Hz, we obtain that a few harmonics of the EM system can be detected in the Doppler data collected by the LISA space mission. This suggests that the EM system gravitational near field could provide an absolute calibration for the LISA sensitivity at very low frequencies.

## Relativistic Suppression of Black Hole Recoils

**Authors:** Kesden, Michael; Sperhake, Ulrich; Berti, Emanuele

**Abstract:** Numerical-relativity simulations indicate that the black hole produced in a binary merger can recoil with a velocity up to $v_{max}$~4,000 km/s with respect to the center of mass of the initial binary. This challenges the paradigm that most galaxies form through hierarchical mergers, yet retain supermassive black holes at their centers despite having escape velocities much less than $v_{max}$. Interaction with a circumbinary disk can align the binary black hole spins with their orbital angular momentum, reducing the recoil velocity of the final black hole produced in the subsequent merger. However, the effectiveness of this alignment depends on highly uncertain accretion flows near the binary black holes. In this Letter, we show that if the spin $S_1$ of the more massive binary black hole is even partially aligned with the orbital angular momentum L, relativistic spin precession on sub-parsec scales can align the binary black hole spins with each other. This alignment significantly reduces the recoil velocity even in the absence of gas. For example, if the angle between $S_1$ and L at large separations is 10 degrees while the second spin $S_2$ is isotropically distributed, the spin alignment discussed in this paper reduces the median recoil from 864 km/s to 273 km/s for maximally spinning black holes with a mass ratio of 9/11. This reduction will greatly increase the fraction of galaxies retaining their supermassive black holes.

## Symplectic Integration of Post-Newtonian Equations of Motion with Spin

**Abstract:** We present a non-canonically symplectic integration scheme tailored to numerically computing the post-Newtonian motion of a spinning black-hole binary. Using a splitting approach we combine the flows of orbital and spin contributions. In the context of the splitting, it is possible to integrate the individual terms of the spin-orbit and spin-spin Hamiltonians analytically, exploiting the special structure of the underlying equations of motion. The outcome is a symplectic, time-reversible integrator, which can be raised to arbitrary order by composition. A fourth-order version is shown to give excellent behavior concerning error growth and conservation of energy and angular momentum in long-term simulations. Favorable properties of the integrator are retained in the presence of weak dissipative forces due to radiation damping in the full post-Newtonian equations.

### Radio observations of NGC 6388: an upper limit on the mass of its central black hole

**Authors:** Cseh, D.; Kaaret, P.; Corbel, S.; Kording, E.; Coriat, M.; Tzioumis, A.; Lanzoni, B.

**Eprint:** http://arxiv.org/abs/1003.4604

**Keywords:** astro-ph.GA; astro-ph.HE; astrophysics; intermediate-mass black holes; observations

**Abstract:** We present the results of deep radio observations with the Australia Telescope Compact Array (ATCA) of the globular cluster NGC 6388. We show that there is no radio source detected (with a r.m.s. noise level of 27 uJy) at the cluster centre of gravity or at the locations of the any of the Chandra X-ray sources in the cluster. Based on the fundamental plane of accreting black holes which is a relationship between X-ray luminosity, radio luminosity and black hole mass, we place an upper limit of $1500 M_\odot$ on the mass of the putative intermediate-mass black hole located at the centre of NGC 6388. We discuss the uncertainties of this upper limit and the previously suggested black hole mass of $5700 M_\odot$ based on surface density profile analysis.

### On the Dynamics and Evolution of Gravitational Instability-Dominated Disks

**Authors:** Krumholz, Mark R.; Burkert, Andreas

**Eprint:** http://arxiv.org/abs/1003.4513

**Keywords:** accretion discs; astro-ph.CO; astro-ph.GA; astrophysics

**Abstract:** We present a first-principles derivation of the evolution equations describing a thin axisymmetric disk of gas and stars with an arbitrary rotation curve that is kept in a state of marginal gravitational instability and energy equilibrium due to the balance between energy released by accretion and energy lost due to decay of turbulence. Unlike previous analyses of this problem, our results do not depend on an assumed model for the rate of mass and angular momentum transport due to gravitational instability, or on an order-of-magnitude energy equilibrium argument. Instead, we self-consistently determine the position- and time-dependent transport rates from the fluid dynamical equations. We show that there is a steady-state configuration for disks dominated by gravitational instability, and for disks in this state we analytically determine the velocity dispersion, surface density, and rates of mass and angular momentum transport as a function of the gas mass fraction, the rotation curve, and the rate of external accretion onto the disk edge. We show that disks that are initially out of steady state will evolve into it on timescales comparable to the orbital period if the accretion rate is high. Finally, we discuss the implications of these results for the structure of disks in a broad range of environments, including high redshift galaxies, the outer gaseous disks of local galaxies, and accretion disks around protostars.

## Dynamical shift condition for unequal mass black hole binaries

**Authors:** Mueller, Doreen; Grigsby, Jason; Bruegmann, Bernd

**Eprint:** http://arxiv.org/abs/1003.4681

**Keywords:** gr-qc; massive binaries of black holes; numerical relativity; supermassive black holes; waveforms

**Abstract:** Certain numerical frameworks used for the evolution of binary black holes make use of a gamma driver, which includes a damping factor. Such simulations typically use a constant value for damping. However, it has been found that very specific values of the damping factor are needed for the calculation of unequal mass binaries. We examine carefully the role this damping plays, and provide two explicit, non-constant forms for the damping to be used with mass-ratios further from one. Our analysis of the resultant waveforms compares well against the constant damping case.

## Formation of Supermassive Black Holes

**Authors:** Volonteri, Marta

**Eprint:** http://arxiv.org/abs/1003.4404

**Keywords:** astro-ph.CO; cosmology; supermassive black holes

How stars distribute around a massive black hole

**Abstract:** Evidence shows that massive black holes reside in most local galaxies. Studies have also established a number of relations between the MBH mass and properties of the host galaxy such as bulge mass and velocity dispersion. These results suggest that central MBHs, while much less massive than the host ($\sim 0.1\%$), are linked to the evolution of galactic structure. In hierarchical cosmologies, a single big galaxy today can be traced back to the stage when it was split up in hundreds of smaller components. Did MBH seeds form with the same efficiency in small proto-galaxies, or did their formation had to await the buildup of substantial galaxies with deeper potential wells? I briefly review here some of the physical processes that are conducive to the evolution of the massive black hole population. I will discuss black hole formation processes for 'seed' black holes that are likely to place at early cosmic epochs, and possible observational tests of these scenarios.

## A General Formula for Black Hole Gravitational Wave Kicks

**Authors:** van Meter, James R.; Miller, M. Coleman; Baker, John G.; Boggs, William D.; Kelly, Bernard J.

**Eprint:** `http://arxiv.org/abs/1003.3865`

**Keywords:** astro-ph.HE; gr-qc; gravitational recoil; kicks/recoil; massive binaries of black holes; numerical relativity

**Abstract:** Although the gravitational wave kick velocity in the orbital plane of coalescing black holes has been understood for some time, apparently conflicting formulae have been proposed for the dominant out-of-plane kick, each a good fit to different data sets. This is important to resolve because it is only the out-of-plane kicks that can reach more than 500 km/s and can thus eject merged remnants from galaxies. Using a different ansatz for the out-of-plane kick, we show that we can fit almost all existing data to better than 5 %. This is good enough for any astrophysical calculation, and shows that the previous apparent conflict was only because the two data sets explored different aspects of the kick parameter space.

## Discovery of Four kpc-Scale Binary AGNs

**Authors:** Liu, Xin; Greene, Jenny E.; Shen, Yue; Strauss, Michael A.

**Eprint:** `http://arxiv.org/abs/1003.3467`

**Keywords:** astro-ph.CO; astrophysics; massive binaries of black holes; observations; supermassive black holes

**Abstract:** We report the discovery of four kpc-scale binary AGNs. These objects were originally selected from the Sloan Digital Sky Survey based on double-peaked [O III] 4959,5007 emission lines in their fiber spectra. The double peaks could result from pairing active supermassive black holes (SMBHs) in a galaxy merger, or could be due to bulk motions of narrow-line region gas around a single SMBH. Deep near-infrared (NIR) images and optical slit spectra obtained from the Magellan 6.5 m and the APO 3.5 m telescopes strongly support the binary SMBH scenario for the four objects. In each system, the NIR images reveal tidal features and double stellar bulges with a projected separation of several kpc, while optical slit spectra show two Seyfert 2 nuclei spatially coincident with the stellar bulges, with line-of-sight velocity offsets of a few hundred km/s. These objects were drawn from a sample of only 43 objects, demonstrating the efficiency of this technique to find kpc-scale binary AGNs.

## Intermediate-Mass Black Holes in Early Globular Clusters

**Authors:** Vesperini, Enrico; McMillan, Stephen L. W.; D'Ercole, Annibale; D'Antona, Francesca

**Abstract:** Spectroscopic and photometric observations show that many globular clusters host multiple stellar populations, challenging the common paradigm that globular clusters are "simple stellar populations" composed of stars of uniform age and chemical composition. The chemical abundances of second-generation (SG) stars constrain the sources of gas out of which these stars must have formed, indicating that the gas must contain matter processed through the high-temperature CNO cycle. First-generation massive Asymptotic Giant Branch (AGB) stars have been proposed as the source of this gas. In a previous study, by means of hydrodynamical and N-body simulations, we have shown that the AGB ejecta collect in a cooling flow in the cluster core, where the gas reaches high densities, ultimately forming a centrally concentrated subsystem of SG stars. In this Letter we show that the high gas density can also lead to significant accretion onto a pre-existing seed black hole. We show that gas accretion can increase the black hole mass by up to a factor of 100. The details of the gas dynamics are important in determining the actual black hole growth. Assuming a near-universal seed black hole mass and small cluster-to-cluster variations in the duration of the SG formation phase, the outcome of our scenario is one in which the present intermediate-mass black hole (IMBH) mass may have only a weak dependence on the current cluster properties. The scenario presented provides a natural mechanism for the formation of an IMBH at the cluster center during the SG star-formation phase.

## The gravitational-wave memory effect

**Authors:** Favata, Marc

**Eprint:** http://arxiv.org/abs/1003.3486

**Keywords:** astro-ph.HE; general relativity; gr-qc; post-Newtonian theory; waveforms

**Abstract:** The nonlinear memory effect is a slowly-growing, non-oscillatory contribution to the gravitational-wave amplitude. It originates from gravitational waves that are sourced by the previously emitted waves. In an ideal gravitational-wave interferometer a gravitational-wave with memory causes a permanent displacement of the test masses that persists after the wave has passed. Surprisingly, the nonlinear memory affects the signal amplitude starting at leading (Newtonian-quadrupole) order. Despite this fact, the nonlinear memory is not easily extracted from current numerical relativity simulations. After reviewing the linear and nonlinear memory I summarize some recent work, including: (1) computations of the memory contribution to the inspiral waveform amplitude (thus completing the waveform to third post-Newtonian order); (2) the first calculations of the nonlinear memory that include all phases of binary black hole coalescence (inspiral, merger, ringdown); and (3) realistic estimates of the detectability of the memory with LISA.

## Eccentric orbital motion of compact binaries with aligned spins and angular momentum under higher order spin coupling

**Authors:** Tessmer, Manuel; Hartung, Johannes; Schaefer, Gerhard

**Eprint:** http://arxiv.org/abs/1003.2735

**Keywords:** gr-qc; massive binaries of black holes; post-Newtonian theory; waveforms

**Abstract:** A quasi-Keplerian parameterisation for the solutions of second post-Newtonian (PN) accurate equations of motion for spinning compact binaries is obtained including leading order spin-spin and next-to-leading order spin-orbit interactions. Rotational deformation of the compact objects is incorporated. For arbitrary mass ratios the spin orientations are taken to be parallel or anti-parallel to the orbital angular momentum vector. The emitted gravitational wave forms are given in analytic form up to 2PN point particle, 1.5PN spin orbit and 1PN spin-spin contributions, where the spins are counted of 0PN order.

## The Underluminous Nature of Sgr A*

**Authors:** Yusef-Zadeh, F.; Wardle, M.

**Eprint:** http://arxiv.org/abs/1003.1519

**Keywords:** astro-ph.GA; astrophysics; observations; Sagittarius A*; supermassive black holes

**Abstract:** In the last several years, a number of observing campaigns of the massive black hole Sgr A* has been carried out in order to address two important issues: one concerns the underluminous nature of Sgr A* with its bolometric luminosity being several orders of magnitude less than those of its more massive counterparts. It turns out that the angular momentum of the ionized stellar winds from orbiting stars in one or two disks orbiting Sgr A* could be a critical factor in estimating accurately the accretion rate unto Sgr A*. A net angular momentum of ionized gas feeding Sgr A* could lower the Bondi rate. Furthermore, the recent time delay picture of the peak flare emission can be understood in the context of adiabatic expansion of hot plasma. The expansion speed of the plasma is estimated to be sub-relativistic. However, relativistic bulk motion of the plasma could lead to outflow from Sgr A*. Significant outflow from Sgr A* could then act as a feedback which could then reduce Bondi accretion rate. These uncertain factors can in part explain the underluminous nature of Sgr A*. The other issue is related to the emission mechanism and the cause of flare activity in different wavelength bands. Modeling of X-ray and near-IR flares suggests that inverse Compton scattering (ICS) of IR flare photons by the energetic electrons responsible for the submm emission can account for the X-ray flares. A time delay of minutes to tens of minutes is predicted between the peak flaring in the near-IR and X-rays, NOT due to adiabatic expansion of optically thick hot plasma, but to the time taken for IR flare photons to cross the accretion flow before being upscattered.

## Understanding the "anti-kick" in the merger of binary black holes

**Authors:** Rezzolla, Luciano; Macedo, Rodrigo P.; Jaramillo, José Luis

**Eprint:** http://arxiv.org/abs/1003.0873

**Keywords:** gr-qc; gravitational recoil; kicks/recoil; massive binaries of black holes; numerical relativity

**Abstract:** The generation of a large recoil velocity from the inspiral and merger of binary black holes represents one of the most exciting results of numerical-relativity calculations. While many aspects of this process have been investigated and explained, the "anti-kick", namely the sudden deceleration after the merger, has not yet found a simple explanation. We show that the anti-kick can be easily understood in terms of the radiation from a deformed black hole where the intrinsically anisotropic curvature distribution on the horizon determines the direction and intensity of the recoil. Our analysis is focussed on the properties of Robinson-Trautman spacetimes and allows us to measure both the energies and momenta radiated in a gauge-invariant manner. At the same time, this simpler setup provides all the qualitative but also quantitative features of inspiralling black hole binaries, thus opening the way to a deeper understanding of the nonlinear dynamics of black-hole spacetimes.

## Semianalytical estimates of scattering thresholds and gravitational radiation in ultrarelativistic black hole encounters

**Authors:** Berti, Emanuele; Cardoso, Vitor; Hinderer, Tanja; Lemos, Madalena; Pretorius, Frans; Sperhake, Ulrich; Yunes, Nicolas

**Eprint:** http://arxiv.org/abs/1003.0812

**Keywords:** general relativity; geodesic motion; gr-qc; massive binaries of black holes; numerical relativity

**Abstract:** Ultrarelativistic collisions of black holes are ideal gedanken experiments to study the nonlinearities of general relativity. In this paper we use semianalytical tools to better understand the nature of these collisions and the emitted gravitational radiation. We explain many features of the energy spectra extracted from numerical relativity simulations using two complementary semianalytical calculations. In the first calculation we estimate the radiation by a "zero-frequency limit" analysis of the collision of two point particles with finite impact parameter. In the second calculation we replace one of the black holes by a point particle plunging with arbitrary energy and impact parameter into a Schwarzschild black hole, and we explore the multipolar structure of the radiation paying particular attention to the near-critical regime. We also use a geodesic analogy to provide qualitative estimates of the dependence of the scattering threshold on the black hole spin and on the dimensionality of the spacetime.

## The NGC 404 Nucleus: Star Cluster and Possible Intermediate Mass Black Hole

**Authors:** Seth, Anil C.; Cappellari, Michele; Neumayer, Nadine; Caldwell, Nelson; Bastian, Nate; Olsen, Knut; Blum, Robert D.; Debattista, Victor P.; McDermid, Richard; Puzia, Thomas; Stephens, Andrew

**Eprint:** http://arxiv.org/abs/1003.0680

**Keywords:** astro-ph.CO; astro-ph.GA; astrophysics; intermediate-mass black holes; observations

**Abstract:** We examine the nuclear morphology, kinematics, and stellar populations in nearby S0 galaxy NGC 404 using a combination of adaptive optics assisted near-IR integral-field spectroscopy, optical spectroscopy, and HST imaging. These observations enable study of the NGC 404 nucleus at a level of detail possible only in the nearest galaxies. The surface brightness profile suggests the presence of three components, a bulge, a nuclear star cluster, and a central light excess within the cluster at radii <3 pc. These components have distinct kinematics with modest rotation seen in the nuclear star cluster and counter-rotation seen in the central excess. Molecular hydrogen emission traces a disk with rotation nearly orthogonal to that of the stars. The stellar populations of the three components are also distinct, with half of the mass of the nuclear star cluster having ages of ~ 1 Gyr (perhaps resulting from a galaxy merger), while the bulge is dominated by much older stars. Dynamical modeling of the stellar kinematics gives a total nuclear star cluster mass of $1.1 \times 10^7 M_\odot$. Dynamical detection of a possible intermediate mass black hole is hindered by uncertainties in the central stellar mass profile. Assuming a constant mass-to-light ratio, the stellar dynamical modeling suggests a black hole mass of $< 1 \times 10^5 M_\odot$, while the molecular hydrogen gas kinematics are best fit by a black hole with mass of $4.5 \times 10^5 M_\odot$. Unresolved and possibly variable dust emission in the near-infrared and AGN-like molecular hydrogen emission line ratios do suggest the presence of an accreting black hole in this nearby LINER galaxy.

## Binary spinning black hole Hamiltonian in canonical center-of-mass and rest-frame coordinates through higher post-Newtonian order

**Authors:** Rothe, Tilman J.; Schäfer, Gerhard

**Eprint:** http://arxiv.org/abs/1003.0390

**Abstract:** The recently constructed Hamiltonians for spinless binary black holes through third post-Newtonian order and for spinning ones through formal second post-Newtonian order, where the spins are counted of zero post-Newtonian order, are transformed into fully canonical center-of-mass and rest-frame variables. The mixture terms in the Hamiltonians between center-of-mass and rest-frame variables are in accordance with the relation between the total linear momentum and the center-of-mass velocity as demanded by global Lorentz invariance. The various generating functions for the center-of-mass and rest-frame canonical variables are explicitly given in terms of the single-particle canonical variables. The no-interaction theorem does not apply because the world-line condition of Lorentz covariant position variables is not imposed.

## A mass estimate of an intermediate-mass black hole in omega Centauri

**Authors:** Miocchi, P.

**Eprint:** <http://arxiv.org/abs/1002.5037>

**Keywords:** astro-ph.GA; astrophysics; globular clusters; intermediate-mass black holes; observations; stellar dynamics

**Abstract:**

Context. The problem of the existence of intermediate-mass black holes (IMBHs) at the centre of globular clusters is a hot and controversial topic in current astrophysical research with important implications in stellar and galaxy formation.

Aims. In this paper, we aim at giving further support to the presence of an IMBH in omega Centauri and at providing an independent estimate of its mass.

Methods. We employed a self-consistent spherical model with anisotropic velocity distribution. It consists in a generalisation of the King model by including the Bahcall-Wolf distribution function in the IMBH vicinity.

Results. By the parametric fitting of the model to recent HST/ACS data for the surface brightness profile, we found an IMBH to cluster total mass ratio of $M_{BH}/M = 5.8(+0.9 - 1.2) \times 10^{-3}$. It is also found that the model yields a fit of the line-of-sight velocity dispersion profile that is better without mass segregation than in the segregated case. This confirms the current thought of a non-relaxed status for this peculiar cluster. The best fit model to the kinematic data leads, moreover, to a cluster total mass estimate of M = (3.1 +/- 0.3) ×$10^6 M_\odot$, thus giving an IMBH mass in the range 13,000 < $M_{BH}$ 12') is required to match the outer surface brightness profile.

## On the angular momentum transport due to vertical convection in accretion discs

**Authors:** Lesur, G.; Ogilvie, G. I.

**Eprint:** http://arxiv.org/abs/1002.4621

**Keywords:** accretion discs; astro-ph.EP; astro-ph.SR; astrophysics; EM counterparts

**Abstract:** The mechanism of angular momentum transport in accretion discs has long been debated. Although the magnetorotational instability appears to be a promising process, poorly ionized regions of accretion discs may not undergo this instability. In this letter, we revisit the possibility of transporting angular momentum by turbulent thermal convection. Using high-resolution spectral methods, we show that strongly turbulent convection can drive outward angular momentum transport at a rate that is, under certain conditions, compatible with observations of discs. We find however that the angular momentum transport is always much weaker than the vertical heat transport. These results indicate that convection might be another way to explain global disc evolution, provided that a sufficiently unstable vertical temperature profile can be maintained.

## EM counterparts of recoiling black holes: general relativistic simulations of non-Keplerian discs

**Authors:** Zanotti, Olindo; Rezzolla, Luciano; Del Zanna, Luca; Palenzuela, Carlos

**Eprint:** http://arxiv.org/abs/1002.4185

**Keywords:** astro-ph.HE; EM counterparts; gr-qc; massive binaries of black holes; numerical relativity

**Abstract:** We investigate the dynamics of a circumbinary disc that responds to the loss of mass and to the recoil velocity of the black hole produced by the merger of a binary system of supermassive black holes. More specifically, we perform the first two-dimensional general relativistic hydrodynamics simulations of extended non-Keplerian discs and employ a new technique to construct a "shock detector", thus determining the precise location of the shocks produced in the accreting disc by the recoiling black hole. In this way we can study how the properties of the system, such as the spin, mass and recoil velocity of the black hole, affect the mass accretion rate and are imprinted on the electromagnetic emission from these sources. In contrast with what done in similar works, we here question the estimates of the bremsstrahlung luminosity when computed without properly taking into account the radiation transfer, thus yielding cooling times that are unrealistically short. At the same time we show, through an approximation based on the relativistic analogue of the isothermal evolution of Corrales 2009, that the luminosity produced can reach a peak value above $L \simeq 10^{43}$erg/s at about $\sim$ 20d after the merger of a binary with total mass $M \simeq 10^6 M_\odot$ and persist for several days at values which are a factor of a few smaller. If confirmed by more sophisticated calculations such a signal could indeed lead to an electromagnetic counterpart of the merger of binary black-hole system.

### Evolution of massive black hole spins

**Authors:** Volonteri, Marta

**Eprint:** http://arxiv.org/abs/1002.3827

**Keywords:** astro-ph.CO; astro-ph.HE; astrophysics; cosmology; spin; supermassive black holes

**Abstract:** Black hole spins affect the efficiency of the "classical" accretion processes, hence the radiative output from quasars. Spins also determine how much energy is extractable from the hole itself. Recently it became clear that massive black hole spins also affect the retention of black holes in galaxies, be cause of the impulsive "gravitational recoil", up to thousands km/s, due to anisotropic emission of gravitational waves at merger. I discuss here the evolution of massive black hole spins along the cosmic history, due to the combination of mergers and accretion events. I describe recent simulations of accreting black holes in merger remnants, and discuss the implication for the spins of black holes in quasars.

### Chandra and Swift Follow-up Observations of the Intermediate Mass Black Hole in ESO243-49

**Authors:** Webb, N. A.; Barret, D.; Godet, O.; Servillat, M.; Farrell, S. A.; Oates, S. R.

**Eprint:** http://arxiv.org/abs/1002.3625

**Keywords:** astro-ph.HE; astrophysics; IMRI; intermediate-mass black holes; observations

**Abstract:** The brightest Ultra-Luminous X-ray source HLX-1 in the galaxy ESO 243-49 provides strong evidence for the existence of intermediate mass black holes. As the luminosity and thus the mass estimate depend on the association of HLX-1 with ESO 243-49, it is essential to confirm its affiliation. This requires follow-up investigations at wavelengths other than X-rays, which in-turn needs an improved source position. To further reinforce the intermediate mass black hole identification, it is necessary to determine HLX-1's environment to establish whether it could potentially form and nourish a black hole at the luminosities observed. Using the High Resolution Camera onboard Chandra, we determine a source position of RA=01h10m28.3s and Dec=-46d04'22.3". A conservative 95% error of 0.3" was found following a boresight correction by cross-matching the positions of 3 X-ray sources in the field with the 2MASS catalog. Combining all Swift UV/Optical Telescope uvw2 images, we failed to detect a UV source at the Chandra position down to a 3sigma limiting magnitude of 20.25 mag. However, there is evidence that the UV emission is elongated in the direction of HLX-1. This is supported by archival data from GALEX and suggests that the far-UV emission is stronger than the near-UV. This could imply that HLX-1 may be situated near the edge of a star forming region. Using the latest X-ray observations we deduce the mass accretion rate of a 500 Msun black hole with the observed luminosity and show that this is compatible with such an environment.

## Connections Between Local and Global Turbulence in Accretion Disks

**Authors:** Sorathia, Kareem A.; Reynolds, Christopher S.; Armitage, Philip J.

**Abstract:** We analyze a suite of global magnetohydrodynamic (MHD) accretion disk simulations in order to determine whether scaling laws for turbulence driven by the magnetorotational instability, discovered via local shearing box studies, are globally robust. The simulations model geometrically-thin disks with zero net magnetic flux and no explicit resistivity or viscosity. We show that the local Maxwell stress is correlated with the self-generated local vertical magnetic field in a manner that is similar to that found in local simulations. Moreover, local patches of vertical field are strong enough to stimulate and control the strength of angular momentum transport across much of the disk. We demonstrate the importance of magnetic linkages (through the low-density corona) between different regions of the disk in determining the local field, and suggest a new convergence requirement for global simulations – the vertical extent of the corona must be fully captured and resolved. Finally, we examine the temporal convergence of the average stress, and show that an initial long-term secular drift in the local flux-stress relation dies away on a time scale that is consistent with turbulent mixing of the initial magnetic field.

## Further Observations of the Intermediate Mass Black Hole Candidate ESO 243-49 HLX-1

**Authors:** Farrell, S. A.; Servillat, M.; Oates, S. R.; Heywood, I.; Godet, O.; Webb, N. A.; Barret, D.

**Eprint:** http://arxiv.org/abs/1002.3404

**Keywords:** astro-ph.CO; astro-ph.HE; astrophysics; IMRI; intermediate-mass black holes; observations

**Abstract:** The brightest Ultra-Luminous X-ray source HLX-1 in the galaxy ESO 243-49 currently provides strong evidence for the existence of intermediate mass black holes. Here we present the latest multi-wavelength results on this intriguing source in X-ray, UV and radio bands. We have refined the X-ray position to sub-arcsecond accuracy. We also report the detection of UV emission that could indicate ongoing star formation in the region around HLX-1. The lack of detectable radio emission at the X-ray position strengthens the argument against a background AGN.

## Gravitational Wave Signal from Assembling the Lightest Supermassive Black Holes

**Authors:** Holley-Bockelmann, Kelly; Micic, Miroslav; Sigurdsson, Steinn; Rubbo, Louis

**Eprint:** http://arxiv.org/abs/1002.3378

**Keywords:** astro-ph.CO; astrophysics; cosmology; massive binaries of black holes; supermassive black holes

**Abstract:**

We calculate the gravitational wave signal from the growth of 10 million solar mass supermassive black holes (SMBH) from the remnants of Population III stars. The assembly of these lower mass black holes is particularly important because observing SMBHs in this mass range is one of the primary science goals for the Laser Interferometer Space Antenna (LISA), a planned NASA/ESA mission to detect gravitational waves. We use high resolution cosmological N-body simulations to track the merger history of the host dark matter halos, and model the growth of the SMBHs with a semi-analytic approach that combines dynamical friction, gas accretion, and feedback. We find that the most common source in the LISA band from our volume consists of mergers between intermediate mass black holes and SMBHs at redshifts less than 2.

This type of high mass ratio merger has not been widely considered in the gravitational wave community; detection and characterization of this signal will likely require a different technique than is used for SMBH mergers or extreme mass ratio inspirals. We find that the event rate of this new LISA source depends on prescriptions for gas accretion onto the black hole as well as an accurate model of the dynamics on a galaxy scale; our best estimate yields about 40 sources with a signal-to-noise ratio greater than 30 occur within a volume like the Local Group during SMBH assembly – extrapolated over the volume of the universe yields roughly 500 observed events over 10 years, although the accuracy of this rate is affected by cosmic variance.

## Flares from Sgr A* and their emission mechanism

**Authors:** Dodds-Eden, K.; Porquet, D.; Trap, G.; Quataert, E.; Gillessen, S.; Grosso, N.; Genzel, R.; Goldwurm, A.; Yusef-Zadeh, F.; Trippe, S.; Bartko, H.; Eisenhauer, F.; Ott, T.; Fritz, T. K.; Pfuhl, O.

**Eprint:** http://arxiv.org/abs/1002.2885

**Keywords:** astro-ph.GA; astro-ph.HE; astrophysics; observations; Sagittarius A*

**Abstract:** We summarize recent observations and modeling of the brightest Sgr A* flare to be observed simultaneously in (near-)infrared and X-rays to date. Trying to explain the spectral characteristics of this flare through inverse Compton mechanisms implies physical parameters that are unrealistic for Sgr A*. Instead, a "cooling break" synchrotron model provides a more feasible explanation for the X-ray emission. In a magnetic field of about 5-30 Gauss the X-ray emitting electrons cool very quickly on the typical dynamical timescale while the NIR-emitting electrons cool more slowly. This produces a spectral break in the model between NIR and X-ray wavelengths that can explain the differences in the observed spectral indices.

## Final spins from the merger of precessing binary black holes

**Authors:** Kesden, Michael; Sperhake, Ulrich; Berti, Emanuele

**Abstract:** The inspiral of binary black holes is governed by gravitational radiation reaction at binary separations r 10 M. Fortunately, binary evolution between these separations is well described by post-Newtonian equations of motion. We examine how this post-Newtonian evolution affects the distribution of spin orientations at separations r near 10 M where numerical-relativity simulations typically begin. Although isotropic spin distributions at r =1000 M remain isotropic at r = 10 M, distributions that are initially partially aligned with the orbital angular momentum can be significantly distorted during the post-Newtonian inspiral. Spin-orbit resonances tend to align (anti-align) the binary black hole spins with each other if the spins were initially partially aligned (anti-aligned) with respect to the orbital angular momentum, thus increasing (decreasing) the average final spin. Resonant effects are stronger for comparable-mass binaries, and they could produce significant spin alignment in massive black hole mergers at high redshifts and in stellar-mass black hole binaries. We also point out that precession induces an intrinsic accuracy limitation of 0.03 in the dimensionless spin magnitude, and about 20 degrees in the direction in predicting the final spin resulting from widely separated binary configurations.

## HST Palpha Survey of the Galactic Center – Searching the missing young stellar populations within the Galactic Center

**Authors:** Dong, H.; Wang, Q. D.; Cotera, A.; Stolovy, S.; Morris, M. R.; Mauerhan, J.; Mills, E. A.; Schneider, G.; Lang, C.

**Abstract:** We present preliminary results of our HST Paα survey of the Galactic Center (gc), which maps the central 0.65×0.25 degrees around Sgr A*. This survey provides us with a more complete inventory of massive stars within the gc, compared to previous observations. We find 157 Paα emitting sources, which are evolved massive stars. Half of them are located outside of three young massive star clusters near Sgr A*. The loosely spatial distribution of these field sources suggests that they are within less massive star clusters/groups, compared to the three massive ones. Our Paα mosaic not only resolves previously well-known large-scale filaments into fine structures, but also reveals many new extended objects, such as bow shocks and H II regions. In particular, we find two regions with large-scale Pa$\alpha$ diffuse emission and tens of Pa$\alpha$ emitting sources in the negative Galactic longitude suggesting recent star formation activities, which were not known previously. Furthermore, in our survey, we detect ~0.6 million stars, most of which are red giants or AGB stars. Comparisons of the magnitude distribution in 1.90 $\mu$m and those from the stellar evolutionary tracks with different star formation histories suggest an episode of star formation process about 350 Myr ago in the gc .

## Evolution and instabilities of disks harboring super massive black holes

**Authors:** Curir, Anna; de Romeri, Valentina; Murante, Giuseppe

**Eprint:** http://arxiv.org/abs/1002.2562

**Keywords:** accretion discs; astro-ph.CO; astrophysics; EM counterparts; supermassive black holes

**Abstract:** The bar formation is still an open problem in modern astrophysics. In this paper we present numerical simulation performed with the aim of analyzing the growth of the bar instability inside stellar-gaseous disks, where the star formation is triggered, and a central black hole is present. The aim of this paper is to point out the impact of such a central massive black hole on the growth of the bar. We use N-body-SPH simulations of the same isolated disk-to-halo mass systems harboring black holes with different initial masses and different energy feedback on the surrounding gas. We compare the results of these simulations with the one of the same disk without black hole in its center. We make the same comparison (disk with and without black hole) for a stellar disk in a fully cosmological scenario. A stellar bar, lasting 10 Gyrs, is present in all our simulations. The central black hole mass has in general a mild effect on the ellipticity of the bar but it is never able to destroy it. The black holes grow in different way according their initial mass and their feedback efficiency, the final values of the velocity dispersions and of the black hole masses are near to the phenomenological constraints.

## Cover art: issues in the metric-guided and metric-less placement of random and stochastic template banks

**Authors:** Manca, Gian Mario; Vallisneri, Michele

**Eprint:** http://arxiv.org/abs/0909.0563

**Keywords:** data analysis; detectors; gr-qc; instruments; interferometers; Metropolis-Hastings; MLDC; numerical methods; parameter estimation; search algorithms

**Abstract:** The efficient placement of signal templates in source-parameter space is a crucial requisite for exhaustive matched-filtering searches of modeled gravitational-wave sources. Unfortunately, the current placement algorithms based on regular parameter-space meshes are difficult to generalize beyond simple signal models with few parameters. Various authors have suggested that a general, flexible, yet efficient alternative can be found in randomized placement strategies such as random placement and stochastic placement, which enhances random placement by selectively rejecting templates that are too close to others. In this article we explore several theoretical and practical issues in randomized placement: the size and performance of the resulting template banks; the effects of parameter-space boundaries; the use of quasi-random (self avoiding) number sequences; most important, the implementation of these algorithms in curved signal manifolds with and without the use of a Riemannian signal metric, which may be difficult to obtain. Specifically, we show how the metric can be replaced with a discrete triangulation-based representation of local geometry. We argue that the broad class of randomized placement algorithms offers a promising answer to many search problems, but that the specific choice of a scheme and its implementation details will still need to be fine-tuned separately for each problem.

## Gravitational self-force on a particle in eccentric orbit around a Schwarzschild black hole

**Authors:** Barack, Leor; Sago, Norichika

**Eprint:** <http://arxiv.org/abs/1002.2386>

**Keywords:** EMRI; geodesic motion; gr-qc; numerical methods; self force

**Abstract:** We present a numerical code for calculating the local gravitational self-force acting on a pointlike particle in a generic (bound) geodesic orbit around a Schwarzschild black hole. The calculation is carried out in the Lorenz gauge: For a given geodesic orbit, we decompose the Lorenz-gauge metric perturbation equations (sourced by the delta-function particle) into tensorial harmonics, and solve for each harmonic using numerical evolution in the time domain (in 1+1 dimensions). The physical self-force along the orbit is then obtained via mode-sum regularization. The total self-force contains a dissipative piece as well as a conservative piece, and we describe a simple method for disentangling these two pieces in a time-domain framework. The dissipative component is responsible for the loss of orbital energy and angular momentum through gravitational radiation; as a test of our code we demonstrate that the work done by the dissipative component of the computed force is precisely balanced by the asymptotic fluxes of energy and angular momentum, which we extract independently from the wave-zone numerical solutions. The conservative piece of the self force does not affect the time-averaged rate of energy and angular-momentum loss, but it influences the evolution of the orbital phases; this piece is calculated here for the first time in eccentric strong-field orbits. As a first concrete application of our code we recently reported the value of the shift in the location and frequency of the innermost stable circular orbit due to the conservative self-force [Phys. Rev. Lett. **102**, 191101 (2009)]. Here we provide full details of this analysis, and discuss future applications.

## Young massive star clusters

**Authors:** Zwart, Simon Portegies; McMillan, Steve; Gieles, Mark

**Eprint:** http://arxiv.org/abs/1002.1961

**Keywords:** astro-ph.GA; astro-ph.SR; astrophysics; globular clusters; GPU; GRAPE hw; IMRI; intermediate-mass black holes; stellar dynamics

**Abstract:** Young massive clusters are dense aggregates of young stars that form the fundamental building blocks of galaxies. Several examples exist in the Milky Way Galaxy and the Local Group, but they are particularly abundant in starburst and interacting galaxies. The few young massive clusters that are close enough to resolve are of prime interest for studying the stellar mass function and the ecological interplay between stellar evolution and stellar dynamics. The distant unresolved clusters may be effectively used to study the star-cluster mass function, and they provide excellent constraints on the formation mechanisms of young cluster populations. Young massive clusters are expected to be the nurseries for many unusual objects, including a wide range of exotic stars and binaries. So far only a few such objects have been found in young massive clusters, although their older cousins, the globular clusters, are unusually rich in stellar exotica. In this review we focus on star clusters younger than $\sim$ 100 Myr, more than a few current crossing times old, and more massive than $\sim 10^4 M_\odot$, irrespective of cluster size or environment. We describe the global properties of the currently known young massive star clusters in the Local Group and beyond, and discuss the state of the art in observations and dynamical modeling of these systems. In order to make this review readable by observers, theorists, and computational astrophysicists, we also review the cross-disciplinary terminology.

## Growing Massive Black Hole Pairs in Minor Mergers of Disk Galaxies

**Authors:** Callegari, S.; Kazantzidis, S.; Mayer, L.; Colpi, M.; Bellovary, J. M.; Quinn, T.; Wadsley, J.

**Eprint:** http://arxiv.org/abs/1002.1712

**Keywords:** astro-ph.CO; astrophysics; cosmology; massive binaries of black holes

**Abstract:** We perform a suite of high-resolution smoothed particle hydrodynamics simulations to investigate the evolution of massive black hole (MBH) pairs during minor mergers of disk galaxies. Our simulation set includes star formation and accretion onto the MBHs, as well as feedback from both processes. We consider 1:10 merger events occurring around a predicted peak of MBH pair formation at a redshift of $z \sim 3$, in the sensitivity window of the Laser Interferometer Space Antenna. Owing to strong tidal torques acting on its host and orbital circularization inside the disk of the primary galaxy, the companion MBH undergoes distinct episodes of enhanced accretion which cause an increase of the initial 1:10 mass ratio of the MBHs. We also find that the efficiency of MBH pair formation in the nuclei of the remnants correlates with the final mass ratio of the pair itself, so that MBH pairs with larger mass ratios are produced more effectively and promptly. Depending on the initial fraction of cold gas in the galactic disks and the geometry of the encounter, the final mass ratios of the resulting MBH pairs can be as large as 1:2, suggesting that minor galaxy mergers can give rise to MBH pairs with major mass ratios. These findings indicate that the mass ratios of MBH pairs in galactic nuclei do not necessarily trace the mass ratios of their host merging galaxies, but are a consequence of the complex interplay between accretion and merger dynamics.

## Accretion and Outflow in Active Galaxies

**Authors:** King, Andrew

**Eprint:** http://arxiv.org/abs/1002.1808

**Keywords:** accretion discs; astro-ph.CO; astro-ph.HE; EM counterparts; spin; supermassive black holes

**Abstract:**

I review accretion and outflow in active galactic nuclei. Accretion appears to occur in a series of very small–scale, chaotic events, whose gas flows have no correlation with the large–scale structure of the galaxy or with each other. The accreting gas has extremely low specific angular momentum and probably represents only a small fraction of the gas involved in a galaxy merger, which may be the underlying driver.

Eddington accretion episodes in AGN must be common in order for the supermassive black holes to grow. I show that they produce winds with velocities $v \sim 0.1c$ and ionization parameters implying the presence of resonance lines of helium– and hydrogenlike iron. The wind creates a strong cooling shock as it interacts with the interstellar medium of the host galaxy, and this cooling region may be observable in an inverse Compton continuum and lower–excitation emission lines associated with lower velocities. The shell of matter swept up by the shocked wind stalls unless the black hole mass has reached the value $M_\sigma$ implied by the $M - \sigma$ relation. Once this mass is reached, further black hole growth is prevented. If the shocked gas did not cool as asserted above, the resulting ('energy-driven') outflow would imply a far smaller SMBH mass than actually observed. Minor accretion events with small gas fractions can produce galaxy-wide outflows, including fossil outflows in galaxies where there is little current AGN activity.

## The M-Sigma Relation Derived from Sphere of Influence Arguments

**Authors:** Batcheldor, D.

**Eprint:** http://arxiv.org/abs/1002.1705

**Keywords:** astro-ph.CO; astro-ph.GA; astrophysics; cosmology; observations; supermassive black holes

**Abstract:** The observed relation between supermassive black hole (SMBH) mass (M) and bulge stellar velocity dispersion (Sigma) is described by log(M) = alpha + beta*log(Sigma/200 km/s). As this relation has important implications for models of galaxy and SMBH formation and evolution, there continues to be great interest in adding to the M catalog. The "sphere of influence" (r) argument uses spatial resolution to exclude some M estimates and pre-select additional galaxies for further SMBH studies. This Letter quantifies the effects of applying the r argument to a population of galaxies and SMBHs that do not follow the M-Sigma relation. All galaxies with known values of Sigma, closer than 100 Mpc, are given a random M and selected when r is spatially resolved. These random SMBHs produce an M-Sigma relation of alpha=8.3, beta=4.0, consistent with observed values. Consequently, future proposed M estimates should not be justified solely on the basis of resolving r. This Letter shows the observed M-Sigma relation may simply be a result of available spatial resolution. However, it also implies the observed M-Sigma relation defines an upper limit. This potentially provides valuable new insight into the processes of galaxy and SMBH formation and evolution.

## The SMBH mass versus $M_G\sigma^2$ relation: A comparison between real data and numerical models

**Authors:** Feoli, A.; Mancini, L.; Marulli, F.; Bergh, S. van den

**Eprint:** <http://arxiv.org/abs/1002.1703>

**Keywords:** astro-ph.CO; astro-ph.GA; astrophysics; cosmology; observations; supermassive black holes

**Abstract:** The relation between the mass of supermassive black holes located in the center of the host galaxies and the kinetic energy of random motions of the corresponding bulges can be reinterpreted as an age-temperature diagram for galaxies. This relation fits the experimental data better than the $M_{bh}$–$M_G$, $M_{bh}$–$L_G$, and $M_{bh}$–$\sigma$ laws. The validity of this statement has been confirmed by using three samples extracted from different catalogues of galaxies. In the framework of the LambdaCDM cosmology our relation has been compared with the predictions of two galaxy formation models based on the Millennium Simulation.

## The Eccentric Disc Instability: Dependency on Background Stellar Cluster

**Authors:** Madigan, Ann-Marie

**Eprint:** <http://arxiv.org/abs/1002.1277>

**Keywords:** astro-ph.GA; astrophysics; N-body; Sagittarius A*; stellar dynamics

**Abstract:** In this paper we revisit the "eccentric disc instability", an instability which occurs in coherently eccentric discs of stars orbiting massive black holes (MBHs) embedded in stellar clusters, which results in stars achieving either very high or low eccentricities. The preference for stars to attain higher or lower eccentricities depends significantly on the density distribution of the surrounding stellar cluster. Here we discuss its mechanism and the implications for the Galactic Centre, home to at least one circum-MBH stellar disc.

## Mass Segregation in the Galactic Centre

**Authors:** Hopman, Clovis; Madigan, Ann-Marie

**Eprint:** http://arxiv.org/abs/1002.1220

**Keywords:** astro-ph.GA; astrophysics; Sagittarius A*; stellar dynamics

**Abstract:**

Two-body energy exchange between stars orbiting massive black holes (MBHs) leads to the formation of a power-law density distribution $n(r) \sim r^{-a}$ that diverges towards the MBH. For a single mass population, a=7/4 and the flow of stars is much less than $N(< r)/t_r$ (enclosed number of stars per relaxation time). This zero-flow solution is maintained for a multi-mass system for moderate mass ratios or systems where there are many heavy stars, and slopes of 3/2<a<2 are reached, with steeper slopes for the more massive stars. If the heavy stars are rare and massive however, the zero-flow limit breaks down and much steeper distributions are obtained.

We discuss the physics driving mass-segregation with the use of Fokker-Planck calculations, and show that steady state is reached in $0.2 - 0.3 t_r$. Since the relaxation time in the Galactic centre (GC) is $t_r \sim 2-3 \times 10^{10}$ yr, a cusp should form in less than a Hubble time. The absence of a visible cusp of old stars in the GC poses a challenge to these models, suggesting that processes other than two-body relaxation have played a role. We discuss astrophysical processes within the GC that depend crucially on the details of the stellar cusp.

## Discriminating between a Stochastic Gravitational Wave Background and Instrument Noise

**Authors:** Adams, Matthew R.; Cornish, Neil J.

**Eprint:** http://arxiv.org/abs/1002.1291

**Keywords:** back/foreground; cosmology; gr-qc; MLDC; noise: confusion; noise: instrumental

**Abstract:** The detection of a stochastic background of gravitational waves could significantly impact our understanding of the physical processes that shaped the early Universe. The challenge lies in separating the cosmological signal from other stochastic processes such as instrument noise and astrophysical foregrounds. One approach is to build two or more detectors and cross correlate their output, thereby enhancing the common gravitational wave signal relative to the uncorrelated instrument noise. When only one detector is available, as will likely be the case with the Laser Interferometer Space Antenna (LISA), alternative analysis techniques must be developed. Here we show that models of the noise and signal transfer functions can be used to tease apart the gravitational and instrument noise contributions. We discuss the role of gravitational wave insensitive "null channels" formed from particular combinations of the time delay interferometry, and derive a new combination that maintains this insensitivity for unequal arm length detectors. We show that, in the absence of astrophysical foregrounds, LISA could detect signals with energy densities as low as $\Omega_{gw} = 6 \times 10^{-13}$ with just one month of data. We describe an end-to-end Bayesian analysis pipeline that is able to search for, characterize and assign confidence levels for the detection of a stochastic gravitational wave background, and demonstrate the effectiveness of this approach using simulated data from the third round of Mock LISA Data Challenges.

## Numerical Models of Sgr A*

**Authors:** Moscibrodzka, M.; Gammie, C. F.; Dolence, J.; Shiokawa, H.; Leung, P. K.

**Eprint:** http://arxiv.org/abs/1002.1261

**Keywords:** astro-ph.HE; observations; Sagittarius A*; spin

**Abstract:** We review results from general relativistic axisymmetric magnetohydrodynamic simulations of accretion in Sgr A*. We use general relativistic radiative transfer methods and to produce a broad band (from millimeter to gamma-rays) spectrum. Using a ray tracing scheme we also model images of Sgr A* and compare the size of image to the VLBI observations at 230 GHz. We perform a parameter survey and study radiative properties of the flow models for various black hole spins, ion to electron temperature ratios, and inclinations. We scale our models to reconstruct the flux and the spectral slope around 230 GHz. The combination of Monte Carlo spectral energy distribution calculations and 230 GHz image modeling constrains the parameter space of the numerical models. Our models suggest rather high black hole spin ($a_* \approx 0.9$), electron temperatures close to the ion temperature ($T_i/T_e \sim 3$) and high inclination angles ($i \approx 90\,\mathrm{deg}$).

## High-Order Post-Newtonian Fit of the Gravitational Self-Force for Circular Orbits in the Schwarzschild Geometry

**Authors:** Blanchet, Luc; Detweiler, Steven; Tiec, Alexandre Le; Whiting, Bernard F.

**Eprint:** http://arxiv.org/abs/1002.0726

**Keywords:** general relativity; gr-qc; post-Newtonian theory; self force

**Abstract:** We continue a previous work on the comparison between the post-Newtonian (PN) approximation and the gravitational self-force (SF) analysis of circular orbits in a Schwarzschild background. We show that the numerical SF data contain physical information corresponding to extremely high PN approximations. We find that knowing analytically determined appropriate PN parameters helps tremendously in allowing the numerical data to be used to obtain higher order PN coefficients. Using standard PN theory we compute analytically the leading 4PN and the next-to-leading 5PN logarithmic terms in the conservative part of the dynamics of a compact binary system. The numerical perturbative SF results support well the analytic PN calculations through first order in the mass ratio, and are used to accurately measure the 4PN and 5PN non-logarithmic coefficients in a particular gauge invariant observable. Furthermore we are able to give estimates of higher order contributions up to the 7PN level. In our best fit we also confirm with high precision the value of the 3PN coefficient. This interplay between PN and SF efforts is important for the synthesis of template waveforms of extreme mass ratio inspirals to be analysed by the space-based gravitational wave instrument LISA. Our work will also have an impact on efforts that combine numerical results in a quantitative analytical framework so as to generate complete inspiral waveforms for the ground-based detection of gravitational waves by instruments such as LIGO and Virgo.

## The Supermassive Black Hole at the Heart of Centaurus A: Revealed by Gas- and Stellar Kinematics

**Authors:** Neumayer, Nadine

**Eprint:** http://arxiv.org/abs/1002.0965

**Keywords:** astro-ph.CO; astro-ph.IM; astrophysics; observations; supermassive black holes

**Abstract:** At less than 4 Mpc distance the radio galaxy NGC 5128 (Centaurus A) is the prime example to study the supermassive black hole and its influence on the environment in great detail. To model and understand the feeding and feedback mechanisms one needs an accurate determination of the mass of the supermassive black hole. The aim of this review is to give an overview of the recent studies that have been dedicated to measure the black hole mass in Centaurus A from both gas and stellar kinematics. It shows how the advancement in observing techniques and instrumentation drive the field of black hole mass measurements and concludes that adaptive optics assisted integral field spectroscopy is the key to identify the effects of the AGN on the surrounding ionised gas. Using data from SINFONI at the ESO Very Large Telescope, the best-fit black hole mass is $M_{BH} = 4.5 + 1.7/ - 1.0 \times 10^7 M_\odot$ (from $H_2$ kinematics) and $M_{BH} = (5.5 + / - 3.0) \times 10^7 M_\odot$ (from stellar kinematics; both with 3 sigma errors). This is one of the cleanest gas vs star comparison of a $M_{BH}$ determination, and brings Centaurus A into agreement with the $M_{BH} - \sigma$ relation.

## Stellar disc – dynamical evolution in a perturbed potential

**Authors:** Subr, Ladislav

**Eprint:** http://arxiv.org/abs/1002.0718

**Keywords:** astro-ph.GA; astrophysics; observations; Sagittarius A*; stellar dynamics

**Abstract:** Models of the origin of young stars in the Galactic Centre are facing various problems. The most promising scenario of the star formation in a thin self-gravitating disc naturally forms stars on coherently rotating orbits, but it fails to explain origin of several tens of stars that evidently do not belong to any of the disc-like structures in the GC. One possible solution lies in rather complicated initial conditions, assuming at least two infalling and interacting gas clouds. We present alternative solution showing that a single thin stellar disc may have given birth to all young stars in the GC. The outliers are explained as stars that have been stripped from the parent structure due to the gravitational interaction with the gaseous circum-nuclear disc.

## Effects of Interplanetary Dust on the LISA drag-free Constellation

**Authors:** Cerdonio, Massimo; De Marchi, Fabrizio; De Pietri, Roberto; Jetzer, Philippe; Marzari, Francesco; Mazzolo, Giulio; Ortolan, Antonello; Sereno, Mauro

**Eprint:** http://arxiv.org/abs/1002.0489

**Keywords:** detectors; gr-qc; instruments; interferometers

**Abstract:** The analysis of non-radiative sources of static or time-dependent gravitational fields in the Solar System is crucial to accurately estimate the free-fall orbits of the LISA space mission. In particular, we take into account the gravitational effects of Interplanetary Dust (ID) on the spacecraft trajectories. The perturbing gravitational field has been calculated for some ID density distributions that fit the observed zodiacal light. Then we integrated the Gauss planetary equations to get the deviations from the LISA keplerian orbits around the Sun. This analysis can be eventually extended to Local Dark Matter (LDM), as gravitational fields are expected to be similar for ID and LDM distributions. Under some strong assumptions on the displacement noise at very low frequency, the Doppler data collected during the whole LISA mission could provide upper limits on ID and LDM densities.

## The Impact of Stellar Collisions in the Galactic Center

**Authors:** Davies, M. B.; Church, R. P.; Malmberg, D.; Nzoke, S.; Dale, J.; Freitag, M.

**Eprint:** http://arxiv.org/abs/1002.0338

**Keywords:** astro-ph.GA; astrophysics; Sagittarius A*; stellar dynamics

**Abstract:** We consider whether stellar collisions can explain the observed depletion of red giants in the galactic center. We model the stellar population with two different IMFs: 1) the Miller-Scalo and 2) a much flatter IMF. In the former case, low-mass main-sequence stars dominate the population, and collisions are unable to remove red giants out to 0.4 pc although brighter red giants much closer in may be depleted via collisions with stellar-mass black holes. For a much flatter IMF, the stellar population is dominated by compact remnants (ie black holes, white dwarfs and neutron stars). The most common collisions are then those between main-sequence stars and compact remnants. Such encounters are likely to destroy the main-sequence stars and thus prevent their evolution into red giants. In this way, the red-giant population could be depleted out to 0.4 pc matching observations. If this is the case, it implies the galactic center contains a much larger population of stellar-mass black holes than would be expected from a regular IMF. This may in turn have implications for the formation and growth of the central supermassive black hole.

## Massive black holes lurking in Milky Way satellites

**Authors:** Van Wassenhove, S.; Volonteri, M.; Walker, M. G.; Gair, J. R.

**Eprint:** http://arxiv.org/abs/1001.5451

**Keywords:** astro-ph.CO; astro-ph.HE; astrophysics; cosmology; massive binaries of black holes

**Abstract:** As massive black holes (MBHs) grow from lower-mass seeds, it is natural to expect that a leftover population of progenitor MBHs should also exist in the present universe. Dwarf galaxies undergo a quiet merger history, and as a result, we expect that dwarfs observed in the local Universe retain some 'memory' of the original seed mass distribution. Consequently, the properties of MBHs in nearby dwarf galaxies may provide clean indicators of the efficiency of MBH formation. In order to examine the properties of MBHs in dwarf galaxies, we evolve different MBH populations within a Milky Way halo from high-redshift to today. We consider two plausible MBH formation mechanisms: 'massive seeds' formed via gas-dynamical instabilities and a Population III remnant seed model. 'Massive seeds' have larger masses than PopIII remnants, but form in rarer hosts. We dynamically evolve all halos merging with the central system, taking into consideration how the interaction modifies the satellites, stripping their outer mass layers. We compute different properties of the MBH population hosted in these satellites. We find that for the most part MBHs retain the original mass, thus providing a clear indication of what the properties of the seeds were. We derive the black hole occupation fraction (BHOF) of the satellite population at z=0. MBHs generated as 'massive seeds' have large masses that would favour their identification, but their typical BHOF is always below 40 per cent and decreases to less than per cent for observed dwarf galaxy sizes. In contrast, Population III remnants have a higher BHOF, but their masses have not grown much since formation, inhibiting their detection.

### Dynamical Models of the Galactic Center

**Authors:** Merritt, David

**Eprint:** <http://arxiv.org/abs/1001.5435>

**Keywords:** astro-ph.CO; astro-ph.GA; astrophysics; observations; stellar dynamics; supermassive black holes

**Abstract:** The distribution of late-type (old) stars in the inner parsec of the Milky Way is very different than expected for a relaxed population around a supermassive black hole. Instead of a density cusp, there is a 0.5 pc core. This article discusses what sorts of dynamical models might explain this "conundrum of old age." A straightforward interpretation is that the nucleus is unrelaxed, and that the distribution of the old giants reflects the distribution of fainter stars and stellar remnants generally in the core. On the other hand, a density cusp could be present in the unobserved populations, and the deficit of bright giants could be a result of interactions with these objects. At the present time, no model is clearly preferred.

### Magnetic Connection Model for Launching Relativistic Jets from a Kerr Black Hole

**Authors:** Dutan, Ioana

**Abstract:** We present an alternative model for launching relativistic jets in active galactic nuclei (AGN) from an accreting Kerr black hole (BH) by converting the accretion disc energy into jet energy, when the rotational energy of the BH is transferred to the inner disc by closed magnetic field lines which connects the BH to the disc (BH-disc magnetic connection). In this way, the available disc energy is increased by the BH rotational energy. We assume that the BH may undergo recurring episodes of its activity with: (i) a first phase when accretion power dominates, and (ii) a second phase when BH spin-down power dominates. In both cases the jet is driven by a low-luminosity, (geometrically) thin accretion disc, as the disc energy is used to launch the jet. We use the general relativistic conservation laws to calculate the mass flow rate into the jets, the launching power of the jets, and the angular momentum transported by the jets. We consider BHs with a spin parameter $a_* \geqslant 0.95$, so that the jets are launched from the region inside of the BH ergosphere. The angular momentum removed from the accretion disc is carried away by the disc particles that ultimately form the jets. As far as the BH is concerned, it can (i) spin up by accreting matter and (ii) spin down due to the magnetic counter-acting torque on the BH. We found that a stationary state of the BH ($a_*$ = const) can be reached if the mass accretion rate is larger than $\dot{m} \sim 0.001$. The maximum value of the BH spin parameter depends on $\dot{m}$ being less but close to 0.9982 (Thorne's model). In addition, the maximum AGN lifetime can be much longer than $\sim 10^7$ yr when using the BH spin-down power. This result is consistent with the estimation of the maximum AGN lifetime when the AGN output power is provided by the Blandford–Znajek mechanism.

## Composition of the galactic center star cluster

**Abstract:** We present a population analysis of the nuclear stellar cluster of the Milky Way based on adaptive optics narrow band spectral energy distributions. We find strong evidence for the lack of a stellar cusp and a similarity of the late type luminosity function to the bulge KLF.

# The search for spinning black hole binaries in mock LISA data using a genetic algorithm

**Authors:** Petiteau, Antoine; Shang, Yu; Babak, Stanislav; Feroz, Farhan

**Eprint:** <http://arxiv.org/abs/1001.5380>

**Keywords:** data analysis; gr-qc; massive binaries of black holes; MLDC; parameter estimation; search algorithms; spin; supermassive black holes

**Abstract:** Coalescing massive Black Hole binaries are the strongest and probably the most important gravitational wave sources in the LISA band. The spin and orbital precessions bring complexity in the waveform and make the likelihood surface richer in structure as compared to the non-spinning case. We introduce an extended multimodal genetic algorithm which utilizes the properties of the signal and the detector response function to analyze the data from the third round of mock LISA data challenge (MLDC 3.2). The performance of this method is comparable, if not better, to already existing algorithms. We have found all five sources present in MLDC 3.2 and recovered the coalescence time, chirp mass, mass ratio and sky location with reasonable accuracy. As for the orbital angular momentum and two spins of the Black Holes, we have found a large number of widely separated modes in the parameter space with similar maximum likelihood values.

# The Current Status of Binary Black Hole Simulations in Numerical Relativity

**Authors:** Hinder, Ian

**Eprint:** <http://arxiv.org/abs/1001.5161>

**Keywords:** gr-qc; massive binaries of black holes; numerical relativity

**Abstract:** Since the breakthroughs in 2005 which have led to long term stable solutions of the binary black hole problem in numerical relativity, much progress has been made. I present here a short summary of the state of the field, including the capabilities of numerical relativity codes, recent physical results obtained from simulations, and improvements to the methods used to evolve and analyse binary black hole spacetimes.

## Towards Tests of Alternative Theories of Gravity with LISA

**Authors:** Sopuerta, Carlos F.; Yunes, Nicolas

**Eprint:** http://arxiv.org/abs/1001.4899

**Keywords:** EMRI; general relativity; gr-qc; hep-th; IMRI; tests of alternative theories

**Abstract:** The inspiral of stellar compact objects into massive black holes, usually known as extreme-mass-ratio inspirals (EMRIs), is one of the most important sources of gravitational-waves for the future Laser Interferometer Space Antenna (LISA). Intermediate-mass-ratio inspirals (IMRIs are also of interest to advance ground-based gravitational-wave observatories. We discuss here how modifications to the gravitational interaction can affect the signals emitted by these systems and their detectability by LISA. We concentrate in particular on Chern-Simons modified gravity, a theory that emerges in different quantum gravitational approaches.

## Dependence of inner accretion disk stress on parameters: the Schwarzschild case

**Authors:** Noble, Scott C.; Krolik, Julian H.; Hawley, John F.

**Eprint:** http://arxiv.org/abs/1001.4809

**Keywords:** accretion discs; astro-ph.CO; astro-ph.HE; astrophysics; EM counterparts; supermassive black holes

**Abstract:** We explore the parameter dependence of inner disk stress in black hole accretion by contrasting the results of a number of simulations, all employing 3-d general relativistic MHD in a Schwarzschild spacetime. Five of these simulations were performed with the intrinsically conservative code HARM3D, which allows careful regulation of the disk aspect ratio, H/R; our simulations span a range in H/R from 0.06 to 0.17. We contrast these simulations with two previously reported simulations in a Schwarzschild spacetime in order to investigate possible dependence of the inner disk stress on magnetic topology. In all cases, much care was devoted to technical issues: ensuring adequate resolution and azimuthal extent, and averaging only over those time-periods when the accretion flow is in approximate inflow equilibrium. We find that the time-averaged radial-dependence of fluid-frame electromagnetic stress is almost completely independent of both disk thickness and poloidal magnetic topology. It rises smoothly inward at all radii (exhibiting no feature associated with the ISCO) until just outside the event horizon, where the stress plummets to zero. Reynolds stress can also be significant near the ISCO and in the plunging region; the magnitude of this stress, however, depends on both disk thickness and magnetic topology. The two stresses combine to make the net angular momentum accreted per unit rest-mass 7-15% less than the angular momentum of the ISCO.

## Modelling Extreme-Mass-Ratio Inspirals using Pseudospectral Methods

**Authors:** Canizares, Priscilla; Sopuerta, Carlos F.

**Eprint:** http://arxiv.org/abs/1001.4697

**Keywords:** EMRI; general relativity; gr-qc; self force; waveforms

**Abstract:** We introduce a new time-domain method for computing the self-force acting on a scalar particle in a Schwarzschild geometry. The principal feature of our method consists in the division of the spatial domain into several subdomains and locating the particle at the interface betweem two them. In this way, we avoid the need of resolving a small length scale associated with the presence of a particle in the computational domain and, at the same time, we avoid numerical problems due to the low differentiability of solutions of equations with point-like singular behaviour.

## The Milky Way Nuclear Star Cluster in Context

**Authors:** Schoedel, Rainer

**Eprint:** http://arxiv.org/abs/1001.4238

**Keywords:** astro-ph.GA; astrophysics; observations; Sagittarius A*

**Abstract:** Nuclear star clusters are located at the dynamical centers of the majority of galaxies. They are usually the densest and most massive star cluster in their host galaxy. In this article, I will give a brief overview of our current knowledge on nuclear star clusters and their formation. Subsequently, I will introduce the nuclear star cluster at the center of the Milky Way, that surrounds the massive black hole, Sagittarius A*. This cluster is a unique template for understanding nuclear star clusters in general because it is the only one of its kind which we can resolve into individual stars. Thus, we can study its structure, dynamics, and population in detail. I will summarize our current knowledge of the Milky Way nuclear star cluster, discuss its relation with nuclear clusters in other galaxies, and point out where further research is needed.

### Massive Young Stars in the Galactic Center

**Authors:** Bartko, H.

**Eprint:** http://arxiv.org/abs/1001.4232

**Keywords:** astro-ph.GA; astrophysics; observations; Sagittarius A*

**Abstract:** We summarize our latest observations of the nuclear star cluster in the central parsec of the Galaxy with the adaptive optics assisted, integral field spectrograph SINFONI on the ESO/VLT, which result in a total sample of 177 bona fide early-type stars. We find that most of these Wolf Rayet (WR), O- and B- stars reside in two strongly warped eccentric ( = 0.36+/-0.06) disks between 0.8" and 12" from Sgr$A^*$, as well as a central compact concentration (the S-star cluster) centered on Sgr$A^*$. The later type B stars (mK&;15) in the radial interval between 0.8" and 12" seem to be in a more isotropic distribution outside the disks. We observe a dearth of late-type stars in the central few arcseconds, which is puzzling. The stellar mass function of the disk stars is extremely top-heavy with a best fit power law of $dN/dm \sim m^{-0.45+/-0.3}$. Since at least the WR/O-stars were formed in situ in a single star formation event ~ 6 Myrs ago, this mass function probably reflects the initial mass function (IMF). The mass functions of the S-stars inside 0.8" and of the early-type stars at distances beyond 12" differ significantly from the disk IMF; they are compatible with a standard Salpeter/Kroupa IMF (best fit power law of $dN/dm \sim m^{-2.15+/-0.3}$.

### On the role of supernovae-driven turbulence in the feeding of supermassive black holes

**Authors:** Hobbs, Alexander; Nayakshin, Sergei; Power, Chris; King, Andrew

**Eprint:** http://arxiv.org/abs/1001.3883

**Keywords:** accretion discs; astro-ph.CO; astro-ph.HE; EM counterparts; supermassive black holes

**Abstract:** It has long been recognised that the main obstacle to accretion of gas onto supermassive black holes (SMBHs) is large specific angular momentum. However, while the mean angular momentum in the bulge is very likely to be large, the deviations from the mean can also be significant. Indeed, inside bulges the gas velocity distribution can be randomised by the velocity kicks due to feedback from star formation. Here we perform hydrodynamical simulations of gaseous rotating shells infalling onto an SMBH, attempting to quantify the importance of velocity dispersion in the gas at relatively large distances from the black hole. We implement this dispersion by means of a supersonic turbulent velocity spectrum. We find that, while in the purely rotating case the circularisation process leads to efficient mixing of gas with different angular momentum, resulting in a low accretion rate, the inclusion of turbulence increases this accretion rate by up to several orders of magnitude. We show that this can be understood based on the notion of "ballistic" accretion, whereby dense filaments, created by convergent turbulent flows, travel through the ambient gas largely unaffected by hydrodynamical drag. This prevents the efficient gas mixing that was found in the simulations without turbulence, and allows a fraction of gas to impact the innermost boundary of the simulations directly. Using the ballistic approximation, we derive a simple analytical formula that captures the numerical results to within a factor of a few. Rescaling our results to astrophysical bulges, we argue that this "ballistic" mode of accretion could provide the SMBHs with a sufficient supply of fuel without the need to channel the gas via large-scale discs or bars. We therefore argue that star formation in bulges can be a strong catalyst for SMBH accretion.

## High redshift formation and evolution of central massive objects I: model description

**Authors:** Devecchi, B.; Volonteri, M.; Colpi, M.; Haardt, F.

**Eprint:** http://arxiv.org/abs/1001.3874

**Keywords:** accretion discs; astro-ph.CO; astrophysics; cosmology; supermassive black holes

**Abstract:** Galactic nuclei host central massive objects either in the form of supermassive black holes or nuclear stellar clusters. Recent investigations have shown that both components co-exist in at least a few galaxies. In this paper we explore the possibility of a connection between nuclear star clusters and black holes that establishes at the moment of their formation. We here model the evolution of high redshift discs, hosted in dark matter halos with virial temperatures $10^4$ K, whose gas has been polluted with metals just above the critical metallicity for fragmentation. A nuclear cluster forms as a result of a central starburst from gas inflowing from the unstable disc. The nuclear stellar cluster provides a suitable environment for the formation of a black hole seed, ensuing from runaway collisions among the most massive stars. Typical masses for the nuclear stellar clusters at the time of black hole formation (z~ 10) are in the range $10^4 - 10^6$ solar masses and have half mass radii < 0.5 pc. The black holes forming in these dense, high redshift clusters can have masses in the range ~ 300-2000 solar masses.

## Advances in Simulations of Generic Black-Hole Binaries

**Authors:** Campanelli, Manuela; Lousto, Carlos O.; Mundim, Bruno C.; Nakano, Hiroyuki; Zlochower, Yosef; Bischof, Hans-Peter

**Eprint:** http://arxiv.org/abs/1001.3834

**Keywords:** gr-qc; massive binaries of black holes; numerical relativity; post-Newtonian theory

**Abstract:** We review some of the recent dramatic developments in the fully non-linear simulation of generic, highly-precessing, black-hole binaries, and introduce a new approach for generating hybrid post-Newtonian / Numerical waveforms for these challenging systems.

## Toward Precision Measurement of Central Black Hole Masses

**Authors:** Peterson, Bradley M.

**Eprint:** http://arxiv.org/abs/1001.3675

**Keywords:** astro-ph.CO; astro-ph.GA; astrophysics; observations; supermassive black holes

**Abstract:** We review briefly direct and indirect methods of measuring the masses of black holes in galactic nuclei, and then focus attention on supermassive black holes in active nuclei, with special attention to results from reverberation mapping and their limitations. We find that the intrinsic scatter in the relationship between the AGN luminosity and the broad-line region size is very small, ∼ 0.11 dex, comparable to the uncertainties in the better reverberation measurements. We also find that the relationship between reverberation-based black hole masses and host-galaxy bulge luminosities also seems to have surprisingly little intrinsic scatter, ∼ 0.17 dex. We note, however, that there are still potential systematics that could affect the overall mass calibration at the level of a factor of a few.

# Numerical modeling of gravitational wave sources accelerated by OpenCL

**Authors:** Khanna, Gaurav; McKennon, Justin

**Eprint:** <http://arxiv.org/abs/1001.3631>

**Keywords:** data analysis; EMRI; GPU; gr-qc; physics.comp-ph

**Abstract:** In this work, we make use of the OpenCL framework to accelerate an EMRI modeling application using the hardware accelerators – Cell BE and Tesla CUDA GPU. We describe these compute technologies and our parallelization approach in detail, present our performance results, and then compare them with those from our previous implementations based on the native CUDA and Cell SDKs. The OpenCL framework allows us to execute identical source-code on both architectures and yet obtain strong performance gains that are comparable to what can be derived from the native SDKs.

# Mass function of binary massive black holes in Active Galactic Nuclei

**Authors:** Hayasaki, Kimitake; Ueda, Yoshihiro; Isobe, Naoki

**Eprint:** <http://arxiv.org/abs/1001.3612>

**Keywords:** astro-ph.CO; astro-ph.GA; astrophysics; massive binaries of black holes

**Abstract:** If the activity of active galactic nuclei (AGNs) is predominantly induced by major galaxy mergers, then a significant fraction of AGNs should harbor binary massive black holes in their centers. We study the mass function of binary massive black holes in nearby AGNs based on the theory of evolution of binary massive black holes interacting with ambient gaseous disks proposed by Hayasaki (2009). The timescale of orbital decay is estimated as the order of $10^8 yr$, being independent of the black hole mass but only dependent on the mass ratio and Eddington ratio. This makes it possible for any binary massive black holes to merge within a Hubble time. We find that 1.3% − −1.7% of the total number of nearby AGNs have close, binary massive black-holes with orbital period less than ten-years, detectable with on-going highly sensitive X-ray monitors such as Monitor of All-sky X-ray Image and/or Swift/Burst Alert Telescope. Close binaries with total black-hole masses of $10^{6.5-7} M_\odot$ are the most frequent in massive binary black-hole populations of nearby AGNs.

## Two-dimensional hydrodynamical simulation of hot accretion flows with radiative cooling

**Authors:** Yuan, Feng; Bu, Defu

**Eprint:** http://arxiv.org/abs/1001.3571

**Keywords:** accretion discs; astro-ph.CO; astro-ph.HE; EM counterparts; spin

**Abstract:** The most important finding of two-dimensional hydrodynamical simulations of hot accretion flows is that the flow is convectively unstable, because of its inward increase of entropy. As a result, the profile of the mass accretion rate is a function of radius, i.e., only a small fraction of accretion gas available at the outer boundary can finally fall onto the black hole, while the rest is lost in the convective outflows. Radiation is usually neglected in these simulations. When the radiative cooling becomes more and more important, the entropy will increase slower inward. The entropy can even decrease when the radiation becomes stronger than the viscous heating, i.e, the flow enters into the luminous hot accretion flow regime. In the present paper, we investigate the convective instability and correspondingly the profile of accretion rate in the presence of strong radiative cooling by performing two-dimensional hydrodynamical numerical simulation. This problem is important because the profile of the mass accretion rate determines the observational appearance of accretion flows, the growth of black hole, and the evolution of black hole spin. We find that the flow is still strongly convectively unstable, and the radial profile of accretion rate changes little compared to the case of non-radiative flow. This is because the gradient of entropy in the gravitational direction still increases inward although the gradient of entropy decreases.

## Measuring the dark energy equation of state with LISA

**Authors:** Broeck, Chris Van Den; Trias, M.; Sathyaprakash, B. S.; Sintes, A. M.

**Eprint:** http://arxiv.org/abs/1001.3099

**Keywords:** astrophysics; cosmology; gr-qc; massive binaries of black holes

**Abstract:** The Laser Interferometer Space Antenna's (LISA's) observation of supermassive binary black holes (SMBBH) could provide a new tool for precision cosmography. Inclusion of sub-dominant signal harmonics in the inspiral signal allows for high-accuracy sky localization, dramatically improving the chances of finding the host galaxy and obtaining its redshift. Combined with the measurement of the luminosity distance, this could allow us to significantly constrain the dark energy equation-of-state parameter $w$ even with a single SMBBH merger at $z \lesssim 1$. Such an event can potentially have component masses from a wide range ($10^5 - 10^8 M_\odot$) over which parameter accuracies vary considerably. We perform an in-depth study in order to understand (i) what fraction of possible SMBBH mergers allow for sky localization, depending on the parameters of the source, and (ii) how accurately $w$ can be measured when the host galaxy can be identified. We also investigate how accuracies on all parameters improve when a knowledge of the sky position can be folded into the estimation of errors. We find that $w$ can be measured to within a few percent in most cases, if the only error in measuring the luminosity distance is due to LISA's instrumental noise and the confusion background from Galactic binaries. However, weak lensing-induced errors will severely degrade the accuracy with which $w$ can be obtained, emphasizing that methods to mitigate weak lensing effects would be required to take advantage of LISA's full potential.

## Persistent junk solutions in time-domain modeling of extreme mass ratio binaries

**Authors:** Field, Scott E.; Hesthaven, Jan S.; Lau, Stephen R.

**Eprint:** http://arxiv.org/abs/1001.2578

**Keywords:** EMRI; gr-qc; numerical relativity; self force

**Abstract:** In the context of metric perturbation theory for non-spinning black holes, extreme mass ratio binary (EMRB) systems are described by distributionally forced master wave equations. Numerical solution of a master wave equation as an initial boundary value problem requires initial data. However, because the correct initial data for generic-orbit systems is unknown, specification of trivial initial data is a common choice, despite being inconsistent and resulting in a solution which is initially discontinuous in time. As is well known, this choice leads to a "burst" of junk radiation which eventually propagates off the computational domain. We observe another unintended consequence of trivial initial data: development of a persistent spurious solution, here referred to as the Jost junk solution, which contaminates the physical solution for long times. This work studies the influence of both types of junk on metric perturbations, waveforms, and self-force measurements, and it demonstrates that smooth modified source terms mollify the Jost solution and reduce junk radiation.

## Intermediate Mass Ratio Black Hole Binaries: Numerical Relativity meets Perturbation Theory

**Authors:** Lousto, Carlos O.; Nakano, Hiroyuki; Zlochower, Yosef; Campanelli, Manuela

**Eprint:** http://arxiv.org/abs/1001.2316

**Keywords:** gr-qc; massive binaries of black holes; numerical methods; numerical relativity

**Abstract:** We study black-hole binaries in the intermediate-mass-ratio regime $0.01 < q < 0.1$ with a new technique that makes use of nonlinear numerical trajectories and efficient perturbative evolutions to compute waveforms at large radii for the leading and nonleading modes. As a proof-of-concept, we compute waveforms for $q=1/10$. We discuss applications of these techniques for LIGO/VIRGO data analysis and the possibility that our technique can be extended to produce accurate waveform templates from a modest number of fully-nonlinear numerical simulations.

## A Separable Solution for the Oscillatory Structure of Plasma in Accretion Disks

**Authors:** Lattanzi, Massimiliano; Montani, Giovanni

**Eprint:** http://arxiv.org/abs/1001.2430

**Keywords:** accretion discs; astro-ph.SR; astrophysics; EM counterparts

**Abstract:** We provide a new analysis of the system of partial differential equations describing the radial and vertical equilibria of the plasma in accretion disks. In particular, we show that the partial differential system can be separated once a definite, oscillatory (or hyperbolic) form for the radial dependence of the relevant physical quantities is assumed. The system is thus reduced to an ordinary differential system in the vertical dimensionless coordinate. The resulting equations can be integrated analytically in the limit of small magnetic pressure. We complete our analysis with a direct numerical integration of the more general case. The main result is that a ring-like density profile (i.e., radial oscillations in the mass density) can appear even in the limit of small magnetic pressure.

## Gravitational recoil: effects on massive black hole occupation fraction over cosmic time

**Authors:** Volonteri, Marta; Gultekin, Kayhan; Dotti, Massimo

**Eprint:** http://arxiv.org/abs/1001.1743

**Keywords:** astro-ph.CO; astrophysics; cosmology; kicks/recoil; massive binaries of black holes

How stars distribute around a massive black hole

**Abstract:** We assess the influence of massive black hole (MBH) ejections from galaxy centres, due to the gravitational radiation recoil, along the cosmic merger history of the MBH population. We discuss the 'danger' of the recoil for MBHs as a function of different MBH spin/orbit configurations and of the host halo cosmic bias, and on how that reflects on the 'occupation fraction' of MBHs. We assess ejection probabilities for mergers occurring in a gas-poor environment, where the MBH binary coalescence is driven by stellar dynamical processes, and the spin/orbit configuration is expected to be isotropically distributed. We contrast this case with the 'aligned' case. The latter is the most realistic situation for 'wet', gas-rich mergers, which are the expectation for high-redshift galaxies. We find that if all halos at z>5-7 host a MBH, the probability of the Milky Way (or similar size galaxy) to host a MBH today is less than 50%, unless MBHs form continuously in galaxies. The 'occupation fraction' of MBHs, intimately related to halo bias and MBH formation efficiency, plays a crucial role in increasing the retention fraction. Small halos, with shallow potential wells and low escape velocities, have a high ejection probability, but the MBH merger rate is very low along their galaxy formation merger hierarchy: MBH formation processes are likely inefficient in such shallow potential wells. Recoils can decrease the overall frequency of MBHs in small galaxies to ∼ 60%, while they have little effect on the frequency of MBHs in large galaxies (at most a 20% effect).

## The extreme luminosity states of Sagittarius A*

**Authors:** Sabha, N.; Witzel, G.; Eckart, A.; Buchholz, R. M.; Bremer, M.; Giessuebel, R.; Garcia-Marin, M.; Kunneriath, D.; Muzic, K.; Schoedel, R.; Straubmeier, C.; Zamaninasab, M.; Zernickel, A.

**Abstract:** We discuss mm-wavelength radio, 2.2-11.8um NIR and 2-10 keV X-ray light curves of the super massive black hole (SMBH) counterpart of Sagittarius A* (SgrA*) near its lowest and highest observed luminosity states. The luminosity during the low state can be interpreted as synchrotron emission from a continuous or even spotted accretion disk. For the high luminosity state SSC emission from THz peaked source components can fully account for the flux density variations observed in the NIR and X-ray domain. We conclude that at near-infrared wavelengths the SSC mechanism is responsible for all emission from the lowest to the brightest flare from SgrA*. For the bright flare event of 4 April 2007 that was covered from the radio to the X-ray domain, the SSC model combined with adiabatic expansion can explain the related peak luminosities and different widths of the flare profiles obtained in the NIR and X-ray regime as well as the non detection in the radio domain.

## Testing MOND/TEVES with LISA Pathfinder

**Authors:** Trenkel, Christian; Kemble, Steve; Bevis, Neil; Magueijo, Joao

**Eprint:** <http://arxiv.org/abs/1001.1303>

**Keywords:** astro-ph.CO; instruments; interferometers; tests of alternative theories

**Abstract:** We suggest that LISA Pathfinder could be used to subject TEVES, and in particular the non-relativistic MOND phenomenology it incorporates, to a direct, controlled experimental test, in just a few years' time. The basic concept is to fly LISA Pathfinder through the region around the Sun-Earth saddle point, following its nominal mission, in order to look for anomalous gravity gradients. We examine various strategies to reach the saddle point, and conclude that the preferred strategy, resulting in relatively short transfer times of order one year, probably involves a lunar fly-by. We present robust estimates of the MOND gravity gradients that LISA Pathfinder should be exposed to, and conclude that if the gradiometer on-board the spacecraft achieves its nominal performance, these gradients will not just be detected, but measured and characterised in some detail, should they exist. Conversely, given the large predicted signal based on standard assumptions, a null result would most likely spell the end of TEVES/MOND.

## Measuring Black Hole Spin in OJ287

**Authors:** Valtonen, M.; Mikkola, S.; Lehto, H. J.; Hyvönen, T.; Nilsson, K.; Merritt, D.; Gopakumar, A.; Rampadarath, H.; Hudec, R.; Basta, M.; Saunders, R.

**Eprint:** <http://arxiv.org/abs/1001.1284>

**Keywords:** astro-ph.CO; astro-ph.HE; astrophysics; massive binaries of black holes; observations; spin

**Abstract:** We model the binary black hole system OJ287 as a spinning primary and a non-spinning secondary. It is assumed that the primary has an accretion disk which is impacted by the secondary at specific times. These times are identified as major outbursts in the light curve of OJ287. This identification allows an exact solution of the orbit, with very tight error limits. Nine outbursts from both the historical photographic records as well as from recent photometric measurements have been used as fixed points of the solution: 1913, 1947, 1957, 1973, 1983, 1984, 1995, 2005 and 2007 outbursts. This allows the determination of eight parameters of the orbit. Most interesting of these are the primary mass of $1.84 \cdot 10^{10} M_\odot$, the secondary mass $1.46 \cdot 10^8 M_\odot$, major axis precession rate $39°.1$ per period, and the eccentricity of the orbit 0.70. The dimensionless spin parameter is 0.28 ± 0.01 (1 sigma). The last parameter will be more tightly constrained in 2015 when the next outburst is due. The outburst should begin on 15 December 2015 if the spin value is in the middle of this range, on 3 January 2016 if the spin is 0.25, and on 26 November 2015 if the spin is 0.31. We have also tested the possibility that the quadrupole term in the Post Newtonian equations of motion does not exactly follow Einstein's theory: a parameter $q$ is introduced as one of the 8 parameters. Its value is within 30% (1 sigma) of the Einstein's value $q = 1$. This supports the $no-hair theorem$ of black holes within the achievable precision. We have also measured the loss of orbital energy due to gravitational waves. The loss rate is found to agree with Einstein's value with the accuracy of 2% (1 sigma).

## An Intermediate-mass Black Hole of Over 500 Solar Masses in the Galaxy ESO 243-49

**Authors:** Farrell, Sean; Webb, Natalie; Barret, Didier; Godet, Olivier; Rodrigues, Joana

**Eprint:** http://arxiv.org/abs/1001.0567

**Keywords:** astro-ph.CO; astro-ph.HE; astrophysics; intermediate-mass black holes; observations

**Abstract:** Ultra-luminous X-ray sources are extragalactic objects located outside the nucleus of the host galaxy with bolometric luminosities $> 10^{39}$ erg $s^{-1}$. These extreme luminosities - if the emission is isotropic and below the theoretical (i.e. Eddington) limit, where the radiation pressure is balanced by the gravitational pressure - imply the presence of an accreting black hole with a mass of $\sim 10^2 - 10^5$ times that of the Sun. The existence of such intermediate mass black holes is in dispute, and though many candidates have been proposed, none are widely accepted as definitive. Here we report the detection of a variable X-ray source with a maximum 0.2-10 keV luminosity of up to $1.2 \times 10^{42}$ erg $s^{-1}$ in the edge-on spiral galaxy ESO 243-49, with an implied conservative lower limit of the mass of the black hole of $\sim$ 500 Msun. This finding presents the strongest observational evidence to date for the existence of intermediate mass black holes, providing the long sought after missing link between the stellar mass and super-massive black hole populations.

## Direct Formation of Supermassive Black Holes via Multi-Scale Gas Inflows in Galaxy Mergers

**Authors:** Mayer, Lucio; Kazantzidis, Stelios; Escala, Andres; Callegari, Simone

**Eprint:** <http://arxiv.org/abs/0912.4262>

**Keywords:** accretion discs; astro-ph.CO; EM counterparts; supermassive black holes

**Abstract:** Observations of distant bright quasars suggest that billion solar mass supermassive black holes (SMBHs) were already in place less than a billion years after the Big Bang. Models in which light black hole seeds form by the collapse of primordial metal-free stars cannot explain their rapid appearance due to inefficient gas accretion. Alternatively, these black holes may form by direct collapse of gas at the center of protogalaxies. However, this requires metal-free gas that does not cool efficiently and thus is not turned into stars, in contrast with the rapid metal enrichment of protogalaxies. Here we use a numerical simulation to show that mergers between massive protogalaxies naturally produce the required central gas accumulation with no need to suppress star formation. Merger-driven gas inflows produce an unstable, massive nuclear gas disk. Within the disk a second gas inflow accumulates more than 100 million solar masses of gas in a sub-parsec scale cloud in one hundred thousand years. The cloud undergoes gravitational collapse, which eventually leads to the formation of a massive black hole. The black hole can grow to a billion solar masses in less than a billion years by accreting gas from the surrounding disk.

## An improved effective-one-body Hamiltonian for spinning black-hole binaries

**Authors:** Barausse, Enrico; Buonanno, Alessandra

**Eprint:** <http://arxiv.org/abs/0912.3517>

**Keywords:** Effective one body; gr-qc; massive binaries of black holes; numerical relativity; spin; waveforms

**Abstract:** Building on a recent paper in which we computed the canonical Hamiltonian of a spinning test particle in curved spacetime, at linear order in the particle's spin, we work out an improved effective-one-body (EOB) Hamiltonian for spinning black-hole binaries. As in previous descriptions, we endow the effective particle not only with a mass m, but also with a spin S*. Thus, the effective particle interacts with the effective Kerr background (having spin $S_{Kerr}$) through a geodesic-type interaction and an additional spin-dependent interaction proportional to S*. When expanded in post-Newtonian (PN) orders, the EOB Hamiltonian reproduces the leading order spin-spin coupling and the spin-orbit coupling through 2.5PN order, for any mass-ratio. Also, it reproduces all spin-orbit couplings in the test-particle limit. Similarly to the test-particle limit case, when we restrict the EOB dynamics to spins aligned or antialigned with the orbital angular momentum, for which circular orbits exist, the EOB dynamics has several interesting features, such as the existence of an innermost stable circular orbit, a photon circular orbit, and a maximum in the orbital frequency during the plunge subsequent to the inspiral. These properties are crucial for reproducing the dynamics and gravitational-wave emission of spinning black-hole binaries, as calculated in numerical relativity simulations.

## Effective-one-body waveforms calibrated to numerical relativity simulations: coalescence of non-precessing, spinning, equal-mass black holes

**Authors:** Pan, Yi; Buonanno, Alessandra; Buchman, Luisa T.; Chu, Tony; Kidder, Lawrence E.; Pfeiffer, Harald P.; Scheel, Mark A.

**Eprint:** http://arxiv.org/abs/0912.3466

**Keywords:** Effective one body; gr-qc; massive binaries of black holes; spin; waveforms

**Abstract:** We present the first attempt at calibrating the effective-one-body (EOB) model to accurate numerical-relativity simulations of spinning, non-precessing black-hole binaries. Aligning the EOB and numerical waveforms at low frequency over a time interval of 1000M, we first estimate the phase and amplitude errors in the numerical waveforms and then minimize the difference between numerical and EOB waveforms by calibrating a handful of EOB-adjustable parameters. In the equal-mass, spin aligned case, we find that phase and fractional amplitude differences between the numerical and EOB (2,2) mode can be reduced to 0.01 radians and 1%, respectively, over the entire inspiral waveforms. In the equal-mass, spin anti-aligned case, these differences can be reduced to 0.13 radians and 1% during inspiral and plunge, and to 0.4 radians and 10% during merger and ringdown. The waveform agreement is within numerical errors in the spin aligned case while slightly over numerical errors in the spin anti-aligned case. Using Enhanced LIGO and Advanced LIGO noise curves, we find that the overlap between the EOB and the numerical (2,2) mode, maximized over the initial phase and time of arrival, is larger than 0.999 for binaries with total mass 30-200Ms. In addition to the leading (2,2) mode, we compare four subleading modes. We find good amplitude and frequency agreements between the EOB and numerical modes for both spin configurations considered, except for the (3,2) mode in the spin anti-aligned case. We believe that the larger difference in the (3,2) mode is due to the lack of knowledge of post-Newtonian spin effects in the higher modes.

## How Do Massive Black Holes Get Their Gas?

**Authors:** Hopkins, Philip F.; Quataert, Eliot

**Eprint:** http://arxiv.org/abs/0912.3257

**Keywords:** accretion discs; astro-ph.CO; astro-ph.GA; astro-ph.HE; astrophysics; EM counterparts; supermassive black holes

**Abstract:** We use multi-scale SPH simulations to follow the inflow of gas from galactic scales to <0.1pc, where the gas begins to resemble a traditional Keplerian accretion disk. The key ingredients are gas, stars, black holes (BHs), self-gravity, star formation, and stellar feedback. We use ~ 100 simulations to survey a large parameter space of galaxy properties and subgrid models for the ISM physics. We generate initial conditions for our simulations of galactic nuclei (<~ 300pc) using galaxy scale simulations, including both major mergers and isolated bar-(un)stable disk galaxies. For sufficiently gas-rich, disk-dominated systems, a series of gravitational instabilities generates large accretion rates of up to $1 - 10 M_\odot$/yr onto the BH (at <~ 10pc, our simulations resemble the 'bars within bars' model, but the gas exhibits diverse morphologies, including spirals, rings, clumps, and bars; their duty cycle is modest, complicating attempts to correlate BH accretion with nuclear morphology. At ~ 1-10pc, the gravitational potential becomes dominated by the BH and bar-like modes are no longer present. However, the gas becomes unstable to a standing, eccentric disk or a single-armed spiral mode (m=1), driving the gas to sub-pc scales. Proper treatment of this mode requires including star formation and the self-gravity of both the stars and gas. We predict correlations between BHAR and SFR at different galactic nuclei: nuclear SF is more tightly coupled to AGN activity, but correlations exist at all scales.

## Inflow-Outflow Solution with Stellar Winds and Conduction near Sgr A*

**Authors:** Shcherbakov, Roman V.; Baganoff, Frederick K.

**Eprint:** http://arxiv.org/abs/0912.3255

**Keywords:** accretion discs; astro-ph.HE; EM counterparts; Sagittarius A*; supermassive black holes

**Abstract:** We propose a 2-temperature radial dynamical model of plasma flow near Sgr A* and fit the bremsstrahlung emission to extensive quiescent X-Ray Chandra data. The model extends from several arcseconds to black hole (BH) gravitational radius, describing the outer accretion flow together with the infalling region. The model incorporates electron heat conduction, relativistic heat capacity of particles and feeding by stellar winds. Stellar winds from each star are considered separately as sources of mass, momentum and energy. Self-consistent search for the stagnation and sonic points is performed. Most of gas is found to outflow from the region. The accretion rate is limited to below 1% of Bondi rate due to the effect of thermal conduction enhanced by entropy production in a turbulent flow. The X-Ray brightness profile proves too steep near the BH, thus a synchrotron self-Compton point source is inferred with luminosity $L = 3x10^{32}$erg/s. We fit the sub-mm emission from the inner flow, thus aiming at a single model of Sgr A* accretion suitable at any radius.

### Toward a dynamical shift condition for unequal mass black hole binary simulations

**Authors:** Mueller, Doreen; Bruegmann, Bernd

**Eprint:** http://arxiv.org/abs/0912.3125

**Keywords:** gr-qc; massive binaries of black holes; numerical methods; numerical relativity

**Abstract:** Moving puncture simulations of black hole binaries rely on a specific gauge choice that leads to approximately stationary coordinates near each black hole. Part of the shift condition is a damping parameter, which has to be properly chosen for stable evolutions. However, a constant damping parameter does not account for the difference in mass in unequal mass binaries. We introduce a position dependent shift damping that addresses this problem. Although the coordinates change, the changes in the extracted gravitational waves are small.

### The EXIST view of Super-Massive Black Holes in the Universe

**Authors:** Della Ceca, Roberto; Ghisellini, Gabriele; Tagliaferri, Gianpiero; Foschini, Luigi; Pareschi, Giovanni; Tavecchio, Fabrizio; Coppi, Paolo; Grindlay, Josh E.; Fiocchi, Maria Teresa; Natalucci, Lorenzo; Panessa, Francesca; Ubertini, Pietro

**Eprint:** http://arxiv.org/abs/0912.3096

**Keywords:** astro-ph.CO; astro-ph.HE; astrophysics; observations; supermassive black holes

**Abstract:** With its large collection area, broad-band energy coverage from optical/NIR (0.3 to 2.2 micron) to soft/hard X-ray (0.1-600 keV), all-sky monitoring capability, and on-board follow-up, the proposed Energetic X-ray Imaging Survey Telescope mission (EXIST, see L. Natalucci contribution at this conference) has been designed to properly tackle the study of the AGN phenomenon and the role that SMBH play in the Universe. In particular EXIST will carry out an unprecedented survey above 10 keV (a factor ∼ 20 increase in hard X-ray sensitivity compared to current and prior X-ray missions) of SMBH activity, not just in space but also in time and over a significant expanded energy range; this strategy will overcome previous selection biases, will break the "multi-wavelength" identification bottleneck and will dramatically increase the number of AGN detected above 10 keV that are amenable to detailed follow-up studies (∼ 50000 AGN are expected). We discuss here on few selected AGN science topics enabled by the unique combination of EXIST's instruments. In particular EXIST will enable major progress in understanding: i) when and where SMBH are active in the Universe (by revealing and measuring heavily obscured accretion in the local - z<0.5 - Universe), ii) the physics of how SMBH accrete (by studying the broad-band X-ray spectra and variability properties of an unbiased and significant sample of AGN), and iii) the link between accretion power and jet/outflow power (by using observations of blazars). Last but not least EXIST's ability to find powerful, but very rare blazars, enables it to probe the appearance of the very first SMBH in the Universe allowing us to set strong constraints on the models of SMBH formation and early growth in the Universe.

## Constraining the evolutionary history of Newton's constant with gravitational wave observations

**Authors:** Yunes, Nicolas; Pretorius, Frans; Spergel, David

**Eprint:** http://arxiv.org/abs/0912.2724

**Keywords:** astro-ph.CO; astro-ph.HE; cosmology; general relativity; gr-qc; hep-ph; numerical relativity

**Abstract:** Space-borne gravitational wave detectors, such as the proposed Laser Interferometer Space Antenna, are expected to observe black hole coalescences to high redshift and with large signal-to-noise ratios, rendering their gravitational waves ideal probes of fundamental physics. The promotion of Newton's constant to a time-function introduces modifications to the binary's binding energy and the gravitational wave luminosity, leading to corrections in the chirping frequency. Such corrections propagate into the response function and, given a gravitational wave observation, they allow for constraints on the first time-derivative of Newton's constant at the time of merger. We find that space-borne detectors could indeed place interesting constraints on this quantity as a function of sky position and redshift, providing a *constraint map* over the entire range of redshifts where binary black hole mergers are expected to occur. A LISA observation of an equal-mass inspiral event with total redshifted mass of $10^5$ solar masses for three years should be able to measure $\dot{G}/G$ at the time of merger to better than $10^{-11}$/yr.

### X-ray Polarization from Accreting Black Holes: Coronal Emission

**Authors:** Schnittman, Jeremy; Krolik, Julian

**Eprint:** http://arxiv.org/abs/0912.0907

**Keywords:** EM counterparts; accretion discs

**Abstract:** We present new calculations of X-ray polarization from accreting black holes (BHs), using a Monte-Carlo ray-tracing code in full general relativity. In our model, an optically thick disk in the BH equatorial plane produces thermal seed photons with polarization oriented parallel to the disk surface. These seed photons are then inverse-Compton scattered through a hot (but thermal) corona, producing a hard X-ray power-law spectrum. We consider three different models for the corona geometry: a wedge "sandwich" with aspect ratio H/R and vertically-integrated optical depth $\tau_0$ constant throughout the disk; an inhomogeneous "clumpy" corona with a finite number of hot clouds distributed randomly above the disk within a wedge geometry; and a spherical corona of uniform density, centered on the BH and surrounded by a truncated thermal disk with inner radius $R_{edge}$. In all cases we find a characteristic transition from horizontal polarization at low energies to vertical polarization above the thermal peak; the vertical direction is defined as the projection of the BH spin axis on the plane of the sky. We show how the details of the spectropolarization signal can be used to distinguish between these models and infer various properties of the corona and BH. Although the bulk of this paper focuses on stellar-mass BHs, we also consider the effects of coronal scattering on the X-ray polarization signal from supermassive BHs in active galactic nuclei.

### Post-ISCO Ringdown Amplitudes in Extreme Mass Ratio Inspiral

**Authors:** Hadar, Shahar; Kol, Barak

**Eprint:** http://arxiv.org/abs/0911.3899

**Keywords:** EMRI; gr-qc; waveforms

**Abstract:** An extreme mass ratio inspiral consists of two parts: adiabatic inspiral and plunge. The plunge trajectory from the innermost stable circular orbit (ISCO) is special (somewhat independent of initial conditions). We write an expression for its solution in closed-form and for the emitted waveform. In particular we extract an expression for the associated black-hole ringdown amplitudes, which reduces to solving certain ordinary (radial) differential equations in the Schwarzschild background.

## LISA long-arm interferometry

**Authors:** Thorpe, James Ira

**Eprint:** http://arxiv.org/abs/0911.3175

**Keywords:** gr-qc; instruments; interferometers

**Abstract:** The Laser Interferometer Space Antenna (LISA) will observe gravitational radiation in the milliHertz band by measuring picometer-level fluctuations in the distance between drag-free proof masses over baselines of approximately five million kilometers. The measurement over each baseline will be divided into three parts: two short-arm measurements between the proof masses and a fiducial point on their respective spacecraft, and a long-arm measurement between fiducial points on separate spacecraft. This work focuses on the technical challenges associated with these long-arm measurements and the techniques that have been developed to overcome them.

## Radiation from low-momentum zoom-whirl orbits

**Authors:** Gold, Roman; Bruegmann, Bernd

**Eprint:** http://arxiv.org/abs/0911.3862

**Keywords:** geodesic motion; gr-qc; massive binaries of black holes; numerical relativity

**Abstract:** We study zoom-whirl behaviour of equal mass, non-spinning black hole binaries in full general relativity. The magnitude of the linear momentum of the initial data is fixed to that of a quasi-circular orbit, and its direction is varied. We find a global maximum in radiated energy for a configuration which completes roughly one orbit. The radiated energy in this case exceeds the value of a quasi-circular binary with the same momentum by 15%. The direction parameter only requires minor tuning for the localisation of the maximum. There is non-trivial dependence of the energy radiated on eccentricity (several local maxima and minima). Correlations with orbital dynamics shortly before merger are discussed. While being strongly gauge-dependent, these findings are intuitive from a physical point of view and support basic ideas about the efficiency of gravitational radiation from a binary system.

### Understanding possible electromagnetic counterparts to loud gravitational wave events: Binary black hole effects on electromagnetic fields

**Authors:** Palenzuela, Carlos; Lehner, Luis; Yoshida, Shin

**Eprint:** http://arxiv.org/abs/0911.3889

**Keywords:** astro-ph.HE; EM counterparts; gr-qc; massive binaries of black holes; numerical relativity

**Abstract:** In addition to producing loud gravitational waves (GW), the dynamics of a binary black hole system could induce emission of electromagnetic (EM) radiation by affecting the behavior of plasmas and electromagnetic fields in their vicinity. We here study how the electromagnetic fields are affected by a pair of orbiting black holes through the merger. In particular, we show how the binary's dynamics induce a variability in possible electromagnetically induced emissions as well as an enhancement of electromagnetic fields during the late-merge and merger epochs. These time dependent features will likely leave their imprint in processes generating detectable emissions and can be exploited in the detection of electromagnetic counterparts of gravitational waves.

### Cosmic Evolution of Black Holes and Spheroids. IV. The BH Mass - Spheroid Luminosity Relation

**Authors:** Bennert, Vardha Nicola; Treu, Tommaso; Woo, Jong-Hak; Malkan, Matthew A.; Bris, Alexandre Le; Auger, Matthew W.; Gallagher, Sarah; Blandford, Roger D.

**Eprint:** http://arxiv.org/abs/0911.4107

**Keywords:** astro-ph.CO; astrophysics; observations; supermassive black holes

**Abstract:** From high-resolution images of 23 Seyfert-1 galaxies at z=0.36 and z=0.57 obtained with the Near Infrared Camera and Multi-Object Spectrometer on board the Hubble Space Telescope (HST), we determine host-galaxy morphology, nuclear luminosity, total host-galaxy luminosity and spheroid luminosity. Keck spectroscopy is used to estimate black hole mass ($M_{BH}$). We study the cosmic evolution of the $M_{BH}$-spheroid luminosity ($L_{sph}$) relation. In combination with our previous work, totaling 40 Seyfert-1 galaxies, the covered range in BH mass is substantially increased, allowing us to determine for the first time intrinsic scatter and correct evolutionary trends for selection effects. We re-analyze archival HST images of 19 local reverberation-mapped active galaxies to match the procedure adopted at intermediate redshift. Correcting spheroid luminosity for passive luminosity evolution and taking into account selection effects, we determine that at fixed present-day V-band spheroid luminosity, $M_{BH}/L_{sph} \propto (1+z)^{2.8+/-1.2}$. When including a sample of 44 quasars out to z=4.5 taken from the literature, with luminosity and BH mass corrected to a self-consistent calibration, we extend the BH mass range to over two orders of magnitude, resulting in $M_{BH}/L_{sph} \propto (1+z)^{1.4+/-0.2}$. The intrinsic scatter of the relation, assumed constant with redshift, is 0.3+/-0.1 dex (<0.6 dex at 95% CL). The evolutionary trend suggests that BH growth precedes spheroid assembly. Interestingly, the $M_{BH}$-total host-galaxy luminosity relation is apparently non-evolving. It hints at either a more fundamental relation or that the spheroid grows by a redistribution of stars. However, the high-z sample does not follow this relation, indicating that major mergers may play the dominant role in growing spheroids above z~ 1.

## Third-and-a-half order post-Newtonian equations of motion for relativistic compact binaries using the strong field point particle limit

**Authors:** Itoh, Yousuke

**Eprint:** http://arxiv.org/abs/0911.4232

**Keywords:** gr-qc; massive binaries of black holes; post-Newtonian theory

**Abstract:** We report our rederivation of the equations of motion for relativistic compact binaries through the third-and-a-half post-Newtonian (3.5 PN) order approximation to general relativity using the strong field point particle limit to describe self-gravitating stars instead of the Dirac delta functional. The computation is done in harmonic coordinates. Our equations of motion describe the orbital motion of the binary consisting of spherically symmetric non-rotating stars. The resulting equations of motion fully agree with the 3.5 PN equations of motion derived in the previous works. We also show that the locally defined energy of the star has a simple relation with its mass up to the 3.5 PN order.

### Ring Formation from an Oscillating Black Hole

**Authors:** Lovelace, R. V. E.; Kornreich, D. A.

**Eprint:** http://arxiv.org/abs/0911.4481

**Keywords:** astro-ph.GA; astrophysics; kicks/recoil; supermassive black holes

**Abstract:**

Massive black hole (BH) mergers can result in the merger remnant receiving a "kick", of order 200 km s$^{-1}$ or more, which will cause the remnant to oscillate about the galaxy centre. Here we analyze the case where the BH oscillates through the galaxy centre perpendicular or parallel to the plane of the galaxy for a model galaxy consisting of an exponential disk, a Plummer model bulge, and an isothermal dark matter halo. For the perpendicular motion we find that there is a strong resonant forcing of the disk radial motion near but somewhat less than the "resonant radii" $r_R$ where the BH oscillation frequency is equal one-half, one-fourth, (1/6, etc.) of the radial epicyclic frequency in the plane of the disk. Near the resonant radii there can be a strong enhancement of the radial flow and disk density which can lead to shock formation. In turn the shock may trigger the formation of a ring of stars near $r_R$. As an example, for a BH mass of $10^8 M_\odot$ and a kick velocity of 150 km s$^{-1}$, we find that the resonant radii lie between 0.2 and 1 kpc. For BH motion parallel to the plane of the galaxy we find that the BH leaves behind it a supersonic wake where star formation may be triggered. The shape of the wake is calculated as well as the slow-down time of the BH.

The differential rotation of the disk stretches the wake into ring-like segments.

### Near infrared flares of Sagittarius A*: Importance of near infrared polarimetry

**Authors:** Zamaninasab, M.; Eckart, A.; Witzel, G.; Dovciak, M.; Karas, V.; Giessuebel, R. Schoedel R.; Bremer, M.; Garcia-Marin, M.; Kunneriath, D.; Muzic, K.; Nishiyama, S.; Sabha, N.; Straubmeier, C.; Zensus, A.

**Eprint:** http://arxiv.org/abs/0911.4659

**Keywords:** astro-ph.GA; astrophysics; observations; Sagittarius A*; supermassive black holes

**Abstract:** We report on the results of new simulations of near-infrared (NIR) observations of the Sagittarius A* (Sgr A*) counterpart associated with the super-massive black hole at the Galactic Center. The observations have been carried out using the NACO adaptive optics (AO) instrument at the European Southern Observatory's Very Large Telescope and CIAO NIR camera on the Subaru telescope (13 June 2004, 30 July 2005, 1 June 2006, 15 May 2007, 17 May 2007 and 28 May 2008). We used a model of synchrotron emission from relativistic electrons in the inner parts of an accretion disk. The relativistic simulations have been carried out using the Karas-Yaqoob (KY) ray-tracing code. We probe the existence of a correlation between the modulations of the observed flux density light curves and changes in polarimetric data. Furthermore, we confirm that the same correlation is also predicted by the hot spot model. Correlations between intensity and polarimetric parameters of the observed light curves as well as a comparison of predicted and observed light curve features through a pattern recognition algorithm result in the detection of a signature of orbiting matter under the influence of strong gravity. This pattern is detected statistically significant against randomly polarized red noise. Expected results from future observations of VLT interferometry like GRAVITY experiment are also discussed.

## Testing Properties of the Galactic Center Black Hole Using Stellar Orbits

**Authors:** Merritt, David; Alexander, Tal; Mikkola, Seppo; Will, Clifford M.

**Eprint:** http://arxiv.org/abs/0911.4718

**Keywords:** astro-ph.CO; astro-ph.GA; astrophysics; EMRI; gr-qc; N-body; no-hair conjecture; Sagittarius A*; stellar dynamics

**Abstract:** The spin and quadrupole moment of the supermassive black hole at the Galactic center can in principle be measured via astrometric monitoring of stars orbiting at milliparsec (mpc) distances, allowing tests of general relativistic "no-hair" theorems (Will 2008). One complicating factor is the presence of perturbations from other stars, which may induce orbital precession of the same order of magnitude as that due to general relativistic effects. The expected number of stars in this region is small enough that full N-body simulations can be carried out. We present the results of a comprehensive set of such simulations, which include a post-Newtonian treatment of spin-orbit effects. A number of possible models for the distribution of stars and stellar remnants are considered. We find that stellar perturbations are likely to obscure the signal due to frame-dragging for stars beyond ~ 0.5 mpc from the black hole, while measurement of the quadrupole moment is likely to require observation of stars inside ~ 0.2 mpc. A high fraction of stellar remnants, e.g. 10-Solar-mass black holes, in this region would make tests of GR problematic at all radii. We discuss the possibility of separating the effects of stellar perturbations from those due to GR.

## Effective One Body description of tidal effects in inspiralling compact binaries

**Authors:** Damour, Thibault; Nagar, Alessandro

**Eprint:** <http://arxiv.org/abs/0911.5041>

**Keywords:** Effective one body; gr-qc; massive binaries of black holes; post-Newtonian theory; waveforms

**Abstract:** The late part of the gravitational wave signal of binary neutron star inspirals can in principle yield crucial information on the nuclear equation of state via its dependence on relativistic tidal parameters. In the hope of analytically describing the gravitational wave phasing during the late inspiral (essentially up to contact) we propose an extension of the effective one body (EOB) formalism which includes tidal effects. We compare the prediction of this tidal-EOB formalism to recently computed nonconformally flat quasi-equilibrium circular sequences of binary neutron star systems. Our analysis suggests the importance of higher-order (post-Newtonian) corrections to tidal effects, even beyond the first post-Newtonian order, and their tendency to *significantly* increase the "effective tidal polarizability" of neutron stars. We compare the EOB predictions to some recently advocated, non-resummed, post-Newtonian based ("Taylor-T4") description of the phasing of inspiralling systems. This comparison shows the strong sensitivity of the late-inspiral phasing to the choice of the analytical model, but raises the hope that a sufficiently accurate numerical–relativity–"calibrated" EOB model might give us a reliable handle on the nuclear equation of state

## Hyperaccreting Disks around Magnetars for Gamma-Ray Bursts: Effects of Strong Magnetic Fields

**Authors:** Zhang, Dong; Dai, Z. G.

**Eprint:** <http://arxiv.org/abs/0911.5528>

**Keywords:** astro-ph.HE; astrophysics; EM counterparts

**Abstract:** (Abridged) The hyperaccreting neutron star or magnetar disks cooled via neutrino emission can be a candidate of gamma-ray burst (GRB) central engines. The strong field $\geq 10^{15} - 10^{16}$ G of the magnetar can play a significant role in affecting the disk properties and even lead to the funnel accretion process. We investigate the effects of strong fields on the disks around magnetars, and discuss implications of such accreting magnetar systems for GRB and GRB-like events. We discuss quantum effects of the strong fields on the disk, and use the MHD conservation equations to describe the behavior of the disk flow coupled with a large scale field, which is generated by the star-disk interaction. In general, stronger fields give higher disk densities, pressures, temperatures and neutrino luminosity, and change the electron fraction and degeneracy state significantly. A magnetized disk is always viscously stable outside the Alfvén radius, but will be thermally unstable near the Alfvén radius where the magnetic field plays a more important role in transferring the angular momentum and heating the disk than the viscous stress. The funnel accretion process will be only important for an extremely strong field, which creates a magnetosphere inside the Alfvén radius and truncates the plane disk. Because of higher temperature and more concentrated neutrino emission of the magnetar surface ring-like belt region covered by funnel accretion, the neutrino annihilation rate from the accreting magnetars can be much higher than that from accreting neutron stars without fields. Furthermore, the neutrino annihilation mechanism and the magnetically-driven pulsar wind from the magnetar surface can work together to generate and feed an ultra-relativistic jet along the stellar magnetic poles.

### Estimating the Prompt Electromagnetic Luminosity of a Black Hole Merger

**Authors:** Krolik, Julian H.

**Eprint:** http://arxiv.org/abs/0911.5711

**Keywords:** astro-ph.CO; astro-ph.HE; astrophysics; EM counterparts

**Abstract:** Although recent work in numerical relativity has made tremendous strides in quantifying the gravitational wave luminosity of black hole mergers, very little is known about the electromagnetic luminosity that might occur in immediate conjunction with these events. We show that whenever the heat deposited in the gas near a pair of merging black holes is proportional to its total mass, and the surface density of the gas in the immediate vicinity is greater than the (quite small) amount necessary to make it optically thick, the characteristic scale of the luminosity emitted in direct association with the merger is the Eddington luminosity independent of the gas mass. The duration of the photon signal is proportional to the gas mass, and is generally rather longer than the merger event. At somewhat larger distances, dissipation associated with realigning the gas orbits to the new spin orientation of the black hole can supplement dissipation of the energy gained from orbital adjustment to the mass lost in gravitational radiation; these two heat sources can combine to augment the electromagnetic radiation over longer timescales.

### Asymptotics of black hole perturbations

**Authors:** Zenginoglu, Anil

**Eprint:** http://arxiv.org/abs/0911.2450

**Keywords:** gr-qc; linearized theory; numerical methods; numerical relativity

**Abstract:** We study linear gravitational perturbations of Schwarzschild spacetime by solving numerically Regge-Wheeler-Zerilli equations in time domain using hyperboloidal surfaces and a compactifying radial coordinate. We stress the importance of including the asymptotic region in the computational domain in studies of gravitational radiation. The hyperboloidal approach should be helpful in a wide range of applications employing black hole perturbation theory.

### Variability and stability in blazar jets on time scales of years: Optical polarization monitoring of OJ287 in 2005-2009

**Authors:** Villforth, C.; Nilsson, K.; Heidt, J.; Takalo, L. O.; Pursimo, T.; Berdyugin, A.; Lindfors, E.; Pasanen, M.; Winiarski, M.; Drozdz, M.; Ogloza, W.; Kurpinska-Winiarska, M.; Siwak, M.; Koziel-Wierzbowska, D.; Porowski, C.; Kuzmicz, A.; Krzesinski, J.; Kundera, T.; Wu, J. -H.; Zhou, X.; Efimov, Y.; Sadakane, K.; Kamada, M.; Ohlert, J.; Hentunen, V. -P.; Nissinen, M.; Dietrich, M.; Assef, R. J.; Atlee, D. W.; Bird, J.; DePoy, D. L.; Eastman, J.; Peeples, M. S.; Prieto, J.; Watson, L.; Yee, J. C.; Liakos, A.; Niarchos, P.; Gazeas, K.; Dogru, S.; Donmez, A.; Marchev, D.; Coggins-Hill, S. A.; Mattingly, A.; Keel, W. C.; Haque, S.; Aungwerojwit, A.; Bergvall, N.

**Eprint:** http://arxiv.org/abs/0912.0005

**Keywords:** astro-ph.CO; astrophysics; massive binaries of black holes; observations

**Abstract:** (Abridged) OJ287 is a BL Lac object that has shown double-peaked bursts at regular intervals of ∼ 12 yr during the last ∼ 40 yr. We analyse optical photopolarimetric monitoring data from 2005-2009, during which the latest double-peaked outburst occurred. The aim of this study is twofold: firstly, we aim to analyse variability patterns and statistical properties of the optical polarization light-curve. We find a strong preferred position angle in optical polarization. The preferred position angle can be explained by separating the jet emission into two components: an optical polarization core and chaotic jet emission. The optical polarization core is stable on time scales of years and can be explained as emission from an underlying quiescent jet component. The chaotic jet emission sometimes exhibits a circular movement in the Stokes plane. We interpret these events as a shock front moving forwards and backwards in the jet, swiping through a helical magnetic field. Secondly, we use our data to assess different binary black hole models proposed to explain the regularly appearing double-peaked bursts in OJ287. We compose a list of requirements a model has to fulfil. The list includes not only characteristics of the light-curve but also other properties of OJ287, such as the black hole mass and restrictions on accretion flow properties. We rate all existing models using this list and conclude that none of the models is able to explain all observations. We discuss possible new explanations and propose a new approach to understanding OJ287. We suggest that both the double-peaked bursts and the evolution of the optical polarization position angle could be explained as a sign of resonant accretion of magnetic field lines, a 'magnetic breathing' of the disc.

## Asymptotic expansions of Maximum Likelihood estimators errors, with an application to gravitational waves generated in the inspiral phase of binary mergers

**Authors:** Zanolin, M.; Vitale, S.; Makris, N.

**Eprint:** http://arxiv.org/abs/0912.0065

**Keywords:** gr-qc; massive binaries of black holes; parameter estimation; search algorithms

**Abstract:** In this paper we describe a new methodology to calculate analytically the error for a maximum likelihood estimate (MLE) for physical parameters from Gravitational wave signals. All the existing litterature focuses on the usage of the Cramer Rao Lower bounds (CRLB) as a mean to approximate the errors for large signal to noise ratios. We show here how the variance and the bias of a MLE estimate can be expressed instead in inverse powers of the signal to noise ratios where the first order in the variance expansion is the CRLB. As an application we compute the second order of the variance and bias for MLE of physical parameters from the inspiral phase of binary mergers and for noises of gravitational wave interferometers . We also compare the improved error estimate with existing numerical estimates. The value of the second order of the variance expansions allows to get error predictions closer to what is observed in numerical simulations. It also predicts correctly the necessary SNR to approximate the error with the CRLB and provides new insight on the relationship between waveform properties SNR and estimation errors. For example the timing match filtering becomes optimal only if the SNR is larger than the kurtosis of the gravitational wave spectrum.

### Relativistic Mergers of Supermassive Black Holes and their Electromagnetic Signatures

**Authors:** Bode, Tanja; Haas, Roland; Bogdanovic, Tamara; Laguna, Pablo; Shoemaker, Deirdre

**Eprint:** http://arxiv.org/abs/0912.0087

**Keywords:** astro-ph.CO; EM counterparts; general relativity; gr-qc; massive binaries of black holes; numerical relativity

**Abstract:** Coincident detections of electromagnetic (EM) and gravitational wave (GW) signatures from coalescence events of supermassive black holes are the next observational grand challenge. Such detections will provide the means to study cosmological evolution and accretion processes associated with these gargantuan compact objects. More generally, the observations will enable testing general relativity in the strong, nonlinear regime and will provide independent cosmological measurements to high precision. Understanding the conditions under which coincidences of EM and GW signatures arise during supermassive black hole mergers is therefore of paramount importance. As an essential step towards this goal, we present results from the first fully general relativistic, hydrodynamical study of the late inspiral and merger of equal-mass, spinning supermassive black hole binaries in a gas cloud. We find that variable EM signatures correlated with GWs can arise in merging systems as a consequence of shocks and accretion combined with the effect of relativistic beaming. The most striking EM variability is observed for systems where spins are aligned with the orbital axis and where orbiting black holes form a stable set of density wakes, but all systems exhibit some characteristic signatures that can be utilized in searches for EM counterparts. In the case of the most massive binaries observable by the Laser Interferometer Space Antenna, calculated luminosities imply that they may be identified by EM searches to z = 1, while lower mass systems and binaries immersed in low density ambient gas can only be detected in the local universe.

### Spinning compact binary inspiral: Independent variables and dynamically preserved spin configurations

**Authors:** Gergely, László Árpád

**Eprint:** http://arxiv.org/abs/0912.0459

**Keywords:** gr-qc; massive binaries of black holes; post-Newtonian theory; spin

**Abstract:** We establish the set of independent variables suitable to monitor the complicated evolution of the spinning compact binary during the inspiral. Our approach is valid up to the second post-Newtonian order, including spin and mass quadrupolar effects, for generic (noncircular, nonspherical) orbits. Then we analyze the conservative spin dynamics in terms of these variables. We prove that the only precessing and spinning black hole or neutron star binary configuration which is preserved by the post-Newtonian evolution with spin-spin and quadrupole-monopole contributions included is the equal mass, equal and identically oriented spin configuration. This analytic result puts severe limitations on what particular configurations can be selected in numerical investigations of compact binary evolutions, even in those including only the last orbits of the inspiral.

## The early evolution of massive black holes

**Authors:** Volonteri, Marta

**Eprint:** <http://arxiv.org/abs/0912.0525>

**Keywords:** astro-ph.CO; astro-ph.HE; cosmology; supermassive black holes

**Abstract:** Massive black holes are nowadays believed to reside in most local galaxies. Studies have also established a number of relations between the MBH mass and properties of the host galaxy such as bulge mass and velocity dispersion. These results suggest that central MBHs, while much less massive than the host (∼ 0.1%), are linked to the evolution of galactic structure. When did it all start? In hierarchical cosmologies, a single big galaxy today can be traced back to the stage when it was split up in hundreds of smaller components. Did MBH seeds form with the same efficiency in small proto-galaxies, or did their formation had to await the buildup of substantial galaxies with deeper potential wells? I briefly review here some of the physical processes that are conducive to the evolution of the massive black hole population. I will discuss black hole formation processes for 'seed' black holes that are likely to place at early cosmic epochs, and possible observational tests of these scenarios.

## Identifying Supermassive Black Hole Binaries With Broad Emission Line Diagnosis

**Authors:** Shen, Yue; Loeb, Abraham

**Eprint:** <http://arxiv.org/abs/0912.0541>

**Keywords:** astro-ph.CO; astrophysics; EM counterparts; massive binaries of black holes

**Abstract:** Double-peaked broad lines in Active Galactic Nuclei (AGNs) may indicate the existence of a supermassive black hole (SMBH) binary whose two broad line regions (BLRs) contribute together to the line profile. An alternative interpretation of the double-peaked broad line feature is a disk origin for the line emission. We calculate the expected broad line profiles for a SMBH binary with various separations, using simple BLR models. Under reasonable assumptions that both BLRs are illuminated by the two black holes (BHs) and that the ionizing flux at the BLR position is roughly constant, we confirm the emergence of double-peaked features and radial velocity drifts of the two peaks due to the binary orbital motion when the two BHs are close enough such that the light-of-sight orbital velocity difference is larger than the FWHM of individual broad components. However, when the two BHs are even closer such that the two BLRs are no longer distinct, the line profile becomes more complex and there are no coherent radial velocity drifts in the two peaks with time. We discuss the temporal variations of the broad line profile for binary SMBHs and highlight the different behaviors of reverberation mapping in the binary and disk emitter scenarios.

## The Mock LISA Data Challenges: from Challenge 3 to Challenge 4

**Authors:** Babak, Stanislav; Baker, John G.; Benacquista, Matthew J.; Cornish, Neil J.; Larson, Shane L.; Mandel, Ilya; Petiteau, Antoine; Porter, Edward K.; Robinson, Emma L.; Vallisneri, Michele; Vecchio, Alberto; Adams, Matt; Arnaud, Keith A.; Błaut, Arkadiusz; Bridges, Michael; Cohen, Michael; Cutler, Curt; Feroz, Farhan; Gair, Jonathan R.; Graff, Philip; Hobson, Mike; Key, Joey Shapiro; Królak, Andrzej; Lasenby, Anthony; Prix, Reinhard; Shang, Yu; Trias, Miquel; Veitch, John; Whelan, John T.

**Eprint:** http://arxiv.org/abs/0912.0548

**Keywords:** gr-qc; MLDC

**Abstract:** The Mock LISA Data Challenges are a program to demonstrate LISA data-analysis capabilities and to encourage their development. Each round of challenges consists of one or more datasets containing simulated instrument noise and gravitational waves from sources of undisclosed parameters. Participants analyze the datasets and report best-fit solutions for the source parameters. Here we present the results of the third challenge, issued in Apr 2008, which demonstrated the positive recovery of signals from chirping Galactic binaries, from spinning supermassive–black-hole binaries (with optimal SNRs between ∼ 10 and 2000), from simultaneous extreme-mass-ratio inspirals (SNRs of 10-50), from cosmic-string-cusp bursts (SNRs of 10-100), and from a relatively loud isotropic background with $\Omega_{gw}(f) \sim 10^{-11}$, slightly below the LISA instrument noise.

## Impact of mergers on LISA parameter estimation for nonspinning black hole binaries

**Authors:** McWilliams, Sean T.; Thorpe, James Ira; Baker, John G.; Kelly, Bernard J.

**Eprint:** http://arxiv.org/abs/0911.1078

**Keywords:** astrophysics; data analysis; EM counterparts; general relativity; gr-qc; interferometers; massive binaries of black holes; Metropolis-Hastings

**Abstract:** We investigate the precision with which the parameters describing the characteristics and location of nonspinning black hole binaries can be measured with the Laser Interferometer Space Antenna (LISA). By using complete waveforms including the inspiral, merger and ringdown portions of the signals, we find that LISA will have far greater precision than previous estimates for nonspinning mergers that ignored the merger and ringdown. Our analysis covers nonspinning waveforms with moderate mass ratios, q >= 1/10, and total masses $10^5 < M/M_\odot < 10^7$. We compare the parameter uncertainties using the Fisher matrix formalism, and establish the significance of mass asymmetry and higher-order content to the predicted parameter uncertainties resulting from inclusion of the merger. In real-time observations, the later parts of the signal lead to significant improvements in sky-position precision in the last hours and even the final minutes of observation. For comparable mass systems with total mass $M/M_\odot =\sim 10^6$, we find that the increased precision resulting from including the merger is comparable to the increase in signal-to-noise ratio. For the most precise systems under investigation, half can be localized to within O(10 arcmin), and 10% can be localized to within O(1 arcmin).

## Hydrodynamic Simulations of Oscillating Shock Waves in a Sub-Keplerian Accretion Flow Around Black Holes

**Authors:** Giri, Kinsuk; Chakrabarti, Sandip K.; Samanta, Madan M.; Ryu, Dongsu

**Eprint:** http://arxiv.org/abs/0912.1174

**Keywords:** accretion discs; astro-ph.HE; EM counterparts

**Abstract:** We study the accretion processes on a black hole by numerical simulation. We use a grid based finite difference code for this purpose. We scan the parameter space spanned by the specific energy and the angular momentum and compare the time-dependent solutions with those obtained from theoretical considerations. We found several important results (a) The time dependent flow behaves close to a constant height model flow in the pre-shock region and a flow with vertical equilibrium in the post-shock region. (c) The infall time scale in the post-shock region is several times higher than the free-fall time scale. (b) There are two discontinuities in the flow, one being just outside of the inner sonic point. Turbulence plays a major role in determining the locations of these discontinuities. (d) The two discontinuities oscillate with two different frequencies and behave as a coupled harmonic oscillator. A Fourier analysis of the variation of the outer shock location indicates higher power at the lower frequency and lower power at the higher frequency. The opposite is true when the analysis of the inner shock is made. These behaviours will have implications in the spectral and timing properties of black hole candidates.

## Measuring the spin of the primary black hole in OJ287

**Authors:** Valtonen, M. J.; Mikkola, S.; Merritt, D.; Gopakumar, A.; Lehto, H. J.; Hyvönen, T.; Rampadarath, H.; Saunders, R.; Basta, M.; Hudec, R.

**Eprint:** http://arxiv.org/abs/0912.1209

**Keywords:** astro-ph.CO; astro-ph.HE; astrophysics; massive binaries of black holes; observations; spin

**Abstract:**

The compact binary system in OJ287 is modelled to contain a spinning primary black hole with an accretion disk and a non-spinning secondary black hole. Using Post Newtonian (PN) accurate equations that include 2.5PN accurate non-spinning contributions, the leading order general relativistic and classical spin-orbit terms, the orbit of the binary black hole in OJ287 is calculated and as expected it depends on the spin of the primary black hole. Using the orbital solution, the specific times when the orbit of the secondary crosses the accretion disk of the primary are evaluated such that the record of observed outbursts from 1913 up to 2007 is reproduced. The timings of the outbursts are quite sensitive to the spin value. In order to reproduce all the known outbursts, including a newly discovered one in 1957, the Kerr parameter of the primary has to be 0.28 ± 0.08. The quadrupole-moment contributions to the equations of motion allow us to constrain the 'no-hair' parameter to be 1.0 ± 0.3 where 0.3 is the one sigma error. This supports the 'black hole no-hair theorem' within the achievable precision.

It should be possible to test the present estimate in 2015 when the next outburst is due. The timing of the 2015 outburst is a strong function of the spin: if the spin is 0.36 of the maximal value allowed in general relativity, the outburst begins in early November 2015, while the same event starts in the end of January 2016 if the spin is 0.2

## Detection of IMBHs from microlensing in globular clusters

**Authors:** Safonova, M.; Stalin, C. S.

**Eprint:** http://arxiv.org/abs/0912.1435

**Keywords:** astro-ph.GA; astrophysics; intermediate-mass black holes; observations

**Abstract:** Globular clusters have been alternatively predicted to host intermediate-mass black holes (IMBHs) or nearly impossible to form and retain them in their centres. Over the last decade enough theoretical and observational evidence have accumulated to believe that many galactic globular clusters may host IMBHs in their centres, just like galaxies do. The well-established correlations between the supermassive black holes and their host galaxies do suggest that, in extrapolation, globular clusters (GCs) follow the same relations. Most of the attempts in search of the central black holes (BHs) are not direct and present enormous observational difficulties due to the crowding of stars in the GC cores. Here we propose a new method of detection of the central BH – the microlensing of the cluster stars by the central BH. If the core of the cluster is resolved, the direct determination of the lensing curve and lensing system parameters are possible; if unresolved, the differential imaging technique can be applied. We calculate the optical depth to central BH microlensing for a selected list of Galactic GCs and estimate the average time duration of the events. We present the observational strategy and discuss the detectability of microlensing events using a 2-m class telescope.

## Time-Dependent Models for the Afterglows of Massive Black Hole Mergers

**Authors:** Tanaka, Takamitsu; Menou, Kristen

**Eprint:** http://arxiv.org/abs/0912.2054

**Keywords:** accretion discs; astro-ph.CO; astrophysics; EM counterparts; gr-qc; massive binaries of black holes

**Abstract:** The Laser Interferometer Space Antenna (LISA) will detect gravitational wave signals from coalescing pairs of massive black holes in the total mass range $(10^5-10^7)/M_\odot$ out to cosmological distances. Identifying and monitoring the electromagnetic counterparts of these events would enable cosmological studies and offer new probes of gas physics around well-characterized massive black holes. Milosavljevic & Phinney (2005) proposed that a circumbinary disk around a binary of mass $\sim 10^6 M_\odot$ will emit an accretion-powered X-ray afterglow approximately one decade after the gravitational wave event. We revisit this scenario by using Green's function solutions to calculate the temporal viscous evolution and the corresponding electromagnetic signature of the circumbinary disk. Our calculations suggest that an electromagnetic counterpart may become observable as a rapidly brightening source soon after the merger, i.e. several years earlier than previously thought. The afterglow can reach super-Eddington luminosities without violating the local Eddington flux limit. It is emitted in the soft X-ray by the innermost circumbinary disk, but it may be partially reprocessed at optical and infrared frequencies. We also find that the spreading disk becomes increasingly geometrically thick close to the central object as it evolves, indicating that the innermost flow could become advective and radiatively inefficient, and generate a powerful outflow. We conclude that the mergers of massive black holes detected by LISA offer unique opportunities for monitoring on humanly tractable timescales the viscous evolution of accretion flows and the emergence of outflows around massive black holes with precisely known masses, spins and orientations.

## Evolution of Supermassive Black Holes from Cosmological Simulations

**Authors:** Filloux, Ch.; Durier, F.; Pacheco, J. A. de Freitas; Silk, J.

**Eprint:** <http://arxiv.org/abs/0912.2223>

**Keywords:** astro-ph.CO; cosmology; supermassive black holes

**Abstract:** The correlations between the mass of supermassive black holes and properties of their host galaxies are investigated through cosmological simulations. Black holes grow from seeds of 100 solar masses inserted into density peaks present in the redshift range 12-15. Seeds grow essentially by accreting matter from a nuclear disk and also by coalescences resulting from merger episodes. At z=0, our simulations reproduce the black hole mass function and the correlations of the black hole mass both with stellar velocity dispersion and host dark halo mass. Moreover, the evolution of the black hole mass density derived from the present simulations agrees with that derived from the bolometric luminosity function of quasars, indicating that the average accretion history of seeds is adequately reproduced . However, our simulations are unable to form black holes with masses above $10^9 M_\odot$ at $z \sim 6$, whose existence is inferred from the bright quasars detected by the Sloan survey in this redshift range.

## Binary Black Hole Mergers in Gaseous Environments: "Binary Bondi" and "Binary Bondi-Hoyle-Lyttleton" Accretion

**Authors:** Farris, Brian D.; Liu, Yuk Tung; Shapiro, Stuart L.

**Eprint:** http://arxiv.org/abs/0912.2096

**Keywords:** accretion discs; astro-ph.HE; astrophysics; EM counterparts; gr-qc; massive binaries of black holes; numerical relativity

**Abstract:** Merging supermassive black hole-black hole (BHBH) binaries produced in galaxy mergers are promising sources of detectable gravitational waves. If such a merger takes place in a gaseous environment, there is a possibility of a simultaneous detection of electromagnetic and gravitational radiation, as the stirring, shock heating and accretion of the gas may produce variability and enhancements in the electromagnetic flux. Such a simultaneous detection can provide a wealth of opportunities to study gravitational physics, accretion physics, and cosmology. We investigate this scenario by performing fully general relativistic, hydrodynamic simulations of merging, equal-mass, nonspinning BHBH binaries embedded in gas clouds. We evolve the metric using the BSSN formulation with standard moving puncture gauge conditions and handle the hydrodynamics via a high-resolution shock-capturing (HRSC) scheme. We consider both "binary Bondi accretion" in which the binary is at rest relative to the ambient gas cloud, as well as "binary Bondi-Hoyle-Lyttleton accretion" in which the binary moves relative to the gas cloud. The gas cloud is assumed to be homogeneous far from the binary and governed by a $\Gamma$-law equation of state. We vary $\Gamma$ between 4/3 and 5/3. For each simulation, we compute the gas flow and accretion rate and estimate the electromagnetic luminosity due to bremsstrahlung and synchrotron emission. We find evidence for significant enhancements in both the accretion rate and luminosity over values for a single black hole of the same mass as the binary. We estimate that this luminosity enhancement should be detectable by LSST for a $10^6 M_\odot$ binary in a hot gas cloud of density n$\sim 10/cm^3$ and temperature T$\sim 10^6$ K at z=1, reaching a maximum of L$\sim 3 \times 10^{43}$ erg/s, with the emission peaking in the visible band.

### Filling the disk hollow following binary black hole merger: The transient accretion afterglow

**Authors:** Shapiro, Stuart L.

**Eprint:** http://arxiv.org/abs/0912.2345

**Keywords:** accretion discs; astro-ph.CO; astro-ph.HE; astrophysics; EM counterparts; gr-qc; massive binaries of black holes

**Abstract:** Tidal torques from a binary black hole (BHBH) empty out the central regions in any circumbinary gaseous accretion disk. The balance between tidal torques and viscosity maintain the inner edge of the disk at a radius r ~ 1.5a – 2a, where a is the binary semimajor axis. Eventually, the inspiraling binary decouples from disk and merges, leaving behind a central hollow ("donut hole") in the disk orbiting the remnant black hole. We present a simple, time-dependent, Newtonian calculation that follows the secular (viscous) evolution of the disk as it fills up the hollow down to the black hole innermost stable circular orbit and then relaxes to stationary equilibrium. We use our model to calculate the electromagnetic radiation ("afterglow") spectrum emitted during this transient accretion epoch. Observing the temporal increase in the total electromagnetic flux and the hardening of the spectrum as the donut hole fills may help confirm a BHBH merger detected by a gravitational wave interferometer. We show how the very existence of the initial hollow can lead to super-Eddington accretion during this secular phase if the rate is not very far below Eddington prior to decoupling. Our model, though highly idealized, may be useful in establishing some of the key parameters, thermal emission features and scalings that characterize this transient. It can serve as a guide in the design and calibration of future radiation-magnetohydrodynamic simulations in general relativity.

### Asymptotic expansions of Maximum Likelihood estimators errors, with an application to gravitational waves generated in the inspiral phase of binary mergers

**Authors:** Zanolin, M.; Vitale, S.; Makris, N.

**Eprint:** http://arxiv.org/abs/0912.0065

**Keywords:** gr-qc; massive binaries of black holes; parameter estimation; search algorithms

**Abstract:** In this paper we describe a new methodology to calculate analytically the error for a maximum likelihood estimate (MLE) for physical parameters from Gravitational wave signals. All the existing litterature focuses on the usage of the Cramer Rao Lower bounds (CRLB) as a mean to approximate the errors for large signal to noise ratios. We show here how the variance and the bias of a MLE estimate can be expressed instead in inverse powers of the signal to noise ratios where the first order in the variance expansion is the CRLB. As an application we compute the second order of the variance and bias for MLE of physical parameters from the inspiral phase of binary mergers and for noises of gravitational wave interferometers . We also compare the improved error estimate with existing numerical estimates. The value of the second order of the variance expansions allows to get error predictions closer to what is observed in numerical simulations. It also predicts correctly the necessary SNR to approximate the error with the CRLB and provides new insight on the relationship between waveform properties SNR and estimation errors. For example the timing match filtering becomes optimal only if the SNR is larger than the kurtosis of the gravitational wave spectrum.

## The quasar $M_{bh} - M_{host}$ relation through Cosmic Time II - Evidence for evolution from z=3 to the present age

**Authors:** Decarli, R.; Falomo, R.; Treves, A.; Labita, M.; Kotilainen, J. K.; Scarpa, R.

**Abstract:** We study the dependence of the $M_{bh} - M_{host}$ relation on the redshift up to z=3 for a sample of 96 quasars the host galaxy luminosities of which are known. Black hole masses were estimated assuming virial equilibrium in the broad line regions (Paper I), while the host galaxy masses were inferred from their luminosities. With this data we are able to pin down the redshift dependence of the $M_{bh}-M_{host}$ relation along 85 per cent of the Universe age. We show that, in the sampled redshift range, the $M_{bh} - L_{host}$ relation remains nearly unchanged. Once we take into account the aging of the stellar population, we find that the $M_{bh}/M_{host}$ ratio (Gamma) increases by a factor ~ 7 from z=0 to z=3. We show that Gamma evolves with z regardless of the radio loudness and of the quasar luminosity. We propose that most massive black holes, living their quasar phase at high-redshift, become extremely rare objects in host galaxies of similar mass in the Local Universe.

### The quasar $M_{bh} - M_{host}$ relation through Cosmic Time I - Dataset and black hole masses

**Authors:** Decarli, R.; Falomo, R.; Treves, A.; Kotilainen, J. K.; Labita, M.; Scarpa, R.

**Eprint:** http://arxiv.org/abs/0911.2983

**Keywords:** astro-ph.CO; astrophysics; cosmology; supermassive black holes

**Abstract:** We study the $M_{bh} - M_{host}$ relation as a function of Cosmic Time in a sample of 96 quasars from z=3 to the present epoch. In this paper we describe the sample, the data sources and the new spectroscopic observations. We then illustrate how we derive $M_{bh}$ from single-epoch spectra, pointing out the uncertainties in the procedure. In a companion paper, we address the dependence of the ratio between the black hole mass and the host galaxy luminosity and mass on Cosmic Time.

### Disk-outflow coupling: Energetics around spinning black holes

**Authors:** Bhattacharya, Debbijoy; Ghosh, Shubhrangshu; Mukhopadhyay, Banibrata

**Eprint:** http://arxiv.org/abs/0911.3049

**Keywords:** accretion discs; astro-ph.HE; astrophysics; EM counterparts; spin; supermassive black holes

**Abstract:** The mechanism by which outflows and plausible jets are driven from black hole systems, still remains observationally elusive. Notwithstanding, several observational evidences and deeper theoretical insights reveal that accretion and outflow/jet are strongly correlated. Here, we model an advective disk-outflow coupled dynamics, incorporating explicitly the vertical flux. Inter-connecting dynamics of outflow and accretion essentially upholds the conservation laws. We investigate the properties of the disk-outflow surface and its strong dependence on the rotation parameter of the black hole. The energetics of disk-outflow strongly depend on mass, accretion rate and spin of the black holes. The model clearly shows that the outflow power extracted from the disk increases strongly with the spin of the black hole, inferring that the power of the observed astrophysical jets has a proportional correspondence with the spin of the central object. In case of blazars (BL Lacs and Flat Spectrum Radio Quasars), most of their emission are believed to be originated from their jets. It is observed that BL Lacs are relatively low luminous than Flat Spectrum Radio Quasars (FSRQs). The luminosity might be linked to the power of the jet, which in turn reflects that the nuclear regions of the BL Lac objects have a relatively low spinning black hole compared to that in the case of FSRQ. If the extreme gravity is the source to power strong outflows and jets, then spin of the black hole, perhaps, might be the fundamental parameter to account for the observed astrophysical processes in an accretion powered system.

## Mock LISA Data Challenge for the galactic white dwarf binaries

**Authors:** Błaut, Arkadiusz; Babak, Stanislav; Królak, Andrzej

**Eprint:** http://arxiv.org/abs/0911.3020

**Keywords:** back/foreground; gr-qc; MLDC

**Abstract:** We present data analysis methods used in detection and the estimation of parameters of gravitational wave signals from the white dwarf binaries in the Mock LISA Data Challenge. Our main focus is on the analysis of Challenge 3.1, where the gravitational wave signals from more than 50 mln. Galactic binaries were added to the simulated Gaussian instrumental noise. Majority of the signals at low frequencies are not resolved individually. The confusion between the signals is strongly reduced at frequencies above 5 mHz. Our basic data analysis procedure is the maximum likelihood detection method. We filter the data through the template bank at the first step of the search, then we refine parameters using the Nelder-Mead algorithm, we remove the strongest signal found and we repeat the procedure. We detect reliably and estimate parameters accurately of more than ten thousand signals from white dwarf binaries.

## Black Hole Spin and the Radio Loud/Quiet Dichotomy of Active Galactic Nuclei

**Authors:** Tchekhovskoy, Alexander; Narayan, Ramesh; McKinney, Jonathan C.

**Eprint:** http://arxiv.org/abs/0911.2228

**Keywords:** astro-ph.GA; astro-ph.HE; astrophysics; spin; supermassive black holes

**Abstract:** The inferred power of radio loud active galactic nuclei (AGN) on average exceeds the power of similar radio quiet AGN by a factor of 1000. We investigate whether this dichotomy can be due to differences in the spin of the central black holes that power the radio-emitting jets in these sources. Using general relativistic magnetohydrodynamic simulations, we construct steady state axisymmetric numerical models of such systems for a wide range of spins (dimensionless spin parameter 0.1<= a <= 0.9999) and a variety of magnetic field geometries. We assume that the total magnetic flux through the hole horizon $r = r_H(a)$ is held constant. We find that, if the black hole is surrounded by a thin accretion disk, the total black hole power output depends approximately quadratically on the hole angular frequency, $P \propto \Omega_H^2 \propto (a/r_H)^2$, and we conclude that in this scenario the spin alone can produce power variations of only a few tens at most. However, if the disk is thick, such that the jet subtends a narrow solid angle around the polar axis, then the power dependence can become much steeper, $P \propto \Omega_H^4$ or even $\propto \Omega_H^6$, and does produce power variations of 1000 for realistic black hole spin distributions. We derive an analytic solution that accurately reproduces this steeper scaling of power, and we provide a numerical fitting formula that accurately reproduces all our simulated results. We discuss other physical effects that might contribute to the observed radio loud/quiet dichotomy of AGN.

## The importance of precession in modelling the direction of the final spin from a black-hole merger

**Authors:** Barausse, Enrico

**Eprint:** <http://arxiv.org/abs/0911.1274>

**Keywords:** astro-ph.CO; astro-ph.GA; gr-qc; kicks/recoil; massive binaries of black holes; post-Newtonian theory; spin

**Abstract:** The prediction of the spin of the black hole resulting from the merger of a generic black-hole binary system is of great importance to study the cosmological evolution of supermassive black holes. Several attempts have been recently made to model the spin via simple expressions exploiting the results of numerical-relativity simulations. Here I compare the results of all the simulations appeared so far in the literature with various formulas for the final spin magnitude and direction. I show that although all the formulas give reasonable results for the final spin magnitude, only the formula that I recently proposed in (Barausse & Rezzolla, Apj 704 L40) accurately predicts the final spin direction when applied to binaries with separations of hundred or thousands of gravitational radii. This makes my formula particularly suitable for cosmological merger-trees and N-body simulations, which provide the spins and angular momentum of the two black holes when their separation is of thousands of gravitational radii, and happens because my formula takes into account the post-Newtonian precession of the spins in a consistent manner.

# Hyper Velocity Stars and the Restricted Parabolic 3-body Problem

**Authors:** Sari, Re'em; Kobayashi, Shiho; Rossi, Elena M.

**Eprint:** <http://arxiv.org/abs/0911.1136>

**Keywords:** astro-ph.GA; astro-ph.HE; astrophysics; EMRI; N-body

**Abstract:** Motivated by detections of hypervelocity stars that may originate from the Galactic Center, we revist the problem of a binary disruption by a passage near a much more massive point mass. The six order of magnitude mass ratio between the Galactic Center black hole and the binary stars allows us to formulate the problem in the restricted parabolic three-body approximation. In this framework, results can be simply rescaled in terms of binary masses, its initial separation and binary-to-black hole mass ratio. Consequently, an advantage over the full three-body calculation is that a much smaller set of simulations is needed to explore the relevant parameter space. Contrary to previous claims, we show that, upon binary disruption, the lighter star does not remain preferentially bound to the black hole. In fact, it is ejected exactly in 50% of the cases. Nonetheless, lighter objects have higher ejection velocities, since the energy distribution is independent of mass. Focusing on the planar case, we provide the probability distributions for disruption of circular binaries and for the ejection energy. We show that even binaries that penetrate deeply into the tidal sphere of the black hole are not doomed to disruption, but survive in 20% of the cases. Nor do these deep encounters produce the highest ejection energies, which are instead obtained for binaries arriving to 0.1-0.5 of the tidal radius in a prograde orbit. Interestingly, such deep-reaching binaries separate widely after penetrating the tidal radius, but always approach each other again on their way out from the black hole.[shortened]

# Classifying LISA gravitational wave burst signals using Bayesian evidence

**Authors:** Feroz, Farhan; Gair, Jonathan R.; Graff, Philip; Hobson, Michael P; Lasenby, Anthony

**Eprint:** <http://arxiv.org/abs/0911.0288>

**Keywords:** data analysis; gr-qc; MLDC; parameter estimation; search algorithms

**Abstract:** We consider the problem of characterisation of burst sources detected with the Laser Interferometer Space Antenna (LISA) using the multi-modal nested sampling algorithm, MultiNest. We use MultiNest as a tool to search for modelled bursts from cosmic string cusps, and compute the Bayesian evidence associated with the cosmic string model. As an alternative burst model, we consider sine-Gaussian burst signals, and show how the evidence ratio can be used to choose between these two alternatives. We present results from an application of MultiNest to the last round of the Mock LISA Data Challenge, in which we were able to successfully detect and characterise all three of the cosmic string burst sources present in the release data set. We also present results of independent trials and show that MultiNest can detect cosmic string signals with signal-to-noise ratio (SNR) as low as ~ 7 and sine-Gaussian signals with SNR as low as ~ 8. In both cases, we show that the threshold at which the sources become detectable coincides with the SNR at which the evidence ratio begins to favour the correct model over the alternative.

## Gravitational waveforms from unequal-mass binaries with arbitrary spins under leading order spin-orbit coupling

**Authors:** Tessmer, Manuel

**Eprint:** http://arxiv.org/abs/0910.5931

**Keywords:** gr-qc; massive binaries of black holes; spin; waveforms

**Abstract:** The paper generalizes the structure of gravitational waves from orbiting spinning binaries under leading order spin-orbit coupling, as given in the work by Königsdörffer and Gopakumar [PRD 71, 024039 (2005)] for single-spin and equal-mass binaries, to unequal-mass binaries and arbitrary spin configurations. The orbital motion is taken to be quasi-circular and the fractional mass difference is assumed to be small against one. The emitted gravitational waveforms are given in analytic form.

## Dual black holes in merger remnants. II: spin evolution and gravitational recoil

**Authors:** Dotti, M.; Volonteri, M.; Perego, A.; Colpi, M.; Ruszkowski, M.; Haardt, F.

**Eprint:** http://arxiv.org/abs/0910.5729

**Keywords:** accretion discs; astro-ph.CO; astro-ph.HE; astrophysics; EM counterparts; kicks/recoil; massive binaries of black holes; spin

**Abstract:** Using high resolution hydrodynamical simulations, we explore the spin evolution of massive dual black holes orbiting inside a circumnuclear disc, relic of a gas-rich galaxy merger. The black holes spiral inwards from initially eccentric co or counter-rotating coplanar orbits relative to the disc's rotation, and accrete gas that is carrying a net angular momentum. As the black hole mass grows, its spin changes in strength and direction due to its gravito-magnetic coupling with the small-scale accretion disc. We find that the black hole spins loose memory of their initial orientation, as accretion torques suffice to align the spins with the angular momentum of their orbit on a short timescale (<1-2 Myr). A residual off-set in the spin direction relative to the orbital angular momentum remains, at the level of <10 degrees for the case of a cold disc, and <30 degrees for a warmer disc. Alignment in a cooler disc is more effective due to the higher coherence of the accretion flow near each black hole that reflects the large-scale coherence of the disc's rotation. If the massive black holes coalesce preserving the spin directions set after formation of a Keplerian binary, the relic black hole resulting from their coalescence receives a relatively small gravitational recoil. The distribution of recoil velocities inferred from a simulated sample of massive black hole binaries has median <70 km/s much smaller than the median resulting from an isotropic distribution of spins.

## Effective Inner Radius of Tilted Black Hole Accretion Disks

**Authors:** Fragile, P. Chris

**Eprint:** http://arxiv.org/abs/0910.5721

**Keywords:** accretion discs; astro-ph.HE; astrophysics; EM counterparts; spin; supermassive black holes

**Abstract:** One of the primary means of determining the spin of an astrophysical black hole is by actually measuring the inner radius of a surrounding accretion disk and using that to infer the spin. By comparing a number of different estimates of the inner radius from simulations of tilted accretion disks with differing black-hole spins, we show that such a procedure can give quite wrong answers. Over the range $0 \leq a/M \leq 0.9$, we find that, for moderately thick disks ($H/r \sim 0.2$) with modest tilt (15 degrees), the inner radius is nearly independent of spin. This result is likely dependent on tilt, such that for larger tilts, it may even be that the inner radius would increase with increasing spin. In the opposite limit, we confirm through numerical simulations of untilted disks that, in the limit of zero tilt, the inner radius recovers approximately the expected dependence on spin.

# Gravitational Self Force in a Schwarzschild Background and the Effective One Body Formalism

**Authors:** Damour, Thibault

**Eprint:** http://arxiv.org/abs/0910.5533

**Keywords:** Effective one body; EMRI; gr-qc; post-Newtonian theory

**Abstract:** We discuss various ways in which the computation of conservative Gravitational Self Force (GSF) effects on a point mass moving in a Schwarzschild background can inform us about the basic building blocks of the Effective One-Body (EOB) Hamiltonian. We display the information which can be extracted from the recently published GSF calculation of the first-GSF-order shift of the orbital frequency of the last stable circular orbit, and we combine this information with the one recently obtained by comparing the EOB formalism to high-accuracy numerical relativity (NR) data on coalescing binary black holes. The information coming from GSF data helps to break the degeneracy (among some EOB parameters) which was left after using comparable-mass NR data to constrain the EOB formalism. We suggest various ways of obtaining more information from GSF computations: either by studying eccentric orbits, or by focussing on a special zero-binding zoom-whirl orbit. We show that logarithmic terms start entering the post-Newtonian expansions of various (EOB and GSF) functions at the fourth post-Newtonian (4PN) level, and we analytically compute the first logarithm entering a certain, gauge-invariant "redshift" GSF function (defined along the sequence of circular orbits).

# Post-merger electromagnetic emissions from disks perturbed by binary black holes

**Authors:** Anderson, Matthew; Lehner, Luis; Megevand, Miguel; Neilsen, David

**Eprint:** http://arxiv.org/abs/0910.4969

**Keywords:** accretion discs; astro-ph.HE; astrophysics; EM counterparts; massive binaries of black holes

**Abstract:** We simulate the possible emission from a disk perturbed by a recoiling super-massive black hole. To this end, we study radiation transfer from the system incorporating bremsstrahlung emission from a Maxwellian plasma and absorption given by Kramer's opacity law modified to incorporate blackbody effects. We employ this model in the radiation transfer integration to compute the luminosity at several frequencies, and compare with previous bremsstrahlung luminosity estimations from a transparent limit (in which the emissivity is integrated over the computational domain and over all frequencies) and with a simple thermal emission model. We find close agreement between the radiation transfer results and the estimated bremsstrahlung luminosity from previous work for electromagnetic signals above $10^{14}$ Hz. For lower frequencies, we find a self-eclipsing behavior in the disk, resulting in a strong intensity variability connected to the orbital period of the disk.

## Constraining the initial mass function of stars in the Galactic Centre

**Authors:** Loeckmann, Ulf; Baumgardt, Holger; Kroupa, Pavel

**Eprint:** http://arxiv.org/abs/0910.4960

**Keywords:** astro-ph.CO; astro-ph.GA; astrophysics; N-body; Sagittarius A*; stellar dynamics; supermassive black holes

**Abstract:** (abridged) Here we discuss the question whether the extreme circumstances in the centre of the Milky Way may be the reason for a significant variation of the IMF. By means of stellar evolution models using different codes we show that the observed luminosity in the central parsec is too high to be explained by a long-standing top-heavy IMF, considering the limited amount of mass inferred from stellar kinematics in this region. In contrast, continuous star formation over the Galaxy's lifetime following a canonical IMF results in a mass-to-light ratio and a total mass of stellar black holes (SBHs) consistent with the observations. Furthermore, these SBHs migrate towards the centre due to dynamical friction, turning the cusp of visible stars into a core as observed in the Galactic Centre. For the first time here we explain the luminosity and dynamical mass of the central cluster and both the presence and extent of the observed core, since the number of SBHs expected from a canonical IMF is just enough to make up for the missing luminous mass. We conclude that the Galactic Centre is consistent with the canonical IMF and do not suggest a systematic variation as a result of the region's properties such as high density, metallicity, strong tidal field etc.

## Gravitational-Wave Recoil from the Ringdown Phase of Coalescing Black Hole Binaries

**Authors:** Tiec, Alexandre Le; Blanchet, Luc; Will, Clifford M.

**Eprint:** http://arxiv.org/abs/0910.4594

**Keywords:** gr-qc; gravitational recoil; massive binaries of black holes; post-Newtonian theory

**Abstract:** The gravitational recoil or "kick" of a black hole formed from the merger of two orbiting black holes, and caused by the anisotropic emission of gravitational radiation, is an astrophysically important phenomenon. We combine (i) an earlier calculation, using post-Newtonian theory, of the kick velocity accumulated up to the merger of two non-spinning black holes, (ii) a "close-limit approximation" calculation of the radiation emitted during the ringdown phase, and based on a solution of the Regge-Wheeler and Zerilli equations using initial data accurate to second post-Newtonian order. We prove that ringdown radiation produces a significant "anti-kick". Adding the contributions due to inspiral, merger and ringdown phases, our results for the net kick velocity agree with those from numerical relativity to 10-15 percent over a wide range of mass ratios, with a maximum velocity of 180 km/s at a mass ratio of 0.38.

## The Close-limit Approximation for Black Hole Binaries with Post-Newtonian Initial Conditions

**Authors:** Tiec, Alexandre Le; Blanchet, Luc

**Eprint:** http://arxiv.org/abs/0910.4593

**Keywords:** gr-qc; massive binaries of black holes; post-Newtonian theory; waveforms

**Abstract:** The ringdown phase of a black hole formed from the merger of two orbiting black holes is described by means of the close-limit (CL) approximation starting from second-post-Newtonian (2PN) initial conditions. The 2PN metric of point-particle binaries is formally expanded in CL form and identified with that of a perturbed Schwarzschild black hole. The multipolar coefficients describing the even-parity (or polar) and odd-parity (axial) components of the linear perturbation consistently satisfy the 2PN-accurate perturbative field equations. We use these coefficients to build initial conditions for the Regge-Wheeler and Zerilli wave equations, which we then evolve numerically. The ringdown waveform is obtained in two cases: head-on collision with zero-angular momentum, composed only of even modes, and circular orbits, for which both even and odd modes contribute. In a separate work, this formalism is applied to the study of the gravitational recoil produced during the ringdown phase of coalescing binary black holes.

## Transition from radiatively inefficient to cooling dominated phase in two temperature accretion discs around black holes

**Authors:** Sinha, Monika; Rajesh, S. R.; Mukhopadhyay, Banibrata

**Eprint:** http://arxiv.org/abs/0910.4818

**Keywords:** accretion discs; astro-ph.HE; astrophysics; supermassive black holes

**Abstract:** We investigate the transition of a radiatively inefficient phase of viscous two temperature accreting flow to a cooling dominated phase and vice versa around black holes. Based on a global sub-Keplerian accretion disc model in steady state, including explicit cooling processes self-consistently, we show that general advective accretion flow passes through various phases during its infall towards a black hole. Bremsstrahlung, synchrotron and inverse Comptonization of soft photons are considered as possible cooling mechanisms. Hence the flow governs much lower electron temperature $\sim 10^8 - 10^{9.5}$K compared to the hot protons of temperature $\sim 10^{10.2} - 10^{11.8}$K in the range of accretion rate in Eddington units 0.01 - 100. Therefore, the solutions may potentially explain the hard X-rays and gamma-rays emitted from AGNs and X-ray binaries. We finally compare the solutions for two different regimes of viscosity and conclude that a weakly viscous flow is expected to be cooling dominated compared to its highly viscous counterpart which is radiatively inefficient. The flow is successfully able to reproduce the observed luminosities of the under-fed AGNs and quasars (e.g. Sgr A*), ultra-luminous X-ray sources (e.g. SS433), as well as the highly luminous AGNs and ultra-luminous quasars (e.g. PKS 0743-67) at different combinations of mass accretion rate, ratio of specific heats.

## Two temperature accretion around rotating black holes: Description of general advective flow paradigm in presence of various cooling processes to explain low to high luminous sources

**Authors:** Rajesh, S. R.; Mukhopadhyay, Banibrata

**Eprint:** http://arxiv.org/abs/0910.4502

**Keywords:** accretion discs; astro-ph.HE; astrophysics; EM counterparts; gr-qc; Sagittarius A*; supermassive black holes

How stars distribute around a massive black hole

**Abstract:** We investigate the viscous two temperature accretion discs around rotating black holes. We describe the global solution of accretion flows with a sub-Keplerian angular momentum profile, by solving the underlying conservation equations including explicit cooling processes selfconsistently. Bremsstrahlung, synchrotron and inverse Comptonization of soft photons are considered as possible cooling mechanisms, for sub-Eddington, Eddington and super-Eddington mass accretion rates around Schwarzschild and Kerr black holes with a Kerr parameter 0.998. It is found that the flow, during its infall from the Keplerian to sub-Keplerian transition region to the black hole event horizon, passes through various phases of advection – general advective paradigm to radiatively inefficient phase and vice versa. Hence the flow governs much lower electron temperature $\sim 10^8 - 10^{9.5}$ K, in the range of accretion rate in Eddington units $0.01 <\sim \dot{M} <\sim 100$, compared to the hot protons of temperature $\sim 10^{10.2} - 10^{11.8}$K. Therefore, the solution may potentially explain the hard X-rays and $\gamma$-rays emitted from AGNs and X-ray binaries. We then show that a weakly viscous flow is expected to be cooling dominated, particularly at the inner region of the disc, compared to its highly viscous counterpart which is radiatively inefficient. With all the solutions in hand, we finally reproduce the observed luminosities of the under-fed AGNs and quasars (e.g. Sgr $A^*$) to ultra-luminous X-ray sources (e.g. SS433), at different combinations of input parameters such as mass accretion rate, ratio of specific heats. The set of solutions also predicts appropriately the luminosity observed in the highly luminous AGNs and ultra-luminous quasars (e.g. PKS 0743-67).

### Alternative derivation of the response of interferometric gravitational wave detectors

**Authors:** Cornish, Neil J.

**Eprint:** http://arxiv.org/abs/0910.4372

**Keywords:** gr-qc; instruments; interferometers

**Abstract:** It has recently been pointed out by Finn that the long-standing derivation of the response of an interferometric gravitational wave detector contains several errors. Here I point out that a contemporaneous derivation of the gravitational wave response for spacecraft doppler tracking and pulsar timing avoids these pitfalls, and when adapted to describe interferometers, recovers a simplified version of Finn's derivation. This simplified derivation may be useful for pedagogical purposes.

## X-Ray Localization of the Intermediate-Mass Black Hole in the Globular Cluster G1 with Chandra

**Authors:** Kong, A. K. H.; Heinke, C. O.; Di Stefano, R.; Barmby, P.; Lewin, W. H. G.; Primini, F. A.

**Eprint:** http://arxiv.org/abs/0910.3944

**Keywords:** astro-ph.HE; astrophysics; globular clusters; intermediate-mass black holes; observations

**Abstract:** We report the most accurate X-ray position of the giant globular cluster G1 in M31 by using the Chandra X-ray Observatory, Hubble Space Telescope (HST), and Canada-France-Hawaii Telescope (CFHT). G1 is clearly detected with Chandra and by cross-registering with HST and CFHT images, we derive a 1sigma error radius of 0.15", significantly smaller than the previous measurement by XMM-Newton. We conclude that the X-ray emission of G1 comes from within the core radius of the cluster. There are two possibilities for the origin of the X-ray emission: it could be due to either accretion of a central intermediate-mass black hole, or ordinary low-mass X-ray binaries. Based on the ratio of X-ray to the Eddington luminosity, an intermediate-mass black hole accreting from the cluster gas seems unlikely and we suggest that the X-rays are due to accretion from a companion. We also find that the X-ray emission may be offset from the radio emission. Future high-resolution and high-sensitivity radio imaging observations will reveal whether there is an intermediate-mass black hole at the center of G1.

## High accuracy binary black hole simulations with an extended wave zone

**Authors:** Pollney, Denis; Reisswig, Christian; Schnetter, Erik; Dorband, Nils; Diener, Peter

**Eprint:** http://arxiv.org/abs/0910.3803

**Keywords:** gr-qc; massive binaries of black holes; numerical methods; numerical relativity

**Abstract:** We present results from a new code for binary black hole evolutions using the moving-puncture approach, implementing finite differences in generalised coordinates, and allowing the spacetime to be covered with multiple communicating non-singular coordinate patches. Here we consider a regular Cartesian near zone, with adapted spherical grids covering the wave zone. The efficiencies resulting from the use of adapted coordinates allow us to maintain sufficient grid resolution to an artificial outer boundary location which is causally disconnected from the measurement. For the well-studied test-case of the inspiral of an equal-mass non-spinning binary (evolved for more than 8 orbits before merger), we determine the phase and amplitude to numerical accuracies better than 0.010% and 0.090% during inspiral, respectively, and 0.003% and 0.153% during merger. The waveforms, including the resolved higher harmonics, are convergent and can be consistently extrapolated to $r \to \infty$ throughout the simulation, including the merger and ringdown. Ringdown frequencies for these modes (to $(\ell, m) = (6, 6)$) match perturbative calculations to within 0.01%, providing a strong confirmation that the remnant settles to a Kerr black hole with irreducible mass $M_{\text{irr}} = 0.884355 \pm 20 \times 10^{-6}$ and spin $S_f/M_f^2 = 0.686923 \pm 10 \times 10^{-6}$.

## Emergent Spectra From Disks Surrounding Kerr Black Holes: Effect of Photon Trapping and Disk Self-Shadowing

**Authors:** Li, Guang-Xing; Yuan, Ye-Fei; Cao, Xinwu

**Eprint:** <http://arxiv.org/abs/0910.3530>

**Keywords:** accretion discs; astro-ph.GA; astro-ph.HE; astrophysics; EM counterparts; spin; supermassive black holes

**Abstract:** Based on a new estimation of their thickness, the global properties of relativistic slim accretion disks are investigated in this work. The resulting emergent spectra are calculated using relativistic ray-tracing method. The angular dependence of the disk luminosity, the effects of the heat advection and the effect of disk thickness on the estimation of the black hole spin and accretion rate are discussed. The improvements compared to previous works are that we use self-consistent disk equations and we consider the disk self-shadowing effect. We find that at moderate accretion rate, with inclusion of the heat advection effect, radiation trapped in the outer region of the accretion disks will escape in the inner region of the accretion disk and contribute to the emergent spectra. At high accretion rate, large inclination and large black hole spin, both the disk thickness and the heat advection have significant influence on the emergent spectra. Consequently, these effects will influence the measurement of the black hole spin based on the spectral fitting and influence the angular dependence of luminosity.

## On the spatial distribution and the origin of hypervelocity stars

**Authors:** Lu, Youjun; Zhang, Fupeng; Yu, Qingjuan

**Eprint:** http://arxiv.org/abs/0910.3260

**Keywords:** astro-ph.GA; astrophysics; EMRI; observations; Sagittarius A*; stellar dynamics

**Abstract:** Hypervelocity stars (HVSs) escaping away from the Galactic halo are dynamical products of interactions of stars with the massive black hole(s) (MBH) in the Galactic Center (GC). They are mainly B-type stars with their progenitors unknown. OB stars are also populated in the GC, with many being hosted in a clockwise-rotating young stellar (CWS) disk within half a parsec from the MBH and their formation remaining puzzles. In this paper, we demonstrate that HVSs can well memorize the injecting directions of their progenitors using both analytical arguments and numerical simulations, i.e., the ejecting direction of an HVS is almost anti-parallel to the injecting direction of its progenitor. Therefore, the spatial distribution of HVSs maps the spatial distribution of the parent population of their progenitors directly. We also find that almost all the discovered HVSs are spatially consistent with being located on two thin disk planes. The orientation of one plane is consistent with that of the (inner) CWS disk, which suggests that most of the HVSs originate from the CWS disk or a previously existed disk-like stellar structure with an orientation similar to it. The rest of HVSs may be correlated with the plane of the northern arm of the mini-spiral in the GC or the plane defined by the outer warped part of the CWS disk. Our results not only support the GC origin of HVSs but also imply that the central disk (or the disk structure with a similar orientation) should persist or be frequently rejuvenated over the past 200 Myr, which adds a new challenge to the stellar disk formation and provides insights to the longstanding problem of gas fueling into massive black holes.

## On strong mass segregation around a massive black hole: Implications for lower-frequency gravitational-wave astrophysics

**Authors:** Preto, Miguel; Amaro-Seoane, Pau

**Eprint:** http://arxiv.org/abs/0910.3206

**Keywords:** astro-ph.CO; astro-ph.GA; astrophysics; EMRI; N-body; stellar dynamics; supermassive black holes

How stars distribute around a massive black hole

**Abstract:** We present, for the first time, a clear *N*-body realization of the *strong mass segregation* solution for the stellar distribution around a massive black hole. We compare our *N*-body results with those obtained by solving the orbit-averaged Fokker-Planck (FP) equation in energy space. The *N*-body segregation is slightly stronger than in the FP solution, but both confirm the *robustness* of the regime of strong segregation when the number fraction of heavy stars is a (realistically) small fraction of the total population. In view of recent observations revealing a dearth of giant stars in the sub-parsec region of the Milky Way, we show that the time scales associated with cusp re-growth are not longer than $(0.1 - 0.25) \times T_{rlx}(r_h)$. These time scales are shorter than a Hubble time for black holes masses $M_\bullet \lesssim 4 \times 10^6 M_\odot$ and we conclude that quasi-steady, mass segregated, stellar cusps may be common around MBHs in this mass range. Since EMRI rates scale as $M_\bullet^{-\alpha}$, with $\alpha \in [1/4, 1]$, a good fraction of these events should originate from strongly segregated stellar cusps.

## The orbit of the star S2 around SgrA* from VLT and Keck data

**Authors:** Gillessen, S.; Eisenhauer, F.; Fritz, T. K.; Bartko, H.; Dodds-Eden, K.; Pfuhl, O.; Ott, T.; Genzel, R.

**Eprint:** http://arxiv.org/abs/0910.3069

**Keywords:** astro-ph.GA; astrophysics; observations; Sagittarius A*; stellar dynamics; supermassive black holes

**Abstract:** Two recent papers (Ghez et al. 2008, Gillessen et al. 2009) have estimated the mass of and the distance to the massive black hole in the center of the Milky Way using stellar orbits. The two astrometric data sets are independent and yielded consistent results, even though the measured positions do not match when simply overplotting the two sets. In this letter we show that the two sets can be brought to excellent agreement with each other when allowing for a small offset in the definition of the reference frame of the two data sets. The required offsets in the coordinates and velocities of the origin of the reference frames are consistent with the uncertainties given in Ghez et al. (2008). The so combined data set allows for a moderate improvement of the statistical errors of mass of and distance to Sgr A*, but the overall accuracies of these numbers are dominated by systematic errors and the long-term calibration of the reference frame. We obtain R0 = 8.28 +- 0.15(stat) +- 0.29(sys) kpc and M(MBH) = 4.30 +- 0.20(stat) +- 0.30(sys) $\times 10^6 M_\odot$ as best estimates from a multi-star fit.

## Statistical studies of Spinning Black-Hole Binaries

**Authors:** Lousto, Carlos O.; Nakano, Hiroyuki; Zlochower, Yosef; Campanelli, Manuela

**Eprint:** <http://arxiv.org/abs/0910.3197>

**Keywords:** general relativity; gr-qc; massive binaries of black holes; spin

**Abstract:** We study the statistical distributions of the spins of generic black-hole binaries during the inspiral and merger, as well as the distributions of the remnant mass, spin, and recoil velocity. For the inspiral regime, we start with a random uniform distribution of spin directions S1 and S2 and magnitudes S1=S2=0.97 for different mass ratios. Starting from a fiducial initial separation of ri=50m, we perform 3.5PN evolutions down to rf=5m. At this final fiducial separation, we compute the angular distribution of the spins with respect to the final orbital angular momentum, L. We perform $16^4$ simulations for six mass ratios between q=1 and q=1/16 and compute the distribution of the angles between L and Delta and L and S, directly related to recoil velocities and total angular momentum. We find a small but statistically significant bias of the distribution towards counter-alignment of both scalar products. To study the merger of black-hole binaries, we turn to full numerical techniques. We introduce empirical formulae to describe the final remnant black hole mass, spin, and recoil velocity for merging black-hole binaries with arbitrary mass ratios and spins. We then evaluate those formulae for randomly chosen directions of the individual spins and magnitudes as well as the binary's mass ratio. We found that the magnitude of the recoil velocity distribution decays as $P(v) \exp{-v/2500 km/s}$, =630km/s, and sqrt - $\hat{2}$= 534km/s, leading to a 23% probability of recoils larger than 1000km/s, and a highly peaked angular distribution along the final orbital axis. The final black-hole spin magnitude show a universal distribution highly peaked at $Sf/mf^2 = 0.73$ and a 25 degrees misalignment with respect to the final orbital angular momentum.

## Post-Newtonian methods: Analytic results on the binary problem

**Authors:** Schaefer, Gerhard

**Eprint:** <http://arxiv.org/abs/0910.2857>

**Keywords:** gr-qc; massive binaries of black holes; post-Newtonian theory; spin

**Abstract:** A detailed account is given on approximation schemes to the Einstein theory of general relativity where the iteration starts from the Newton theory of gravity. Two different coordinate conditions are used to represent the Einstein field equations, the generalized isotropic ones of the canonical formalism of Arnowitt, Deser, and Misner and the harmonic ones of the Lorentz-covariant Fock-de Donder approach. Conserved quantities of isolated systems are identified and the Poincaré algebra is introduced. Post-Newtonian expansions are performed in the near and far (radiation) zones. The natural fitting of multipole expansions to post-Newtonian schemes is emphasized. The treated matter models are ideal fluids, pure point masses, and point masses with spin and mass-quadrupole moments modelling rotating black holes. Various Hamiltonians of spinning binaries are presented in explicit forms to higher post-Newtonian orders. The delicate use of black holes in post-Newtonian expansion calculations and of the Dirac delta function in general relativity find discussions.

## The evolution of Black Hole scaling relations in galaxy mergers

**Authors:** Johansson, Peter H.; Burkert, Andreas; Naab, Thorsten

**Eprint:** http://arxiv.org/abs/0910.2232

**Keywords:** astro-ph.CO; astrophysics; cosmology; supermassive black holes

**Abstract:** We study the evolution of black holes (BHs) on the $M_{BH} - \sigma$ and $M_{BH} - M_{bulge}$ planes as a function of time in disk galaxies undergoing mergers. We begin the simulations with the progenitor black hole masses being initially below (Delta log $M_{BH}$=-2), on (Delta log $M_{BH}$=0) and above (Delta log $M_{BH}$=0.5) the observed local relations. The final relations are rapidly established after the final coalescense of the galaxies and their BHs. Progenitors with low initial gas fractions ($f_{gas}$=0.2) starting below the relations evolve onto the relations (Delta log $M_{BH}$=-0.18), progenitors on the relations stay there (Delta log $M_{BH}$=0) and finally progenitors above the relations evolve towards the relations, but still remaining above them (Delta log $M_{BH}$=0.35). Mergers in which the progenitors have high initial gas fractions ($f_{gas}$=0.8) evolve above the relations in all cases (Delta log $M_{BH}$=0.5). We find that the initial gas fraction is the prime source of scatter in the observed relations, dominating over the scatter arising from the evolutionary stage of the merger remnants. The fact that BHs starting above the relations do not evolve onto the relations, indicates that our simulations rule out the scenario in which overmassive BHs evolve onto the relations through gas-rich mergers. By implication our simulations thus disfavor the picture in which supermassive BHs develop significantly before their parent bulges.

## Triplets of supermassive black holes: Astrophysics, Gravitational Waves and Detection

**Authors:** Amaro-Seoane, Pau; Sesana, Alberto; Hoffman, Loren; Benacquista, Matthew; Eichhorn, Christoph; Makino, Junichiro; Spurzem, Rainer

**Eprint:** http://arxiv.org/abs/0910.1587

**Keywords:** astro-ph.CO; astro-ph.GA; astrophysics; bursts; cosmology; GRAPE hw; N-body; stellar dynamics; supermassive black holes

**Abstract:** Supermassive black holes (SMBHs) found in the centers of many galaxies have been recognized to play a fundamental active role in the cosmological structure formation process. In hierarchical formation scenarios, SMBHs are expected to form binaries following the merger of their host galaxies. If these binaries do not coalesce before the merger with a third galaxy, the formation of a black hole triple system is possible. Numerical simulations of the dynamics of triples within galaxy cores exhibit phases of very high eccentricity (as high as $e \sim 0.99$). During these phases, intense bursts of gravitational radiation can be emitted at orbital periapsis. This produces a gravitational wave signal at frequencies substantially higher than the orbital frequency. The likelihood of detection of these bursts with pulsar timing and the Laser Interferometer Space Antenna (*LISA*) is estimated using several population models of SMBHs with masses $\gtrsim 10^7 M_\odot$. Assuming a fraction of binaries $\geq 0.1$ in triple system, we find that few to few dozens of these bursts will produce residuals $> 1$ ns, within the sensitivity range of forthcoming pulsar timing arrays (PTAs). However, most of such bursts will be washed out in the underlying confusion noise produced by all the other 'standard' SMBH binaries emitting in the same frequency window. A detailed data analysis study would be required to assess resolvability of such sources. Implementing a basic resolvability criterion, we find that the chance of catching a resolvable burst at a one nanosecond precision level is 2-50%, depending on the adopted SMBH evolution model. On the other hand, the probability of detecting bursts produced by massive binaries (masses $\gtrsim 10^7 M_\odot$) with *LISA* is negligible.

## Estimating Black Hole Masses in Triaxial Galaxies

**Authors:** Bosch, Remco C. E. van den; de Zeeuw, P. Tim

**Eprint:** http://arxiv.org/abs/0910.0844

**Keywords:** astro-ph.CO; astro-ph.GA; astrophysics; supermassive black holes

How stars distribute around a massive black hole

**Abstract:** Most of the super massive black hole mass estimates based on stellar kinematics use the assumption that galaxies are axisymmetric oblate spheroids or spherical. Here we use fully general triaxial orbit-based models to explore the effect of relaxing the axisymmetric assumption on the previously studied galaxies M32 and NGC 3379. We find that M32 can only be modeled accurately using an axisymmetric shape viewed nearly edge-on and our black hole mass estimate is identical to previous studies. When the observed 5 degrees kinematical twist is included in our model of NGC 3379, the best shape is mildly triaxial and we find that our best-fitting black hole mass estimate doubles with respect to the axisymmetric model. This particular black hole mass estimate is still within the errors of that of the axisymmetric model and consistent with the M-sigma relationship. However, this effect may have a pronounced impact on black hole demography, since roughly a third of the most massive galaxies are strongly triaxial.

## Black Hole Growth and Starburst Activity at z=0.6-4 in the Chandra Deep Field South

**Authors:** Brusa, M.; Fiore, F.; Santini, P.; Grazian, A.; Comastri, A.; Zamorani, G.; Hasinger, G.; Merloni, A.; Civano, F.; Fontana, A.; Mainieri, V.

**Abstract:** The co-evolution of host galaxies and the active black holes which reside in their centre is one of the most important topics in modern observational cosmology. Here we present a study of the properties of obscured Active Galactic Nuclei (AGN) detected in the CDFS 1Ms observation and their host galaxies. We limited the analysis to the MUSIC area, for which deep K-band observations obtained with ISAAC@VLT are available, ensuring accurate identifications of the counterparts of the X-ray sources as well as reliable determination of photometric redshifts and galaxy parameters, such as stellar masses and star formation rates. In particular, we: 1) refined the X-ray/infrared/optical association of 179 sources in the MUSIC area detected in the Chandra observation; 2) studied the host galaxies observed and rest frame colors and properties. We found that X-ray selected ($L_X > 10^{42} ergs^{-1}$) AGN show Spitzer colors consistent with both AGN and starburst dominated infrared continuum; the latter would not have been selected as AGN from infrared diagnostics. The host galaxies of X-ray selected obscured AGN are all massive ($M_* > 10^{10} M_\odot$) and, in 50% of the cases, are also actively forming stars (1/SSFR1 and $M_* > 3x10^{11} M_\odot$, a fraction significantly higher than in the local Universe for AGN of similar luminosities.

## Advanced drag-free concepts for future space-based interferometers: acceleration noise performance

**Authors:** Gerardi, D.; Allen, G.; Conklin, J. W.; Sun, K-X.; DeBra, D.; Buchman, S.; Gath, P.; Fichter, W.; Byer, R. L.; Johann, U.

**Eprint:** http://arxiv.org/abs/0910.0758

**Keywords:** gr-qc; instruments; interferometers

**Abstract:** Future drag-free missions for space-based experiments in gravitational physics require a Gravitational Reference Sensor with extremely demanding sensing and disturbance reduction requirements. A configuration with two cubical sensors is the current baseline for the Laser Interferometer Space Antenna (LISA) and has reached a high level of maturity. Nevertheless, several promising concepts have been proposed with potential applications beyond LISA and are currently investigated at HEPL, Stanford, and EADS Astrium, Germany. The general motivation is to exploit the possibility of achieving improved disturbance reduction, and ultimately understand how low acceleration noise can be pushed with a realistic design for future mission. In this paper, we discuss disturbance reduction requirements for LISA and beyond, describe four different payload concepts, compare expected strain sensitivities in the 'low-frequency' region of the frequency spectrum, dominated by acceleration noise, and ultimately discuss advantages and disadvantages of each of those concepts in achieving disturbance reduction for space-based detectors beyond LISA.

## Compact Binaries in Star Clusters I - Black Hole Binaries Inside Globular Clusters

**Authors:** Downing, J. M. B.; Benacquista, M. J.; Giersz, M.; Spurzem, R.

**Eprint:** http://arxiv.org/abs/0910.0546

**Keywords:** astro-ph.HE; astro-ph.SR; astrophysics; globular clusters; stellar dynamics

**Abstract:** We study the compact binary population in star clusters, focusing on binaries containing neutron stars and black holes, using a self-consistent Monte Carlo treatment of dynamics and full stellar evolution. We find that the black holes experience strong mass segregation and become centrally concentrated. In the core the black holes interact strongly with each other and black hole-black hole binaries are formed very efficiently. The strong interactions, however, also destroy or eject the black hole-black hole binaries. We find no black hole-black hole mergers within our simulations but produce many hard escapers that will merge in the galactic field within a Hubble time. We also find two highly eccentric black hole-black hole binaries that are potential LISA sources, suggesting that star clusters are interesting targets for space-based detectors. We conclude that star clusters must be taken into account when predicting compact binary population statistics.

### Soft gamma-ray constraints on a bright flare from the Galactic Center supermassive black hole

**Authors:** Trap, G.; Goldwurm, A.; Terrier, R.; Dodds-Eden, K.; Gillessen, S.; Genzel, R.; Pantin, E.; Lagage, P. O.; Ferrando, P.; Belanger, G.; Porquet, D.; Grosso, N.; Yusef-Zadeh, F.; Melia, F.

**Eprint:** http://arxiv.org/abs/0910.0399

**Keywords:** astro-ph.HE; astrophysics; observations; Sagittarius A*; supermassive black holes

**Abstract:** Sagittarius A* (Sgr A*) is the supermassive black hole residing at the center of the Milky Way. It has been the main target of an extensive multiwavelength campaign we carried out in April 2007. Herein, we report the detection of a bright flare from the vicinity of the horizon, observed simultaneously in X-rays (XMM/EPIC) and near infrared (VLT/NACO) on April 4th for 1-2 h. For the first time, such an event also benefitted from a soft gamma-rays (INTEGRAL/ISGRI) and mid infrared (VLT/VISIR) coverage, which enabled us to derive upper limits at both ends of the flare spectral energy distribution (SED). We discuss the physical implications of the contemporaneous light curves as well as the SED, in terms of synchrotron, synchrotron self-Compton and external Compton emission processes.

### Mass of black holes: The State of the Art

**Authors:** Czerny, B.; Nikolajuk, M.

**Eprint:** http://arxiv.org/abs/0910.0313

**Keywords:** astro-ph.HE; astrophysics; intermediate-mass black holes; observations; supermassive black holes

**Abstract:** In this small review we present the actual state the knowledge about weighting black holes. Black holes can be found in stellar binary systems in our Galaxy and in other nearby galaxies, in globular clusters, which we can see in our and nearby galaxies, and in centres of all well-developed galaxies. Range of values of their masses is wide and cover about ten orders of magnitude (not taking into account the hypothetic primordial black holes). Establishing the presence of black holes, and in particular the measurement of their mass is one on the key issues for many branches of astronomy, from stellar evolution to cosmology.

## Detection of IMBHs with ground-based gravitational wave observatories: A biography of a binary of black holes, from birth to death

**Authors:** Amaro-Seoane, Pau; Santamaria, Lucia

**Eprint:** http://arxiv.org/abs/0910.0254

**Keywords:** astro-ph.CO; astro-ph.GA; astrophysics; globular clusters; GRAPE hw; intermediate-mass black holes; massive binaries of black holes; N-body; parameter estimation; stellar dynamics; waveforms

**Abstract:** Even though the existence of intermediate-mass black holes has not yet been corroborated observationally, these objects are of high interest for astrophysics. Our understanding of formation and evolution of supermassive black holes (SMBHs), as well as galaxy evolution modeling and cosmography would dramatically change if an IMBH was observed. The prospect of detection and, possibly, observation and characterization of an IMBH has good chances in lower-frequency gravitational-wave (GW) astrophysics with ground-based detectors such as LIGO, Virgo and the future Einstein Telescope (ET). We present an analysis of the signal of a system of a binary of IMBHs based on a waveform model obtained with numerical relativity simulations coupled with post-Newtonian calculations at the highest available order so as to extend the waveform to lower frequencies. We find that initial LIGO and Virgo are in the position of detecting IMBHs with a signal-to-noise ratio (SNR) of $\sim$ 10 for systems with total mass between 100 and $500 M_\odot$ situated at a distance of 100 Mpc. Nevertheless, the event rate is too low and the possibility that these signals are mistaken with a glitch is, unfortunately, non-negligible. When going to second- and third-generation detectors, such as Advanced LIGO or the proposed ET, the event rate becomes much more promising (tens per year for the first and thousands per year for the latter) and the SNR at 100 Mpc is as high as $100 - 1000$ and $1000 - 10^5$ respectively. The prospects for IMBH detection and characterization with ground-based GW observatories would not only provide us with a robust test of general relativity, but would also corroborate the existence of these systems. Such detections would be a probe to the stellar environments of IMBHs and their formation.

## Post-Newtonian and Numerical Calculations of the Gravitational Self-Force for Circular Orbits in the Schwarzschild Geometry

**Authors:** Blanchet, Luc; Detweiler, Steven; Tiec, Alexandre Le; Whiting, Bernard F.

**Eprint:** http://arxiv.org/abs/0910.0207

**Keywords:** general relativity; gr-qc; post-Newtonian theory; self force

**Abstract:** The problem of a compact binary system whose components move on circular orbits is addressed using two different approximation techniques in general relativity. The post-Newtonian (PN) approximation involves an expansion in powers of v/c«1, and is most appropriate for small orbital velocities v. The perturbative self-force (SF) analysis requires an extreme mass ratio m1/m2«1 for the components of the binary. A particular coordinate-invariant observable is determined as a function of the orbital frequency of the system using these two different approximations. The post-Newtonian calculation is pushed up to the third post-Newtonian (3PN) order. It involves the metric generated by two point particles and evaluated at the location of one of the particles. We regularize the divergent self-field of the particle by means of dimensional regularization. We show that the poles proportional to 1/(d-3) appearing in dimensional regularization at the 3PN order cancel out from the final gauge invariant observable. The 3PN analytical result, through first order in the mass ratio, and the numerical SF calculation are found to agree well. The consistency of this cross cultural comparison confirms the soundness of both approximations in describing compact binary systems. In particular, it provides an independent test of the very different regularization procedures invoked in the two approximation schemes.

## Hydrodynamical Response of a Circumbinary Gas Disk to Black Hole Recoil and Mass Loss

**Authors:** Corrales, Lia R.; Haiman, Zoltán; MacFadyen, Andrew

**Eprint:** http://arxiv.org/abs/0910.0014

**Keywords:** accretion discs; astro-ph.CO; astro-ph.HE; astrophysics; EM counterparts; kicks/recoil; massive binaries of black holes; supermassive black holes

**Abstract:** Finding electromagnetic (EM) counterparts of future gravitational wave (GW) sources would bring rich scientific benefits. A promising possibility, in the case of the coalescence of a super-massive black hole binary (SMBHB), is that prompt emission from merger-induced disturbances in a supersonic circumbinary disk may be detectable. We follow the post-merger evolution of a thin, zero-viscosity circumbinary gas disk with two-dimensional simulations, using the hydrodynamic code FLASH. We analyze perturbations arising from the 530 km/s recoil of a $10^6 M_\odot$ binary, oriented in the plane of the disk, assuming either an adiabatic or a pseudo-isothermal equation of state for the gas. We find that a single-armed spiral shock wave forms and propagates outward, sweeping up about 20% of the mass of the disk. The morphology and evolution of the perturbations agrees well with those of caustics predicted to occur in a collisionless disk. Assuming that the disk radiates nearly instantaneously to maintain a constant temperature, we estimate the amount of dissipation and corresponding post-merger light-curve. The luminosity rises steadily on the time-scale of months, and reaches few times $10^{43}$ erg/s, corresponding to about 10% of the Eddington luminosity of the central SMBHB. We also analyze the case in which gravitational wave emission results in a 5% mass loss in the merger remnant. The mass-loss reduces the shock overdensities and the overall luminosity of the disk by 15-20%, without any other major effects on the spiral shock pattern.

## Black hole mergers: the first light

**Authors:** Rossi, Elena M.; Lodato, G.; Armitage, P. J.; Pringle, J. E.; King, A. R.

**Eprint:** <http://arxiv.org/abs/0910.0002>

**Keywords:** astro-ph.GA; astro-ph.HE; astrophysics; EM counterparts; kicks/recoil; supermassive black holes

**Abstract:** The coalescence of supermassive black hole binaries occurs via the emission of gravitational waves, that can impart a substantial recoil to the merged black hole. We consider the energy dissipation, that results if the recoiling black hole is surrounded by a thin circumbinary disc. Our results differ significantly from those of previous investigations. We show analytically that the dominant source of energy is often potential energy, released as gas in the outer disc attempts to circularize at smaller radii. Thus, dimensional estimates, that include only the kinetic energy gained by the disc gas, underestimate the real energy loss. This underestimate can exceed an order of magnitude, if the recoil is directed close to the disc plane. We use three dimensional Smooth Particle Hydrodynamics (SPH) simulations and two dimensional finite difference simulations to verify our analytic estimates. We also compute the bolometric light curve, which is found to vary strongly depending upon the kick angle. A prompt emission signature due to this mechanism may be observable for low mass ($10^6$ solar mass) black holes whose recoil velocities exceed about 1000 km/s. Emission at earlier times can mainly result from the response of the disc to the loss of mass, as the black holes merge. We derive analytically the condition for this to happen.

### Self-force and motion of stars around black holes

**Authors:** Spallicci, A.; Aoudia, S.

**Eprint:** http://arxiv.org/abs/0909.5558

**Keywords:** astro-ph.HE; EMRI; general relativity; geodesic motion; gr-qc; hep-th; physics.hist-ph; self force

**Abstract:** Through detection by low gravitational wave space interferometers, the capture of stars by supermassive black holes will constitute a giant step forward in the understanding of gravitation in strong field. The impact of the perturbations on the motion of the star is computed via the tail, the back-scattered part of the perturbations, or via a radiative Green function. In the former approach, the self-force acts upon the background geodesic, while in the latter, the geodesic is conceived in the total (background plus perturbations) field. Regularisations (mode-sum and Riemann-Hurwitz $\zeta$ function) intervene to cancel divergencies coming from the infinitesimal size of the particle. The non-adiabatic trajectories require the most sophisticated techniques for studying the evolution of the motion, like the self-consistent approach.

### Evidence for two populations of Galactic globular clusters from the ratio of their half-mass to Jacobi radii

**Authors:** Baumgardt, Holger; Parmentier, Genevieve; Gieles, Mark; Vesperini, Enrico

**Eprint:** http://arxiv.org/abs/0909.5696

**Keywords:** astro-ph.GA; astrophysics; globular clusters; stellar dynamics

**Abstract:** We investigate the ratio between the half-mass radii $r_h$ of Galactic globular clusters and their Jacobi radii $r_J$ given by the potential of the Milky Way and show that clusters with galactocentric distances $R_{GC} > 8$ kpc fall into two distinct groups: one group of compact, tidally-underfilling clusters with $r_h/r_J < 0.05$ and another group of tidally filling clusters which have $0.1 < r_h/r_J < 0.3$. We find no correlation between the membership of a particular cluster to one of these groups and its membership in the old or younger halo population. Based on the relaxation times and orbits of the clusters, we argue that compact clusters and most clusters in the inner Milky Way were born compact with half-mass radii $r_h < 1$ pc. Some of the tidally-filling clusters might have formed compact as well, but the majority likely formed with large half-mass radii. Galactic globular clusters therefore show a similar dichotomy as was recently found for globular clusters in dwarf galaxies and for young star clusters in the Milky Way. It seems likely that some of the tidally-filling clusters are evolving along the main sequence line of clusters recently discovered by Kuepper et al. (2008) and are in the process of dissolution.

> **Intention and purpose of GW Notes**
>
> A succinct explanation

The electronic publishing service **arXiv** is a dynamic, well-respected source of news of recent work and is updated daily. But, perhaps due to the large volume of new work submitted, it is probable that a member of our community might easily overlook relevant material. This new e-journal and its blog, **The LISA Brownbag (http://www.lisa-science.org/brownbag)**, both produced by the AEI, propose to offer scientist of the Gravitational Wave community the opportunity to more easily follow advances in the three areas mentioned: Astrophysics, General Relativity and Data Analysis. We hope to achieve this by selecting the most significant e-prints and list them in abstract form with a link to the full paper in both a single e-journal (GW Newsletter) and a blog (The LISA Brownbag). Of course, *this also implies that the paper will have its impact increased, since it will reach a broader public*, so that we encourage you to not forget submitting your own work

In addition to the abstracts, in each PDF issue of GW Notes, we will offer you a previously unpublished article written by a senior researcher in one of these three domains, which addresses the interests of all readers.

Thus the aim of The LISA Brownbag and GW Notes is twofold:

- Whenever you see an interesting paper on GWs science and LISA, you can submit the **arXiv** number to our **submission page (http://brownbag.lisascience.org)**. This is straightforward: No registration is required (although recommended) to simply type in the number in the entry field of the page, indicate some keywords and that's it

- We will publish a new full article in each issue, if available. This "feature article" will be from the fields of Astrophysics, General Relativity or the Data Analysis of gravitational waves and LISA. We will prepare a more detailed guide for authors, but for now would like to simply remind submitters that they are writing for colleagues in closely related but not identical fields, and that cross-fertilization and collaboration is an important goal of our concept

Subscribers get the issue distributed in PDF form. Additionally, they will be able to submit special announcements, such as meetings, workshops and jobs openings, to the list of registered people. For this, please register at the **registration page (http://lists.aei.mpg.de/cgi-bin/mailman/listinfo/lisa_brownbag)** by filling in your e-mail address and choosing a password.

> ### *The Astro-GR meetings*
> *Past, present and future*

Sixty two scientists attended the **Astro-GR@AEI** meeting, which took place September 18-22 2006 at the **Max-Planck Institut für Gravitationsphysik (Albert Einstein-Institut)** in Golm, Germany. The meeting was the brainchild of an AEI postdoc, who had the vision of bringing together Astrophysicists and experts in General Relativity and gravitational-wave Data Analysis to discuss sources for **LISA**, the planned Laser Interferometer Space Antenna. More specifically, the main topics were EMRIs and IMRIs (Extreme and Intermediate Mass-Ratio Inspiral events), i.e. captures of stellar-mass compact objects by supermassive black holes and coalescence of intermediate-mass black holes with supermassive black holes.

The general consensus was that the meeting was both interesting and quite stimulating. It was generally agreed that someone should step up and host a second round of this meeting. Monica Colpi kindly did so and this led to **Astro-GR@Como**, which was very similar in its informal format, though with a focus on all sources, meant to trigger new ideas, as a kind of brainstroming meeting.

Also, in the same year, in the two first weeks of September, we had another workshop in the Astro-GR series with a new "flavour", namely, the **Two Weeks At The AEI (2W@AEI)**, in which the interaction between the attendees was be even higher than what was reached in the previous meetings. To this end, we reduced the number of talks, allowing participants more opportunity to collaborate. Moreover, participants got office facilities and we combined the regular talks with the so-called "powerpointless" seminar, which will were totally informal and open-ended, on a blackboard. The next one was held in Barcelona in 2009 at the beginning of September, **Astro-GR@BCN** and next 2010 it will be the turn of Paris, at the APC.

**LISA Astro-GR@Paris (Paris, APC Monday 13th to Friday September 17th 2010)**

If you are interested in hosting in the future an Astro-GR meeting, please contact us. We are open to new formats, as long as the *Five Golden Rules* are respected.

A proper Astro-GR meeting **MUST** closely follow the *Five Golden Rules*:

*I.* Bring together Astrophysicists, Cosmologists, Relativists and Data Analysts

*II.* Motivate new collaborations and projects

*III.* Be run in the style of Aspen, ITP, Newton Institute and Modest meetings, with plenty of time for discussions

How stars distribute around a massive black hole

||||. Grant access to the slides in a cross-platform format, such as PDF and, within reason, to the recorded movies of the talks in a free format which everybody can play like **Theora**, for those who could not attend, following the good principles of **Open Access**

卌. Keep It Simple and... Spontaneous

How stars distribute around a massive black hole